\documentclass[acmsmall]{acmart}
\AtBeginDocument{%
  }

\setcopyright{acmlicensed}
\copyrightyear{2018}
\acmYear{2018}
\acmDOI{XXXXXXX.XXXXXXX}
\acmConference[Conference acronym 'XX]{Make sure to enter the correct
  conference title from your rights confirmation email}{June 03--05,
  2018}{Woodstock, NY}
\acmISBN{978-1-4503-XXXX-X/2018/06}

\usepackage{subcaption}
\usepackage{multirow}
\usepackage{soul}
\usepackage{xcolor}

\newcommand{\revision}[1]{\textcolor{black}{#1}}

\definecolor{darkblue}{rgb}{0, 0, 0.5}
\hypersetup{colorlinks=true, citecolor=darkblue, linkcolor=darkblue, urlcolor=darkblue}
\begin{document}

\title{Designing and Evaluating \revision{Chain-of-Hints for Scientific Question Answering}}
% \title{Designing and Evaluating Hint Generation Systems for Science Education}
%SM I added s to Systems

\author{Anubhav Jangra}
\email{anubhav@cs.columbia.edu}
\orcid{0000-0001-5571-6098}
\affiliation{
  \institution{Columbia University}
  \city{New York}
  \state{New York}
  \country{USA}
}

\author{Smaranda Muresan}
\email{smara@columbia.edu}
\affiliation{
  \department{Barnard College}
  \institution{Columbia University}
  \city{New York}
  \state{New York}
  \country{USA}
}

% %%
% %% By default, the full list of authors will be used in the page
% %% headers. Often, this list is too long, and will overlap
% %% other information printed in the page headers. This command allows
% %% the author to define a more concise list
% %% of authors' names for this purpose.
% \renewcommand{\shortauthors}{Jangra and Muresan}

\begin{abstract}
    % LLMs are influencing the education landscape, with students relying on them in their learning process. 
    LLMs are reshaping education, with students increasingly relying on them for learning.
    Implemented using general-purpose models, these systems are likely to give away the answers, potentially undermining conceptual understanding and critical thinking.
    Prior work shows that hints can effectively promote cognitive engagement. Building on this insight, we evaluate 18 open-source LLMs on chain-of-hints generation that scaffold users toward the correct answer.
    We compare two distinct hinting strategies: \textit{static} hints, pre-generated for each problem, and \textit{dynamic} hints, adapted to a learners' progress. We evaluate these systems on five pedagogically grounded automatic metrics for hint quality. Using the best performing LLM as the backbone of a quantitative study with 41 participants, we uncover distinct user preferences across hinting strategies, and identify the limitations of automatic evaluation metrics to capture them. Our findings highlight key design considerations for future research on tutoring systems and contribute toward the development of more learner-centered educational technologies.

\end{abstract}



  

\maketitle

\section{Introduction}

Large language models (LLMs) have reshaped how students interact with digital learning tools. These models are increasingly used for everyday study practices, from gaining conceptual understanding to supporting problem-solving and writing \cite{levine2025students, sublime2024chatgpt, rahma2024potential, syarifah2024exploring}. At the same time, online learning continues to expand at unprecedented scales, with at least 220 million students being enrolled in massive open online courses in 2021 \cite{antoninis2023global}. % According to UNESCO’s Global Education Monitoring Report (2023) \cite{antoninis2023global}, the number of students enrolled in massive open online courses reached at least 220 million in 2021, reflecting both the demand for and the accessibility of scalable educational resources. 
While prior work in education research has established a strong correlation between the student-teacher ratio and learning outcomes \cite{koc2015impact}, teacher recruitment and retention crisis remains a worldwide conundrum \cite{huang2022struggle, williams2022teacher, castro2022tensions}. % ``\textit{Even where general teacher supply and demand are in balance, many countries face shortages of specialist teachers and shortages in schools serving disadvantaged or isolated communities.}" \cite{repec:nos:voprob:2012:i:1:p:74-92}. 
These constraints create a growing reliance on automated support to bridge the gap in access to personalized instruction. 
%, \revision{motivating research to design automated tutors for support.}
% Yet, the availability of expert tutors has not kept pace \todo{find reference}. 
However, despite their rapid adoption, many educational technologies are deployed with limited evaluation before reaching learners, and most evidence of the impact of education technology comes from the richest countries. %\citet{antoninis2023global} reports that 
According to UNESCO’s Global Education Monitoring Report (2023) \cite{antoninis2023global},
in the United Kingdom, only 7\% of education technology companies had conducted randomized controlled trials, while a survey of teaching staff in 17 states within the United States revealed that only 11\% requested peer-reviewed evidence prior to adoption. % This context underscores the need for research that not only develops new forms of automated support, but also studies their effectiveness, limitations, and design trade-offs. 
% The issue is particularly acute in higher education, where students increasingly experiment with LLMs in self-directed ways, yet universities and platforms lack systematic frameworks to assess the pedagogical value of the feedback these models provide. 

\begin{figure*}[t!]
\centering
\includegraphics[width=\textwidth]{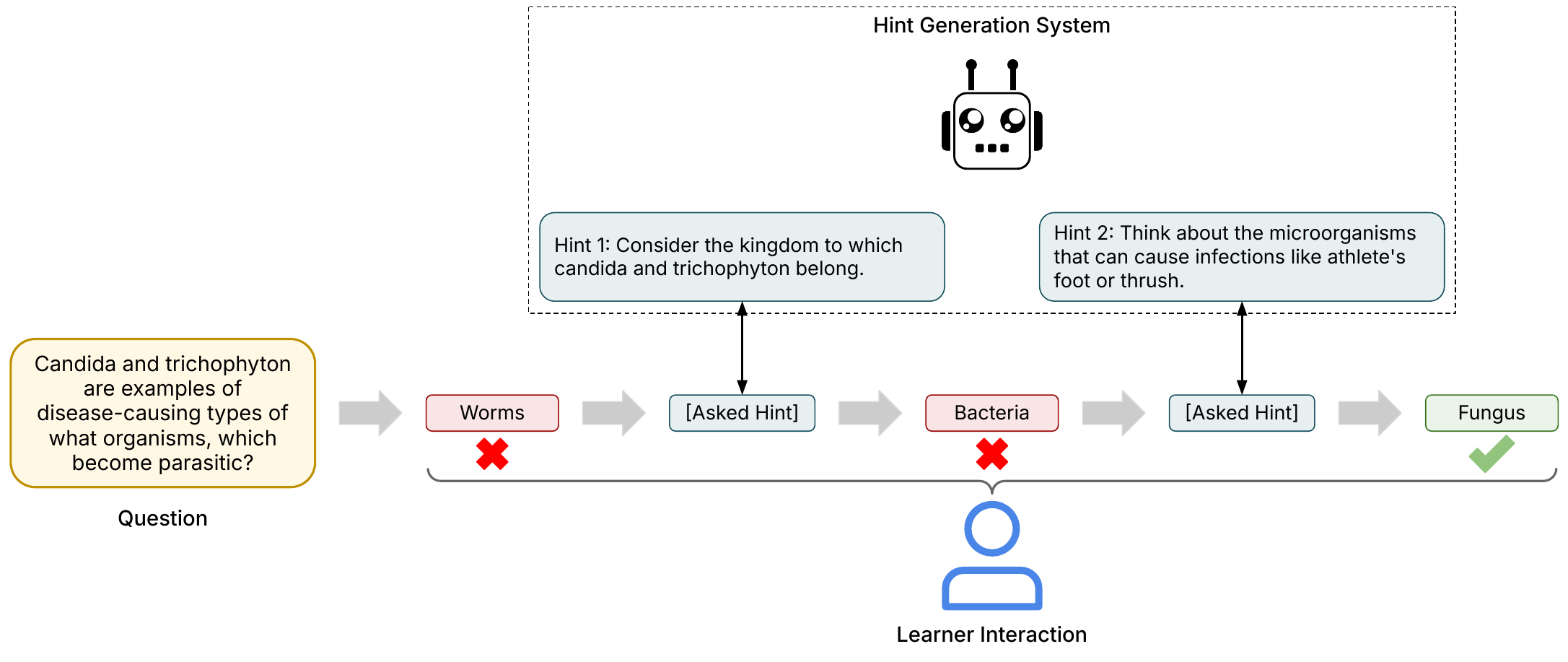}
\caption{Illustration of automatic chain-of-hint generation for scientific question answering \revision{(example interaction from the user study)}. \revision{After the learner’s initial incorrect response (“Worms”), the system provided a first hint directing attention to the kingdom (worms are not a biological kingdom) to which \textit{Candida} and \textit{Trichophyton} belong. When the learner then answered “Bacteria,” the system generated a second hint linking the organisms to familiar infections (e.g., athlete’s foot, thrush) without revealing the answer. Using this scaffolded guidance, the participant arrived at the correct response (“Fungus”), bridging the gap in the question and their prior knowledge.}} \label{fig:intro}
\end{figure*}

% Together, these trends highlight a growing dependence on automated support systems whose effectiveness and risks remain insufficiently understood.
As LLMs become embedded in everyday study practices, there is a pressing need to evaluate not only what they generate for learners, but also how these outputs are assessed and integrated into instructional workflows \cite{chu2025llm, alhafni2024llms, yan2024practical, abd2023large}. However, the majority of research has centered on structured domains such as programming and mathematics, where correctness can be precisely defined and evaluation is straightforward \cite{jangra2024navigating}. In contrast, relatively little attention has been paid to unstructured problem spaces, with open-ended hint and solution space. 
To address this gap, our work focuses on hint generation systems for scientific question answering, specifically designing hints for factoid questions as the first step towards scaffolding learners to enhance understanding and critical thinking in science domains (see Figure \ref{fig:intro}). While factoid questions have well-defined answers that make evaluation tractable, the space of effective hints remains open-ended and diverse, requiring models to provide meaningful conceptual scaffolding. As such, factoid science questions offer a practical yet challenging testbed, bridging structured evaluation with realistic tutoring-style hint generation. 

In this paper, we introduce the concept of a \textit{chain-of-hints}, where information is revealed incrementally (Section \ref{subsec:prob_def}). % to scaffold the learner’s reasoning process \smara{here it can again lead to the expectation we will test how the chain of hints scaffold the reasoning process of the learner, basically asking about learning outcomes, while the things we do is perceived helpfulness. I would cut "to scaffold the learner’s reasoning process ", since you then ground this in the tutoring literature in next sentence}. 
% \smara{this is an important part and Fig 1 does not show the chain of hints}
% problem-solving process. %SM I guess QA can be a type of limited problem solving. 
Inspired by the principles of human tutoring, this approach mirrors how teachers provide just-in-time interventions — offering minimal guidance first, and providing progressively more explicit support as needed \cite{wood1976role}. We systematically compare two %different 
strategies for generating and delivering these hints: %contrasting 
pre-generated \textit{static} hints %with 
and adaptive \textit{dynamic} hints that respond to learner's progress.
% \smara{predeterimed to me sounds like hand-crafted, not generated by a model}
Through this lens, we investigate how learners perceive the helpfulness of different hinting strategies, and derive design principles that can guide the development of pedagogically effective educational systems. We ground our study in three research questions: \textbf{RQ1:} \revision{What characteristics of LLM-generated hint sequences do learners perceive as helpful or unhelpful during scientific question answering?} \textbf{RQ2:} How do learners perceive adaptive (dynamic) versus pre-generated (static) chain-of-hints? \textbf{RQ3:} To what extent do automatic evaluation metrics align with learners’ perceptions of hint quality?

% We draw on Bloom's taxonomy of educational goals \cite{bloom1956taxonomy, anderson2001taxonomy} to discuss the design considerations grounded on the findings from our user study. We express several limitations of using large language models 

% Drawing from our findings from the user study, we propose several design considerations for future work. Using 
% as a north-star direction of future tutoring systems, we present several limitations in existing LLM-driven technologies, and propose towards learner-centered development 

% \todo{Para: Consolidate findings. Focus on 1. Learner's diverging preferences, 2. misaligned automatic metrics, 3. Helped increase understanding/providing context, 4. Unhelpful when lacking familiarity, 5. Preferred diversity in hint chains, 6. with incremental scaffolding.}
In order to answer these questions, we first conducted a comprehensive evaluation of 18 open-source LLMs (Section \ref{subsec:baselines}) using five pedagogically grounded automatic evaluation metrics: \textit{information gain}, \textit{redundancy}, \textit{consistency}, \textit{readability}, and \textit{leakage} (Section \ref{subsec:auto_eval_tldr}). We observed substantial variance across model families and sizes, indicating that current LLMs are not uniformly capable of generating high-quality pedagogical feedback. We investigated the dependencies among metrics and observed notable trade-offs, including between \textit{information gain} and \textit{leakage} (Section \ref{subsec:auto_eval_results}). % We further analyzed the interplay of metrics, observing trade-offs between several metrics like \textit{information gain} and \textit{leakage}.
% Building on this assessment \smara{maybe make it stronger here, you did a very thorough evaluation here, including computing Pareto-optimal fronts to select model. This is an experimental design that showcase how to select models to then be used in user studies, people will find this important and otherwise it gets lost.}, 
We selected the best-performing model at the Pareto-optimal front of the metric space
for an exploratory study with 41 participants\footnote{We intentionally recruited recent graduates rather than current high-school students to mitigate potential risks associated with exposing minors to experimental AI-based instructional systems \cite{yu2025understanding}. This study design choice may limit direct generalizability to in-school population (see Section \ref{sec:limitations}).} (Section \ref{sec:study_design}). % \footnote{For the purposes of this study, we intentionally recruited recent high-school graduates instead of current high-school students. Prior work has shown that exposing young learners to experimental technology can have detrimental effects for their learning \cite{}. We acknowledge that may impact the applicability of our findings. Please see limitations section.}. 
Quantitative results showed that access to hints enabled participants to answer approximately $\sim$22\% more questions compared to a control baseline, with roughly $\sim$27\% of hints rated as informative (Section \ref{subsec:finding1}). Qualitative feedback revealed that participants particularly valued hints that clarified underlying concepts, provided additional context, and offered incremental guidance (Section \ref{subsec:finding2}).
However, their preferences diverged over certain characteristics of \textit{static} and \textit{dynamic} hints, static hints were valued for their contextual richness, whereas dynamic hints were useful for their answer-directed efficiency (Section \ref{subsec:finding3}). 
% \aj{What kind of divergence...}
% We interpret these differences through the lens of cognitive offloading (Section \ref{subsec:cognitive_offloading}). 
Additionally, we highlight critical limitations of the automatic evaluation metrics, that only partially aligned with the participants' perceptions of hint helpfulness (Section \ref{subsec:autoeval_analysis}).
Building on these insights, we discuss design implications for future hint generation and tutoring systems, applicable beyond factoid question answering. 
% We call for learner-centered design practices that identify and account for diverse needs and preferences, rather than using %off-the-shelf 
% general-purpose models. 
We advocate for tutoring frameworks that integrate self-regulated learning with gradually faded support, enabling learners to develop metacognitive awareness and inhibiting over-reliance on automated assistance. In doing so, we position hint generation as a pathway to tackle the rising problem of cognitive debt due to over-reliance on readily available AI assistants designed to provide complete responses (Section \ref{sec:discussion}). % \todo{Add: a tension between satisfaction and cognitive offloading!}
% In order to answer these questions, %SM I replace "towards answering with 'in order to answer"
%This work makes three key contributions. First, we present a systematic exploration of scientific hints as a form of interactive support for learners, benchmarking 18 open-source models. Second, we conduct extensive automatic evaluation of chain-of-hints across five dimensions, establishing a framework for assessing the quality of chain-of-hints. Finally, we carry out a human evaluation study with 41 participants, offering learner-driven insights into the components that make hints helpful and highlighting design implications for future educational systems. %SM shoud we say "user study" intead of "human evaluation study", or remove "study" from the "human evaluation"
%SM I think the introduction will benefit from an example of 'static' and 'dynamic' chain of hints is. maybe a picture with 2 interactions between learner and your system; with static and dynamic hints.

This work makes three key contributions through our exploration of scientific hints as a form of interactive learning support: (1) a comprehensive evaluation of 18 open-source LLM using a five pedagogically-grounded evaluation metrics; (2) a detailed qualitative analysis from an \revision{exploratory} study with 41 participants, aimed at identifying the components that make hints effective; and (3) design implications for the development and evaluation of intelligent tutoring systems, grounded in two distinct hint-generation strategies (static and dynamic).

\section{Related Work}

% \smara{I do not see changes highlighted in related work, and i think reviewer expect more engagement with the literature?}
\subsection{Automatic Hint Generation: Past and Present }

As a core component of intelligent tutoring systems, automatic hint generation has been widely studied across educational domains, with sustained interest in how hints can improve students’ learning outcomes and reasoning skills. 
% , research on hint generation dates back more than two decades \cite{} \todo{cite}. 
Prior work can be broadly categorized into three paradigms based on their underlying framework to provide hints: (i) rule-based models, (ii) data-driven models, and (iii) LLM-based models \cite{stamper2024enhancing}.

\textbf{Rule-based models} formed the backbone of early intelligent tutoring systems. These systems relied heavily on expert-designed learner models, which allowed them to provide effective, problem-specific feedback \cite{mitrovic2001constraint, aleven2009new}. However, their dependence on expert input made them difficult to scale and limited their generalizability across diverse learners and domains.

\textbf{Data-driven models} offered a more scalable alternative, particularly in domains with well-structured problems such as programming \cite{barnes2008toward, rivers2017data, obermuller2021guiding, jin2012program, zimmerman2015automated, paassen2018continuous, price2016generating, rolim2017learning}. These models typically followed a deterministic framework consisting of three components: (1) a corpus of diverse candidate solutions, (2) a matching algorithm to align learner attempts with the closest example, and (3) a path traversal algorithm to guide learners toward the correct solution. A rich body of surveys have documented the innovations in this space \cite{le2013review, crow2018intelligent, mcbroom2021survey, mahdaoui2022comparative}. While effective in programming education, data-driven models were constrained by their dependence on high-quality training data, their limited adaptability beyond formal logic domains, and their difficulty in handling solution diversity.

To overcome these limitations, recent work has increasingly explored \textbf{LLM-based hint generation models}. Large language models could offer various advantages, including improved reasoning ability, adaptability, and domain generalizability \cite{bommasani2021opportunities, naveed2025comprehensive, kasneci2023chatgpt, minaee2024large}. While programming remains a popular testbed for LLM-based systems \cite{rogers2025playing, demirtacs2025generating, barros2025large, borchers2025can, ma2024teach, liang2024towards, kazemitabaar2024codeaid}, the scope has expanded to other subjects with structured solution space, like mathematics. For instance, \citet{wang2023step} curated a benchmark (\texttt{ReMath}) co-developed with teachers to evaluate step-by-step math tutoring, with the aim to use LLMs to assist math tutors in remediating student mistakes. \citet{pal2024autotutor} proposed MWPTutor, a framework for generating math word problem hints by reformulating questions around the learner’s next step. \citet{pardos2023oatutor} introduced \texttt{OATutor}, an open-source ITS platform to support adaptive tutoring research across multiple subjects. Building on this, \citet{pardos2024chatgpt} evaluated ChatGPT-generated hints within \texttt{OATutor}, comparing them to human tutor hints across algebra and statistics \revision{using an in-context learning approach using chain-of-thought reasoning with self-consistency decoding \cite{wang2022self}}. Their results demonstrate statistically significant learning gains from ChatGPT-generated hints. % We direct the readers to the survey by \cite{jangra2024navigating} for a comprehensive overview.

\revision{
While most prior work on LLM-generated hints relies on in-context learning--typically through few-shot examples or chain-of-thought prompting--recent research has begun exploring two complementary directions: synthetic data generation and post-training. In math tutoring, for example, \citet{wang2025training} demonstrate that LLMs can role-play teacher–student interactions to produce synthetic datasets, which can then be used to fine-tune smaller models \citep{zhang2025sefl} or to train tutoring policies via LLM-based reward models \citep{dinucu2025problem}. These post-trained smaller models models can match or even exceed the performance of state-of-the-art systems on  specialized tutoring tasks. \citet{scarlatos2025training} report that a Llama3-8B \citep{grattafiori2024llama} fine-tuned with direct preference optimization \cite{rafailov2023direct} outperformed GPT models on several mathematics-tutoring benchmarks. However, such gains come with substantial costs. Post-training requires significant computational resources and depends heavily on high-quality, pedagogically grounded datasets. As \citet{dinucu2025problem} highlight, even within the same problem-solving domain, different alignment strategies (\textit{e.g.,} supervised fine-tuning versus RL-based optimization) can lead to divergent outcomes.
}

Recent work has also examined automatic feedback in more open-ended contexts. For instance, \citet{lieb2024student} developed \texttt{NewtBot}, a LLM-based tutoring chatbot to support secondary school students in physics in German. They explored several pedagogy-grounded and context-grounded prompting strategies against a general purpose model, however, their findings reveal no statistically significant difference across the prompting strategies, likely due to small group size. \revision{Similarly, another small-scale investigation for conceptual physics problem solving by \citet{wan2024exploring} found students to equally prefer GPT-generated feedback to human-written feedback. Expert human evaluators judged the LLM-generated responses to requiring minimal edits for most feedback.}
% In design education, \citet{duan2024generating} utilized LLMs to generate heuristic feedback on UI mockups, offering constructive but sometimes inconsistent suggestions. In legal education,  \citet{weber2024legalwriter} developed \texttt{LegalWriter}, an intelligent writing support system that can help novice students produce more structured and persuasive case solutions. 
These efforts highlight that while early work concentrated on structured topics, researchers are beginning to explore how LLM-based feedback can support learning in more open-ended domains.

Despite this growing diversity, we identify two key reasons why research on hint generation has focused most heavily on structured domains. First, evaluation is more straightforward: solutions adhere to strict rules, allowing researchers to measure correctness with high reliability. %SM You mean evaluation of the hinting strategy? if so you need to say, so. since evauation of solution is strighforward in your QA setting.
Second, these domains align with the broader aim to improve formal reasoning capabilities of LLMs, developing %and pushing 
benchmarks for establishing state-of-the-art performance \cite{chen2021evaluating, zhuo2024bigcodebench, jiang2024survey, paul2024benchmarks, glazer2024frontiermath, mishra2022lila, li2020isarstep, lu2023mathvista}. By contrast, subjects involving a natural language solution space are underexplored within the intelligent tutoring systems community, due in part to the difficulty of evaluation. Through this work, we aim to contribute toward the development of hint generation methods that extend beyond formal logic settings.

\subsection{Automatic Evaluation of Hints}

Evaluating the quality of the generated hints at scale requires reliable automatic metrics. %that can capture whether generated hints are pedagogically useful. 
Prior work %has %broadly 
%approached this challenge 
has proposed %through 
two families of methods: reference-based and reference-free evaluation.

\textbf{Reference-based} metrics assume access to expert-written hints and compare generated hints against them. For example, \citet{price2019comparison} proposed \texttt{QualityScore}, which evaluates programming hints by measuring similarity to tutor-authored hints using abstract syntax tree (AST) representations \cite{mccarthy1964formal, knuth1968semantics}. While such methods can provide fine-grained comparisons, they are inherently limited by the availability and coverage of expert references—often infeasible to scale across domains.

\textbf{Reference-free} metrics, in contrast, aim to predict hint helpfulness without requiring gold-standard references. These methods either leverage learner interaction data or adopt learner-agnostic success measures. Interaction-based metrics quantify how a learner’s state changes after receiving a hint.
%SM so these also relate to the next section on Learner-centered evaluation?"
For instance, \citet{paassen2018continuous} computed error over two distances: (i) between the predicted post-hint state and the true next state, and (ii) between the predicted post-hint state and the learner’s final state. Learner-agnostic approaches instead assess hints through generalizable outcomes such as correctness or efficiency. For programming hints, \citet{rivers2017data} evaluated hints based on whether they eventually reached a correct solution and the length of the hint sequence. More recently, for natural language hints, \citet{mozafari2025hinteval} proposed several metrics such as relevance, readability, convergence, familiarity, and answer leakage to capture a hint's quality in open-ended domains.

In our work, we extended this line of research by investigating a broader set of pedagogically-grounded evaluation metrics for a chain of hints that better reflect the sequential nature of hint delivery (Section \ref{subsec:auto_eval_tldr}). We also evaluated the alignment of these metrics to learners' needs and experiences, identifying the blind spots of automatic evaluation (Section \ref{subsec:autoeval_analysis}). This analysis surfaces gaps in current metrics, and highlights open challenges for designing metrics that better capture hint utility from a learner-centered perspective.

\subsection{Learner-centered evaluation of generated hints}

Human-centered evaluation is indispensable for building effective hint generation systems. Because learners vary in their needs, pace, and strategies, evaluations must consider how different designs support diverse populations. Prior work has explored a range of evaluation settings, shaped by study design choices and deployment targets. We organize these into four broad approaches: i) third-party annotation, ii) randomized controlled trials, iii) classroom-scale deployments, and iv) longitudinal studies.

\textbf{Third-party annotation.} To assess the pedagogical quality of LLM-generated responses, \citet{tack2022ai} conducted a pairwise human-in-the-loop comparison study, with annotators rating responses along dimensions such as talking like teachers, understanding students, and helping improve student understanding. Their findings revealed that while LLMs often match teachers in conversational uptake, they lag significantly in their ability to foster deeper understanding. \revision{Similar findings were reported by \citet{dai2023can} for LLM-generated evaluations for an introductory data science course; where they found that while the LLM-generated outputs achieved higher readability scores compared to instructor-written responses, they did not translate into pedagogical accuracy. The LLM-generated assessments misaligned with the instructors' evaluations of student work along key pedagogical dimensions of goal, benefit and clarity. For math error detection, expert tutor annotated analysis by \citet{kakarla2024using} revealed that while GPT-4 \cite{achiam2023gpt} was able to accurately recognize the response criteria for student mistakes, it overidentified student errors for instances that human evaluators considered correct. While first-party feedback from learners can help gauge their affective state better, third-party evaluations using pedagogical experts can uncover aspects of learning that might be misrepresented by the learners' feeling of learning \cite{deslauriers2019measuring}.} \looseness=-1

\textbf{Randomized controlled trials (RCTs).} RCTs are among the most common designs in prior work, allowing direct comparisons between feedback strategies in controlled settings. For instance, \citet{lieb2024student} evaluated feedback strategies for secondary physics, while \citet{duan2024generating} examined feedback for UI mockup design. Both highlight how targeted experimental setups provide causal evidence of effectiveness, though typically over shorter time frames.

\textbf{Classroom-scale deployments.} A number of studies have examined hinting systems in authentic classroom contexts, capturing the social and ecological dynamics of learning. For example, \citet{kazemitabaar2024codeaid} deployed \texttt{CodeAid}, a programming assistant designed to provide hints without revealing full solutions, while \citet{rogers2025playing} studied programming support in a classroom setting through \texttt{MatlabTutee}, a tool designed with an emphasis on the learning-by-teaching principle.

\textbf{Longitudinal studies.} Longer-term deployments allow researchers to capture sustained patterns of tool use and their influence on learner development, especially within the nascent stages of prototype development. \citet{rogers2025playing}, for instance, distinguished between high-usage and low-usage groups in a two month-long deployment, finding that heavy users of their proposed system (\texttt{MatlabTutee}) reported the highest self-efficacy gains, underscoring the value of extended exposure in evaluating learning technologies.

Across these evaluation modes, researchers emphasized different outcome dimensions depending on the domain and artifact design. \textit{Helpfulness} or \textit{usefulness} consistently appears as a central measure of perceived utility. Other dimensions included \textit{self-efficacy} \cite{rogers2025playing}, \textit{attitudes toward AI assistance} (measured via instruments such as the godspeed questionnaire \cite{bartneck2009measurement} or short version of the user experience questionnaire \cite{schrepp2014applying}), \textit{enjoyability} and \textit{cognitive load} \cite{lieb2024student}, and task-specific factors such as \textit{analogical reasoning} in the case of bio-inspired design training \cite{chen2024bidtrainer}. These variations underscore how human evaluations illuminate distinct facets of educational support that automatic metrics alone cannot capture.

In our work, we adopted a within-subjects randomized controlled trial to evaluate hints for scientific question answering. Our study emphasized on hint's satisfaction, informativeness, and answer leakage, % usefulness, insightfulness, and oversimplification, % and perceived difficulty, 
while also examining how these perceptions align or diverge from automatic evaluation metrics. We describe the motivation for selecting these dimensions in our user study design (Section \ref{sec:study_design}).

\section{Scientific Hint Generation} \label{sec:sci_hint_gen}

% Expert tutors are able to provide high quality support to students to help them meet their learning goals. These tutors take into consideration each individual's learning objectives \cite{hoque2016three, sonmez2017association} and prior knowledge \cite{orsmond2011feedback, biggs1996enhancing} to provide meticulously crafted \textit{direct instructions} \cite{kozloff1999direct, kim2005direct, rosenshine2008five} to guide the learners to the correct answer, molding them into \textit{deep conceptual thinkers} \cite{rillero2016deep} with an enhanced sense of accomplishment \cite{vanlehn2011relative, merrill1992effective, arroyo2014multimedia}. \aj{should we keep the above bit? It is more learner than user related.}

% Through this work, w
We aim to explore the capabilities of state-of-the-art reasoning models to provide scaffolded guidance to users to reach the correct answer. Limiting the focus to scientific hint generation, in this section, we formally define the \textit{chain-of-hints} generation task (Section \ref{subsec:prob_def}), and describe the dataset we use for the evaluation (Section \ref{subsec:dataset}) of several LLM-based baselines (Section \ref{subsec:baselines}). We also briefly describe the automatic evaluation metrics (Section \ref{subsec:auto_eval_tldr}) we use to select the model for user study, and provide the detailed description and analysis in Appendix \ref{sec:autoeval}.

% \noindent\textbf{Problem Definition.}
\subsection{Problem Definition} \label{subsec:prob_def}

Drawing from the \textit{just-in-time} incremental interventions portrayed by expert human tutors, we formally define the \textit{chain-of-hints} generation task as follows: given a question $q$ and answer $a$, a hint generation system $\mathbf{H}$ generates a sequence of hints $\textbf{h} = \langle h_1, h_2, ... , h_{k}\rangle$, such that: 

\begin{equation*}
    % P(a|q) < P(a|q, h_1) < P(a|q, h_2) < ... < P(a|q, h_k)
    P(a|q,h_i) < 1 \; \forall i \in \{1, 2, ... , k\}
\end{equation*}

\noindent where $P(a|.)$ denotes the probability of a user successfully answering the question. We propose two variants of the task depending on whether the hint generation system takes prior attempts into account. In the \textit{\textbf{static}} setting, the hint generation function is defined as 

% \vspace{-1em}
\begin{equation*}
    \mathbf{H}: \{q, a, \langle h_1, h_2 \ldots h_{i-1}\rangle\} \mapsto h_i
\end{equation*}

\noindent where the next hint $h_i$ only depends on the question, the correct answer and the sequence of previously provided hints. In the \textit{\textbf{dynamic}} setting, the function additionally conditions on the user's past incorrect attempts, \textit{i.e.,}

\begin{equation*}
    \mathbf{H}: \{q, a, \langle h_1, h_2, \ldots h_{i-1}\rangle, \langle a_1^{\times}, a_2^{\times} \ldots\rangle\} \mapsto h_i
\end{equation*}

\begin{table*}[t]
\centering
\caption{Dataset statistics for the automatic evaluation and human evaluation splits.} \label{tab:data_stats}
\resizebox{\textwidth}{!}{
\begin{tabular}{lclccc}
\toprule
\multirow{2}{*}{\textbf{Dataset Split}} & \multirow{2}{*}{\textbf{\#instances}} & \multirow{2}{*}{\textbf{\#instances per Subject}} & \multicolumn{3}{c}{\textbf{Average \#words}} \\
 &  &  & \textbf{Question} & \textbf{Answer} & \textbf{Context} \\\midrule
Automatic Evaluation & 1000 & \begin{tabular}[c]{@{}l@{}}Biology (518), Chemistry (160),\\ Geology (143), Physics (162),\\ Miscellaneous (17)\end{tabular} & 13.23 & 1.52 & 66.89 \\
Human Evaluation & 30 & \begin{tabular}[c]{@{}l@{}}Biology (8),  Chemistry (7), \\ Geology (7), Physics (8)\end{tabular} & 13.20 & 1.27 & 61.30 \\\bottomrule
\end{tabular}
}
\end{table*}

% \noindent\textbf{Dataset.}
\subsection{Dataset} \label{subsec:dataset}

% As the task of scientific hint generation is quite recent, there doesn't exist any relevant dataset for the task of hint generation for scientific questions \aj{contradictory}. 
To evaluate the ability of large language models to generate hints, we use the \texttt{SciQ} dataset \cite{welbl-etal-2017-crowdsourcing} for our evaluation. \texttt{SciQ} dataset is a science exam question answering dataset comprising of 13,679 multiple choice questions spread across biology, chemistry, physics and geology; spanning across elementary level to college introductory level questions. Since having multiple choices for each question hinders the ability to test the quality of generated hints, we discard the distractor choices, and view the task as a free-form short answer question answering task. We use two different dataset splits for the automatic and human evaluation as follows (see Table \ref{tab:data_stats} for detailed statistics) - 

\begin{itemize}
    \item \underline{Dataset for automatic evaluation} To extensively evaluate the hint generation capabilities of the LLM-based baselines, we use the 1,000 examples from the validation split of the \texttt{SciQ} dataset.
    \item \underline{Dataset for human evaluation} As the original \texttt{SciQ} dataset comprised of multiple-choice question answers, we hand-pick 30 questions from the test split of \texttt{SciQ} dataset to ensure valid open-ended questions with unambiguous answers, while ensuring equal representation of all four subjects in the final quiz.
\end{itemize}

% \noindent\textbf{Baselines.} 
\subsection{Models} \label{subsec:baselines} 
% \smara{Why baselines, can you say models? If you say baseline one expects to propose a new model.}
To extensively evaluate the capabilities of large language models to generate static and dynamic hints, we explore a diverse range of open-source model families. We limit our exploration on smaller open-source language models for three reasons - 1) open-source language model weights are constant, unlike API-based models,  2) there are privacy concerns with using closed-source models, and 3) any real-world application for learning feedback will have computational constraints, and thus we do not explore large language models that require multiple GPUs to execute. More specifically, in this work, we explore the following language models for the scientific hint generation task - \texttt{DeepSeek-R1} (1.5B, 7B, 8B, 14B, 32B) \cite{guo2025deepseek}, \texttt{Gemma3} (1B, 4B, 12B, 27B) \cite{team2025gemma}, \texttt{Mistral-Small} (24B)\footnote{\url{https://huggingface.co/mistralai/Mistral-Small-24B-Instruct-2501}}, \texttt{Phi4} (14B) \cite{abdin2024phi}, and \texttt{Qwen3} (0.6B, 1.7B, 4B, 8B, 14B, 30B, 32B) \cite{yang2025qwen3}. 
We selected these model families to capture a diverse range of open-source, instruction-tuned LLMs that vary in scale, training philosophy, and reasoning capability, while remaining practical for controlled experimentation. In particular, \texttt{DeepSeek-R1} and \texttt{Qwen3} include variants explicitly optimized for reasoning-style behavior, making them well-suited for evaluating multi-step hint generation. \texttt{Gemma3}, \texttt{Mistral-Small}, and \texttt{Phi4} serve as strong general-purpose models with competitive performance across instruction-following and educational tasks. Together, these models provide broad coverage across parameter sizes (0.6B–32B) and architectural/modeling choices, enabling us to systematically study how model scale and reasoning specialization affect the quality and pedagogical structure of generated hint chains.
% \smara{Can you say a bit why these spefic ones, some are reasoning models, etc.}
We use an in-context learning approach to generate the static and dynamic chain-of-hints (refer to Figures \ref{fig:static_prompt} and \ref{fig:dynamic_prompt} in Appendix \ref{app:prompts} for the static and dynamic hint generation prompts respectively). % \smara{I think we want the prompsts in a figure here: left for static, right for dynamic. Also connect them back to the problem definition. }

% We limit our exploration on smaller open-source language models for three reasons - 1) open-source language model weights are constant, unlike API-based models,  2) there are privacy concerns with using closed-source models, and 3) any real-world application for learning feedback will have computational constraints, and thus we don't explore really large language models that require multiple GPUs to execute.
% \smara{I would move this paragraph before you the sentence "More specifically, in this worl". Since you want to say why you choose those models upfront}

% We use the \texttt{ollama} library\footnote{\url{https://ollama.com/}} to run the models on a single Nvidia A100-SXM4-40GB GPU. While we acknowledge that generating responses using \texttt{ollama}'s 4-bit quantized models might yield worse performance than original models \todo{cite performance vs quantization papers as evidence}, it allows us to explore the hint generation capabilities of larger LLMs on realistic computational constraints. \aj{Should I state this point? Or is it negative to state this, even though I've justified it ?} \smara{i don't think you need this sentence}

\subsection{Automatic Evaluation Metrics} \label{subsec:auto_eval_tldr}

Reliable automatic evaluation metrics are critical in advancing research on automatic hint generation, and intelligent tutoring systems more broadly \cite{mousavinasab2021intelligent, jangra2024navigating}. %SM I added metrics above 
These automatic evaluation metrics provide a fast feedback cycle in early-stage system development. At the same time, they serve as a safeguard against premature deployment of underdeveloped systems. % , mitigating potential ethical risks.
% These automatic evaluation metrics provide a fast feedback cycle in early-stage system development, while safeguarding deployment of underdeveloped systems, mitigating potential ethical risks.  % to learners, which raises ethical concerns.
Exposing users to low-quality hint generation systems may not only undermine the learning outcomes, but also jeopardize %put at risk 
their trust in educational technologies more broadly. Having robust automatic evaluation metrics also helps %the community to 
ensure comparability and reproducibility of research findings across different studies. %Therefore, 
To automatically evaluate the pedagogical quality of the chain-of-hints in ways that align with user's needs and goals, we %establish
discuss several automatic evaluation metrics, either by %repurposing 
adapting metrics from prior works, or proposing new ones wherever necessary. We briefly describe the evaluation metrics here, and provide detailed definitions in Appendix \ref{subsec:autoeval_metrics}. 

%To automatically evaluate the pedagogical quality of hint chains in ways that align with learner needs and goals, %SM I moved this above and made this a new paragraph that describes now the metrics 
We investigate a suite of evaluation metrics spanning five dimensions: \textit{information gain}, \textit{redundancy}, \textit{consistency}, \textit{readability}, and \textit{leakage}. To capture the degree to which hints help users progress toward correct answers, we propose \textbf{information gain} ($InfoGain$), which leverages a language model to compare question-answering performance with and without hints. We implement two variants: $InfoGain_{mean}$, the average gain across all hints in a chain, and $InfoGain_{comb}$, the maximum gain achieved by the entire hint chain.

Drawing on variation theory \cite{ling2012variation}, which emphasizes the value of presenting learners with diverse perspectives, we assess \textbf{redundancy} by %repurposing 
adapting self-referenced redundancy from summarization research \cite{chen2021training}, with the goal of minimizing repetitive information. To ensure factual correctness, we introduce \textbf{consistency}, computed using \texttt{AlignScore} \cite{zha2023alignscore}, a metric designed to evaluate factual alignment between generated text and the source material.

To prevent hints from oversimplifying tasks or directly revealing answers, we implement \textbf{leakage} with two complementary measures: a string-matching approach ($leakage_{EM}$) and a LLM-based approach ($leakage_{LLM}$) designed to detect subtle forms of answer disclosures. Finally, to evaluate clarity, we apply established \textbf{readability} measures, including the Flesch–Kincaid grade level \cite{kincaid1975derivation}, Flesch reading ease \cite{flesch1979write}, and Dale–Chall readability score \cite{dale1948formula}.

For model selection for our human evaluation study, we combine information gain, redundancy, consistency, and leakage into a single aggregate score to holistically evaluate the content of the generated hints.

\begin{figure*}[t!]
\centering
\includegraphics[width=\textwidth]{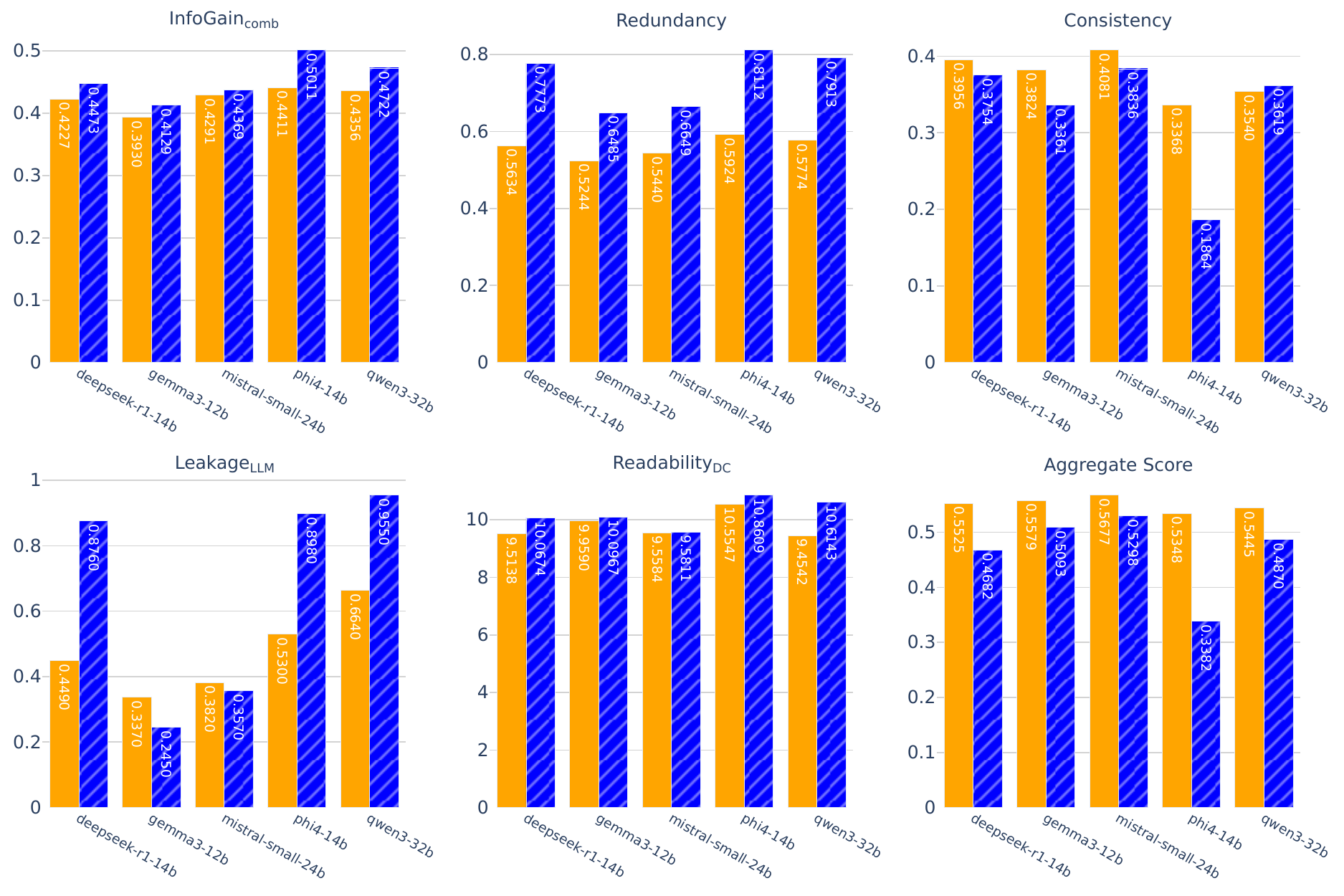}
\caption{Performance of best models from five baseline model families evaluated using automatic evaluation metrics. \textcolor{orange}{Orange} (solid) bars denote the static hint generation results and \textcolor{blue}{blue} (striped) bars denote the dynamic hint generation results. \revision{All metric values lie between 0 and 1, except for $Readability_{DC}$. Tables \ref{tab:static_autoeval} and \ref{tab:dynamic_autoeval} provide detailed automatic evaluation metric results for the static and dynamic hint generation settings respectively.}} \label{fig:autoeval_main}
\end{figure*}

\subsection{Experimental Setup} \label{subsec:exp_setup}
% \underline{Experimental Setup}. \smara{Why is this part of 3.4, it should be its own subsection 3.5 right? 3.4 is the metrics and 3.3 models etc} 
We evaluate all baselines over 1,000 instances from the validation split of the \texttt{SciQ} question answering benchmark \cite{welbl2017crowdsourcing}. We explore two different hint generation settings to quantify the hint generation capabilities of these baselines - \textit{static} and \textit{dynamic}. For the \textit{static} setting, we prompt the LLM baseline to generate 4 hints given a question answer pair. For the \textit{dynamic} setting, \revision{due to unavailability of user interaction data on the validation question-answer pairs, we used the three distractor choices from the \texttt{SciQ} multiple choice questions as the incorrect attempts, and interleaved them within each hint to replicate a learner interaction for the in-context learning prompt. For our human evaluation, however, we use actual responses to generate dynamic hints on-the-fly for better representation of a real-world interaction.
}
% we used the three distractor choices from the \texttt{SciQ} multiple choice questions as the incorrect responses, and interleaved them within each hint to replicate a learner interaction for the in-context learning prompt. 
We provide the prompts for static and dynamic hint generation in Appendix \ref{app:prompts}, Figures \ref{fig:static_prompt} and \ref{fig:dynamic_prompt} respectively.\looseness=-1

% Figure \ref{fig:autoeval_main} presents the performance of best-performing baselines from each model family. Between \textit{static} and \textit{dynamic} hint generation settings, we observe that dynamic hints tend to be more redundant than static hints, lower consistency scores. This lack of content diversity and increased leakage rates in the dynamic hint chains lead to lower aggregate scores compared to static hint chains. Overall, \texttt{Mistral-Small-24b} achieves the highest performance in both \textit{static} and \textit{dynamic} hint settings. 
\subsection{Automatic Evaluation Results} \label{subsec:auto_eval_results}
Figure \ref{fig:autoeval_main} summarizes the performance of the best-performing baselines from each model family. Comparing \textit{static} and \textit{dynamic} hint generation, we find that dynamic hints generally exhibit higher redundancy and lower consistency scores. This reduced content diversity, combined with elevated leakage rates, results in lower aggregate scores for dynamic hint chains relative to static hints. Across both settings, \texttt{Mistral-Small-24b} achieves the highest overall performance.
% \smara{before you say what you have in appendix you need to say the main outcome of figure. Amd then say that is why we selected system X for the user study. This will also allow you to justify why you put all the rest of details in appendix and say you then focus on correlation of these metrics with user feedback } 
Given the large number of baselines and evaluation settings, we report detailed experimental results in Appendix \ref{subsec:autoeval_results}. There, we examine trends across metrics, contrasts between static and dynamic strategies, and the impact of model parameters on hint quality. Later, in Section \ref{subsec:autoeval_analysis}, we analyze the effectiveness of these automatic evaluation metrics to evaluate the quality of hints grounded on user feedback from our study.
% Due to a large number of baselines and evaluation settings, we we provide detailed experimental results in Appendix \ref{subsec:autoeval_results} for brevity. We explore trends within and across evaluation metrics, difference in static and dynamic hint generation strategies, as well as the influence of model parameters on the hint generation quality. We analyze the effectiveness of these automatic evaluation metrics to evaluate the quality of hints grounded on learner feedback later on in Section \ref{subsec:autoeval_analysis}.

\underline{Selecting a Model for User Study:} We select \texttt{Mistral-Small-24b} model to conduct the human evaluation study for two main reasons - i) it achieved the highest aggregate score in both static and dynamic hint generation settings, and ii) due to it's great instruction following capabilities, the hints generated from \texttt{Mistral-Small-24b} followed our desired structure in the output with minimal post-processing, an important property for the human evaluation study with online hint generation for the dynamic hint generation setting.
\section{User Study Design} \label{sec:study_design}

We conducted a within-subjects human evaluation study %in August 2025 %SM do not think you need to say the month you did it?  % I AJ: just looked at some prior qualitative papers as reference when I started writing, so had a monkey-see-monkey-do style of writing initially. :p 
with 41 participants to identify the characteristics of high-quality hints that learners might find most useful. We asked the participants to engage with our user interface in an unmoderated setting at their own convenience to obtain feedback in a more realistic self-study environment preventing any researcher bias. 
% \smara{do you need to say prototype, or just system/model?} AJ: Changed it to user interface...

\begin{figure*}[t]
\centering
\includegraphics[width=0.9\textwidth]{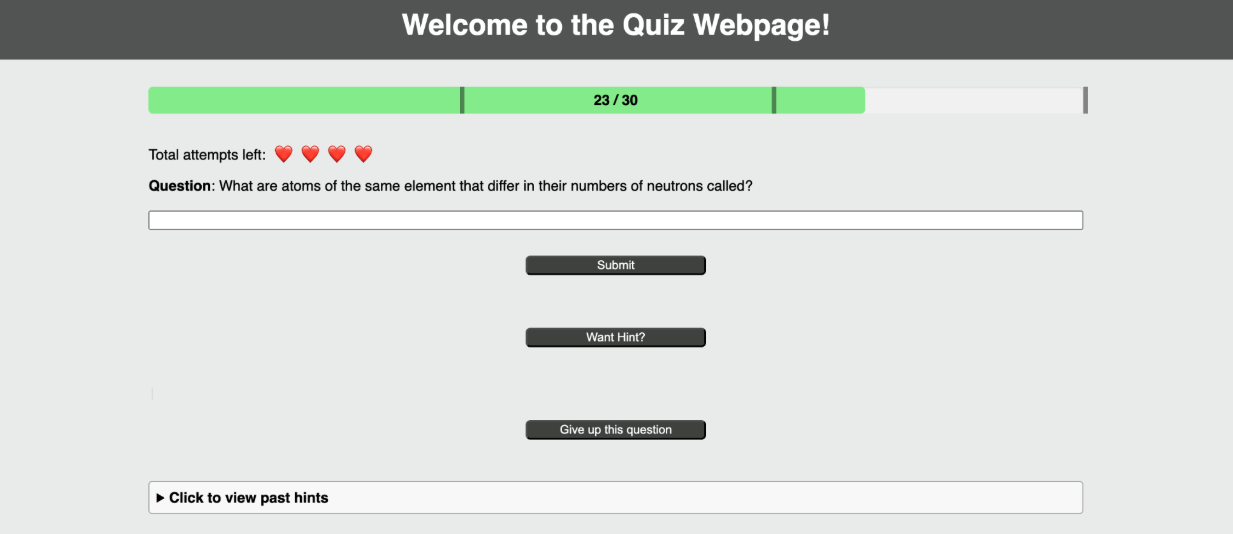}
\caption{Our quiz interface for human evaluation study.} \label{fig:ui}
\end{figure*}

\textbf{Procedure.} 
% We designed an interactive prototype to assess hint effectiveness across multiple science domains. 
% \smara{should you say designed and then interactive system?}
Our study comprised of three parts: pre-quiz briefing, quiz, and post-quiz briefing. In the pre-quiz assessment phase, participants completed a preliminary questionnaire collecting demographic information (gender, ethnicity, age, highest educational attainment) and self-reported familiarity with four science domains: physics, chemistry, biology, and earth science (measured on a 5-point scale from ‘novice’ to ‘expert’).

In the quiz assessment phase, the participants where given 30
% The primary part of the study was a quiz comprising of 30 
free-form short answer questions standardized for all participants (8 physics, 8 biology, 7 chemistry and 7 geology questions). 
% \smara{the above has to be "parallel" with prequiz. So maybe "In the quiz assessment phase, the participants where given 30 ... "} % AJ: I don't quite get this point. Have updated the sentence a bit
The quiz consisted of three sections, each containing 10 questions. With a within-subjects study design in mind, the first section  served as the control condition with no hints provided, while the 2nd and 3rd sections featured either \textit{static} or \textit{dynamic}
% responsive (online) hints or pre-determined (offline) 
hints, assigned randomly to prevent order bias or recency bias in post-quiz briefing survey. 
% \smara{here you use responsive and pre-determined instead of dynamic and static the way you discuss before? i think you need to replace with dynamic and static. ALso I replace Section 1 etc with first section, 2nd section.. you should replace in the entire document like that} % AJ: I like the section name changes, help avoid confusion.
At any point during a question, participants could request a hint or attempt an answer, with up to four hints and five attempts before their response was marked incorrect (see Figure \ref{fig:ui}). This open design allowed us to observe naturalistic help-seeking behavior and prevented the interface from artificially enforcing a fixed hint–attempt sequence.

% For each question, participants had the option to request up to four hints before submitting an answer, and were allowed up to five attempts before their response was marked incorrect (see Figure \ref{fig:ui}). 
% \revision{ To avoid introducing external confounds into how learners engaged with the system, we intentionally did not impose rigid constraints on the interaction flow. For each question, participants were free to request up to four hints at any point before submitting an answer attempt, and they could choose whether to attempt an answer after any given hint. This open design allowed us to observe naturalistic help-seeking behavior and prevented the interface from artificially enforcing a fixed hint–attempt sequence. Both the static and dynamic conditions followed this identical interaction structure, ensuring that any observed differences arose from the hinting mechanism rather than UI restrictions (see Figure \ref{fig:interaction}).}

After each question, participants were presented with the correct answer and asked to evaluate the hints they had received. 
To assess the quality of each hint, we focused on three dimensions: participants’ overall \textit{satisfaction} (five-point Likert scale), whether the hint provided new knowledge (\textit{informativeness}), and whether it directly disclosed the correct answer (\textit{leakage}).
% Specifically, they responded to three questions for each hint: one assessing their satisfaction on a five-point Likert scale, one indicating whether they learned something new, and one indicating whether the hint directly revealed the correct answer. \smara{maybe here is a point where you connect this back to the automatic metrics, if the is a parallel with the information gain and leakage maybe?} 
\revision{We selected these three dimensions based on established principles in learning science and instructional design. Satisfaction reflects learners’ affective responses to support, which influence engagement \cite{ryan2000self} and effective help-seeking \cite{aleven2004toward}. Informativeness corresponds to whether the hint provides actionable scaffolding aligned with theories of feedback \cite{hattie2007power}, cognitive load \cite{sweller1988cognitive}, and the zone of proximal development \cite{vygotsky1978mind}. Leakage captures the extent to which a hint over-assists by revealing the solution itself, directly relating to the assistance dilemma \cite{koedinger2007exploring} and research showing that excessive guidance undermines productive struggle \citep{renkl2004fading, kapur2008productive}. Together, these three dimensions map onto motivational, cognitive, and instructional considerations central to effective learning support, providing a theoretically grounded basis for evaluating hint quality.}

After each section, participants were asked to answer four questions: one assessing the difficulty of the quiz on a five-point Likert scale, one rating the overall quality of hints in previous section on a five-point Likert scale, and two open-ended questions about the positives and negatives of hints shown in previous section. Due to lack of hints in the first Section, we only asked the first question in the 1st section's survey.
Participants progressed through the quiz at their own pace and had the option to take a break after completing each section, allowing them to pause and resume the study as needed.\looseness=-1

Upon completing the quiz questions and the associated surveys,
% Section 3 survey \smara{maybe say the associated surveys because you had 3 sections?}, 
participants responded to a short post-quiz briefing survey, where they were informed about the two different hint strategies used in the 2nd and 3rd sections of the quiz. Their entire quiz interaction, including questions, answers, attempted answers and hints shown was displayed for reference, and based on their experience and perception, we asked three questions - one asking to select the most helpful hint strategy, one asking to select the hint strategy that helped improve the question understanding, and an open-ended question about their observed differences in hints between the two strategies (dynamic vs static). We also provided space for participants to share their thoughts about the study at the end of this post-quiz briefing survey.  The study was reviewed and approved by the IRB at our institution. 

\begin{figure*}[t!]
\centering
\includegraphics[width=0.85\textwidth]{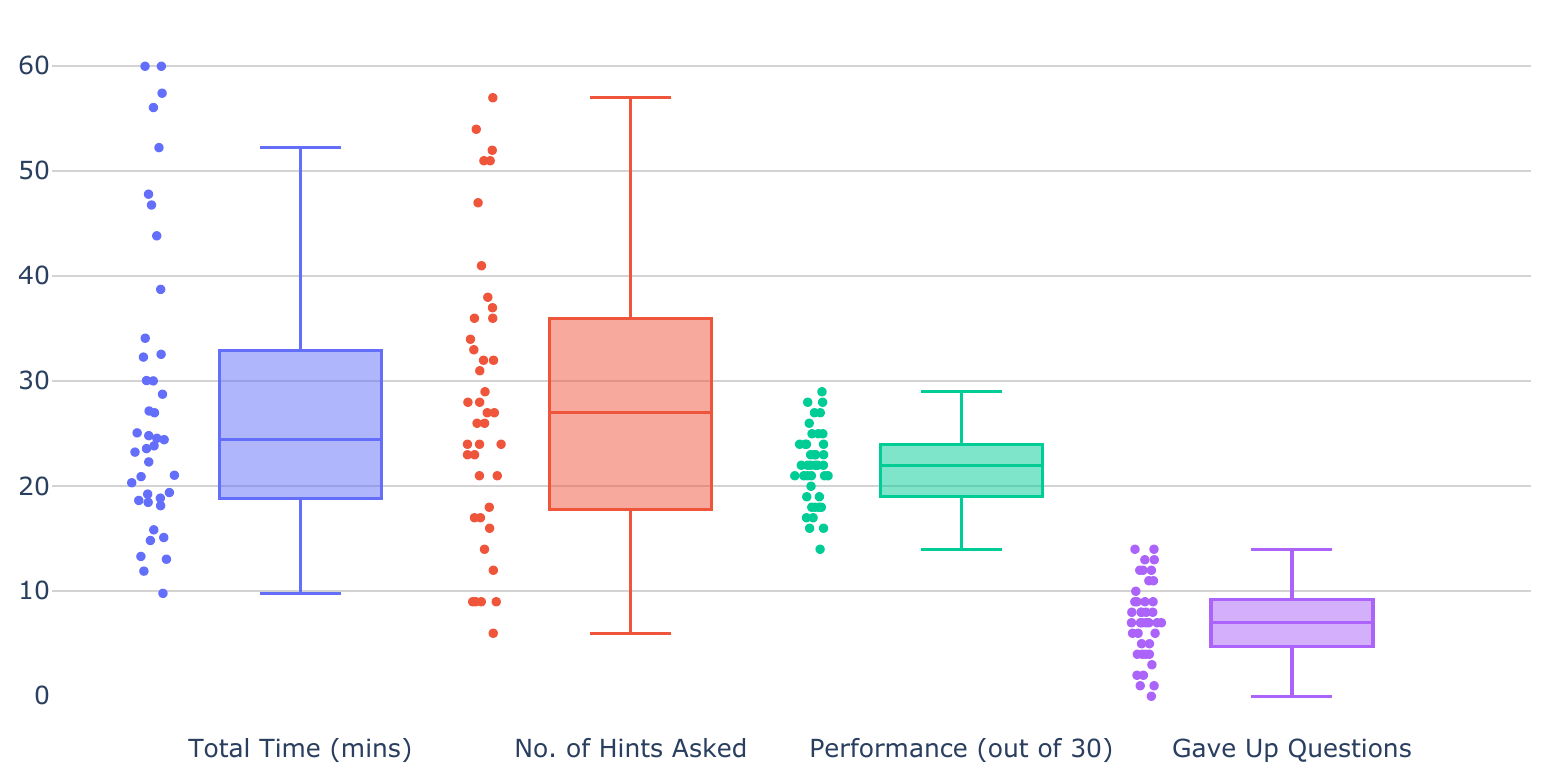}
\caption{Statistics of 41 human evaluation study participants.} \label{fig:participant_stats}
\end{figure*}

\begin{table*}[t!]
\centering % \small

\resizebox{0.8\textwidth}{!}{%
\begin{tabular}{ll}
\toprule
\textbf{Type}                 & \textbf{Count} \\
\midrule
Age                  & 24.81 years (mean), 3.85 years (std)  \\\hline
% Years Post-High School &  years (mean),  years (std)  \\\hline
% Most Recent Job      &   \\\hline
Ethnicity            & \begin{tabular}[c]{@{}l@{}}Asian (29), Multi-racial (3), White (3), \\ Black (2), Middle-eastern (2), Undisclosed (2) \end{tabular}\\\hline
Education Background & Masters (19), Bachelors (10), PhD (9), Undisclosed (2), High school (1) \\\hline
Gender               & Female (23), Male (16), Non-binary (1), Undisclosed (1) \\
\bottomrule
\end{tabular}
}
\caption{Demographics of 41 study participants.}\label{tab:demographics}
\end{table*}

\textbf{Participants.} We recruited 46 participants by circulating our study through university mailing lists and slack channels across several universities. We solicited participants who had studied science subjects in high school, lived in the United States, and were at least 18 years old. Each participant received a 20 dollars gift card for their participation. We do not include in our analysis the data from five participants that asked less than five hints to prevent introducing noise from learners who did not meaningfully engage with the system. From the remaining 41 participants in our study, we obtained feedback over 1,149 hints asked (639 static, 510 dynamic). 20 participants were first shown static hints and then dynamic hints, while for the other 21 participants  the order was reversed to prevent ordering bias. On average, participants took 28.44 minutes to complete the study, where they asked 28.02 hints and successfully answered 21.85/30 questions. \revision{Average self-reported quiz difficulty across all three sections on a 5-point Likert scale (1:very easy and 5:very hard) was 3.47/5, with a standard deviation of 0.47.} We provide the distribution of these statistics in Figure \ref{fig:participant_stats}, and the aggregate participant demographics in Table \ref{tab:demographics}. We redact all identifiable details from the data, and use participant IDs when quoting them in our findings.

\textbf{Analysis.} To identify the effectiveness of a hint in answering questions, we conducted extensive quantitative analysis over three self-reported quality indicators: 
% \smara{what is first-party?}  : changed it to learner-reported
1) \underline{participant satisfaction rating}, an explicit measure of hint quality obtained directly from the learners on a 5-point Likert scale, 
% 2) \underline{Success after hint}, an implicit measure of hint quality, computed as the fraction of hints that yield successful answer submission after, 
2) \underline{insightfulness}, the participant's response to the question, "\textit{Did you learn something new from the hint?}", and 3) \underline{answer disclosure}, the participant's response to the question, "\textit{Did the hint give away the answer?}". We also analyzed the influence of multiple hints, and explore the preference of different hinting strategies across a sequence of hints amongst the participants. To compare effectiveness across the two different hint generation strategies (static vs dynamic), we used an aggregate of these fine-grained indicators in tandem with the participant-reported quality assessment scores in post-section survey. We also analyzed the responses to the open-ended questions using inductive analysis to identify patterns in participants' feedback from post-section and post-quiz surveys \cite{thomas2003general}.\looseness=-1

% \aj{refer to "phenomenography" for motivation of collecting perceived quality survey responses. Ref - \url{https://pubs.rsc.org/en/content/articlehtml/2013/rp/c2rp20145c}}
\section{Findings}

% XX participants took part in our evaluation study, on average successfully answering XX questions (out of 30) in XX minutes, while giving up XX questions (refer to Figure \ref{fig:participant_stats} for the overall distribution). Out of the XX participants, YY participants had \textit{static} hints in Section 2 and \textit{dynamic} hints in Section 3, while the other XX participants had the order reversed \todo{compute statistical significance to show there was no order bias}. We provide the aggregate demographics of our study participants in Table \ref{tab:demographics}. We discarded XX participants from the rest of the analysis that asked less than XX hints overall. \aj{move this to method}

% In this Section, we describe the findings from our analysis, with Section \ref{}, ...

In this section, we present findings from both qualitative and quantitative analyses. We begin with \textbf{RQ\#1}, examining what makes \revision{LLM-generated} hints useful to participants. This includes both the value of a single hint (Section \ref{subsec:finding1}) and the added complexity of chains of hints (Section \ref{subsec:finding2}). For \textbf{RQ\#2}, we compare how learners perceive different hinting strategies, contrasting static and dynamic approaches (Section \ref{subsec:finding3}). Finally, for \textbf{RQ\#3}, we address the alignment between automatic evaluation metrics and learners’ own judgments of hint quality (Section \ref{subsec:autoeval_analysis}).

\subsection{When a Single Hint Helps (or Hurts)} \label{subsec:finding1}

Our analysis reveals that at a sentence-level, hints portrayed several key characteristics of a successful feedback such as clarity and simplicity, yet recurring issues emerged around answer leakage and ambiguity. % Based on inductive analysis from participant responses, we identified several hint characterisitcs
While most characteristics were clearly divided into good and bad across participants, we found that the preference was divided among the participants for certain types of hints. Our inductive analysis of participant feedback highlights several themes around perceived hint quality: % Based on inductive analysis, we identified a
% certain types of hints to be polarizing among the participants. 
% \smara{polarizing sounds strange. maybe say "preference was evenly splitted (are you talking about dynamic and static strategy if yes, say so)} - | AJ: while it is to some degree static vs dynamic, I don't want to use these terms here, and it's more about the characteristics they portrayed than the exact label. 

\begin{figure*}[t]
    \centering
    \begin{subfigure}[]{0.4\textwidth}
    \centering
    \includegraphics[width=\textwidth]{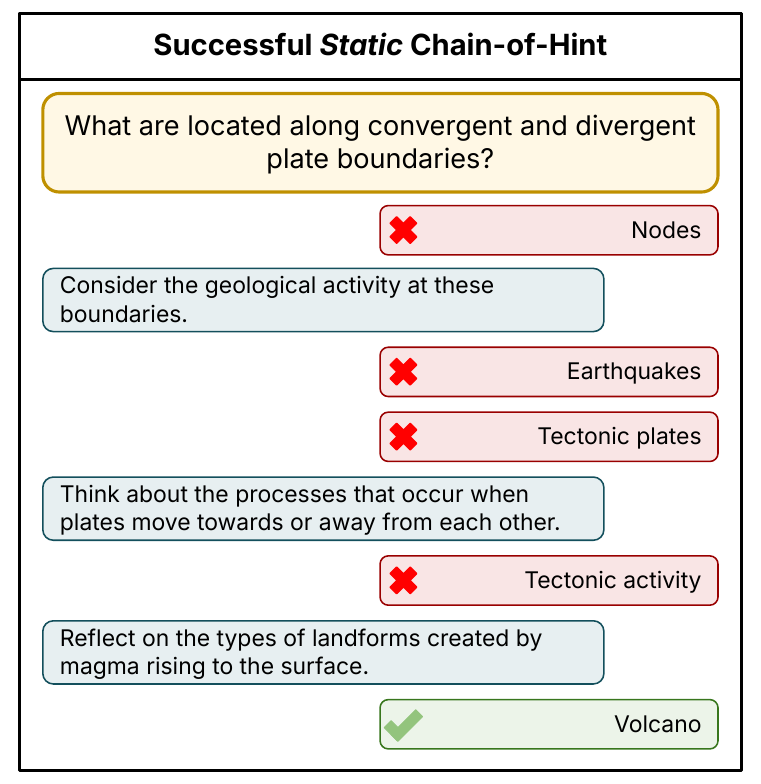}
  \end{subfigure}
  \hspace{2em}
  \begin{subfigure}[]{0.4\textwidth}
    \includegraphics[width=\textwidth]{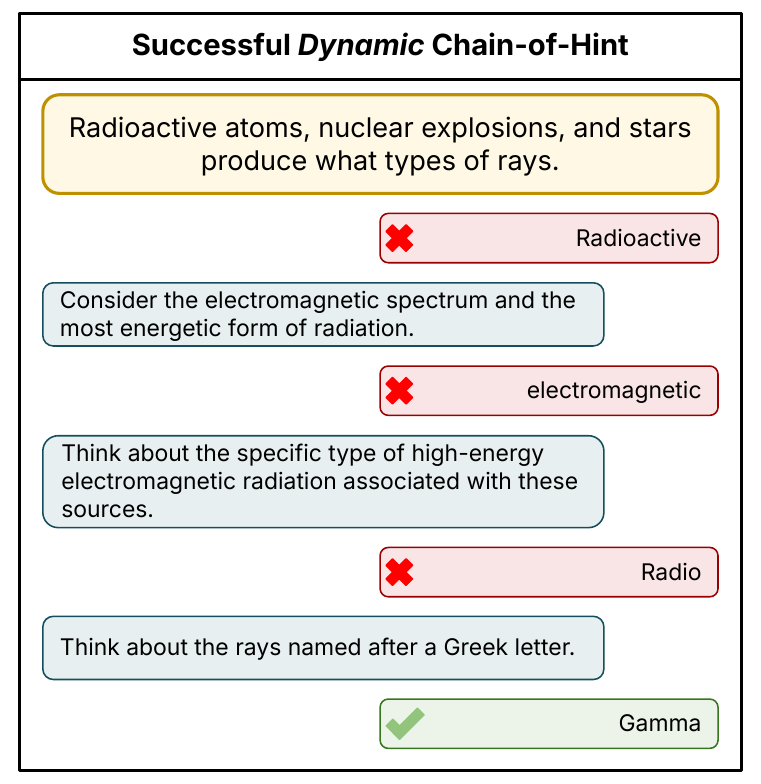}
  \end{subfigure}
  \hfill
  \begin{subfigure}[]{0.4\textwidth}
    \includegraphics[width=\textwidth]{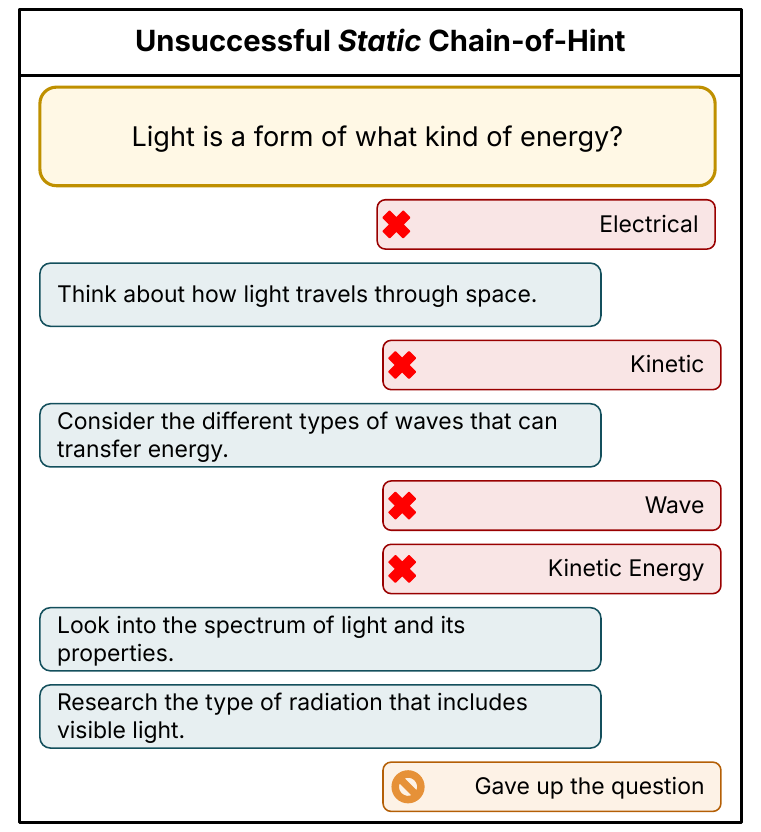}
  \end{subfigure}
  \hspace{2em}
  \begin{subfigure}[]{0.4\textwidth}
    \includegraphics[width=\textwidth]{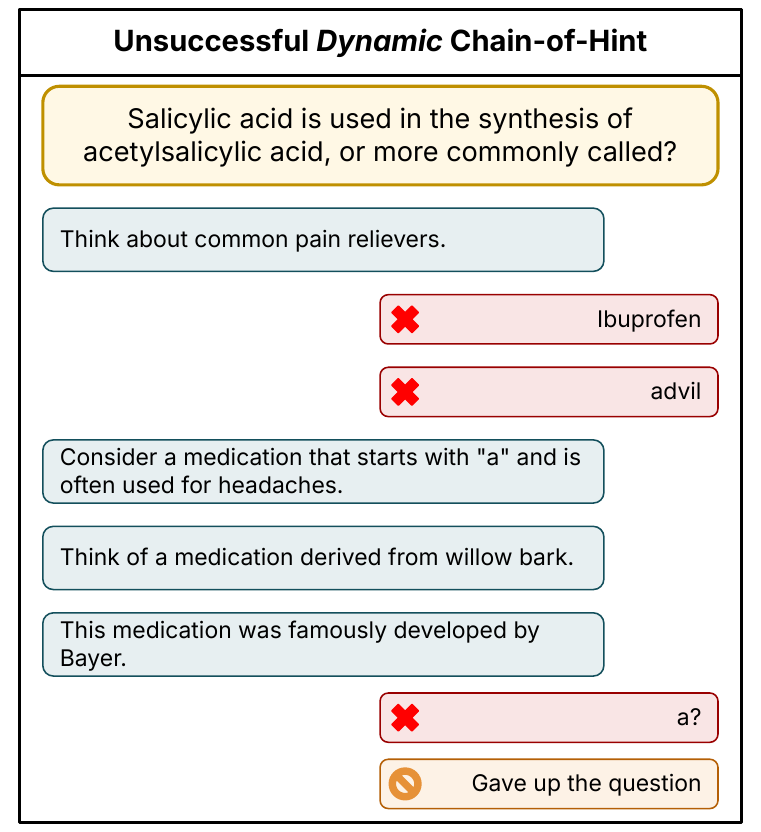}
  \end{subfigure}

  \caption{Interactions between the participants and \textit{static} (left) and \textit{dynamic} (right) hint generation strategies. The top interactions depict a successful journey, whereas the bottom interactions illustrate failure to get participants to reach the answer.}
  \label{fig:case_study}
\end{figure*}

\textbf{Improved conceptual understanding.}
%Hints lead to improved understanding.}
Overall, participants marked 26.63\% (306/1149) hints as informative, with a moderate correlation between the self-reported satisfaction and informativeness ($corr_{pearson}=0.4955$), and several participants explicitly noted that hints helped improve their understanding of the question, not just leading them to the answer. As P-22 described, "\textit{hints encouraged to make associations and engage in reflection from a scientific perspective}". 
\revision{Across all interactions, 109 hints directly assisted participants from an incorrect attempt to a correct submission. 83.4\% of these hints were given high satisfaction ratings (4 or 5), and 59\% of these hints were reported to be informative.}
% From out inductive analysis of qualitative responses, w
We identified three distinct ways in which hints supported this process. \underline{New perspectives:} some hints reframed the questions to open new lines of thoughts. \underline{Linking to prior knowledge:} other hints helped connect the new concepts to their own knowledge, helping them form associations. \underline{Supplementary} \underline{knowledge:} finally, several hints helped enrich participants' knowledge with additional facts. Figure \ref{fig:case_study} (top left example) illustrates an interaction where hints help improve the participant's understanding about the question by indicating a geological activity that involves magma rising to the surface to lead them to the correct answer volcanoes. 
% (see Figure \ref{fig:interaction} for hint strategy-wise statistics).
% Below is an example of participant P-16's interaction where the hints help improve their understanding about the question. 
% \smara{where these categorizations above been derived from your inductive analysis of themes, then say so?}

\begin{figure*}[t]
    \centering
    \begin{subfigure}[]{0.45\textwidth}
    \centering
    \includegraphics[width=\textwidth]{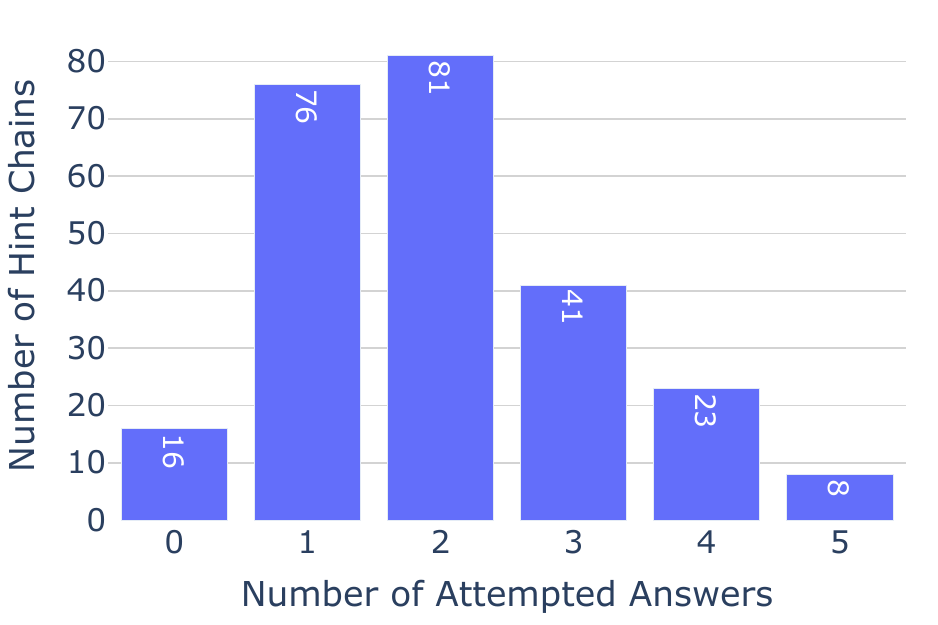}
    \caption{Distribution of number of attempted answers for \textit{static} hints.}
  \end{subfigure}
  \hspace{3em}
  \begin{subfigure}[]{0.45\textwidth}
    \includegraphics[width=\textwidth]{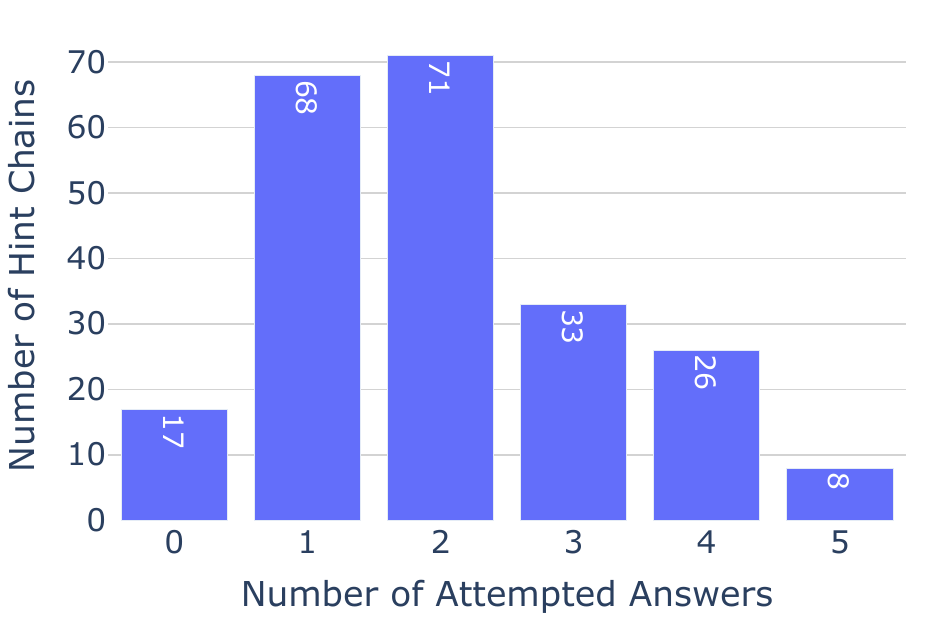}
    \caption{Distribution of number of attempted answers for \textit{dynamic} hints.}
  \end{subfigure}
  \hfill
  \begin{subfigure}[]{0.45\textwidth}
    \includegraphics[width=\textwidth]{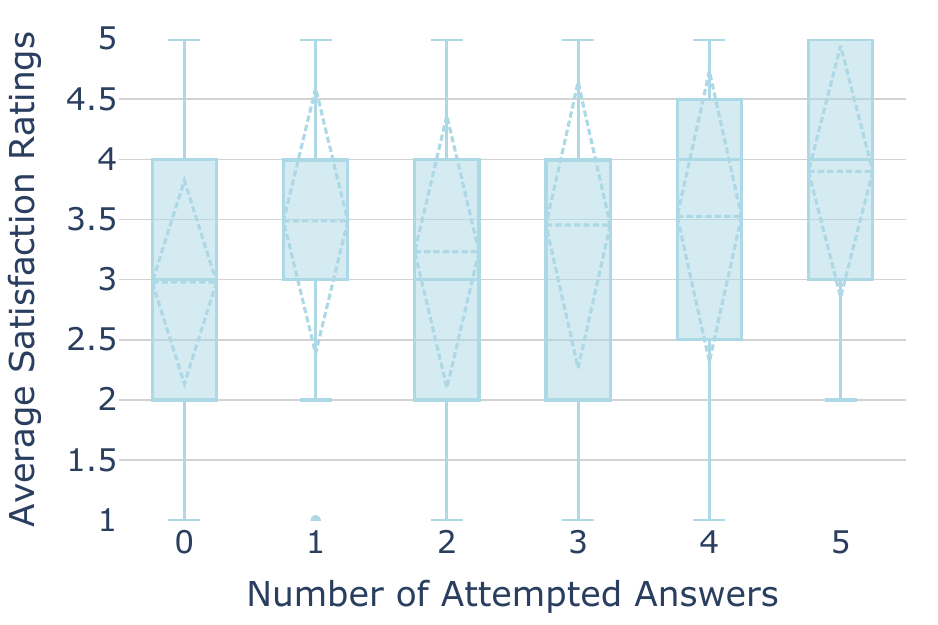}
    \caption{Average satisfaction ratings relative to the number of attempted answers for \textit{static} hints.}
  \end{subfigure}
  \hspace{3em}
  \begin{subfigure}[]{0.45\textwidth}
    \includegraphics[width=\textwidth]{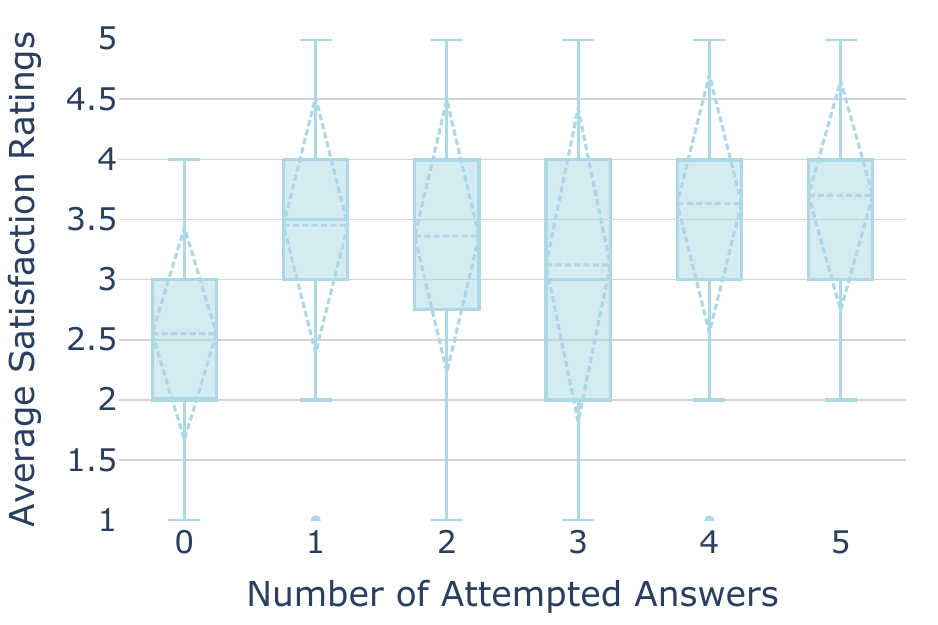}
    \caption{Average satisfaction ratings relative to the number of attempted answers for \textit{dynamic} hints.}
  \end{subfigure}
  \hfill
  \begin{subfigure}[]{0.45\textwidth}
    \includegraphics[width=\textwidth]{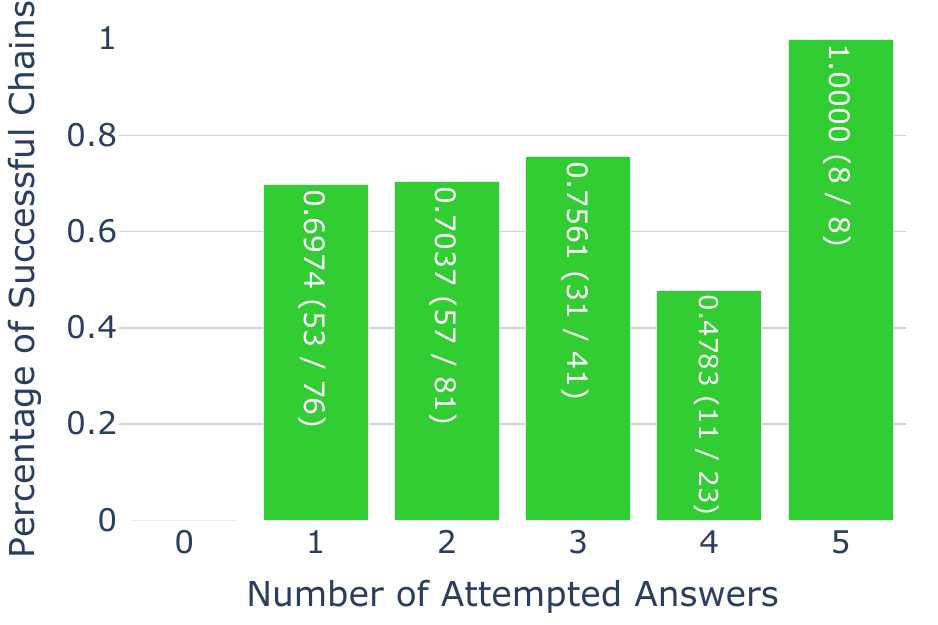}
    \caption{\revision{Percentage of successful \textit{static} hint chains grouped by number of attempted answers.}}
  \end{subfigure}
  \hspace{3em}
  \begin{subfigure}[]{0.45\textwidth}
    \includegraphics[width=\textwidth]{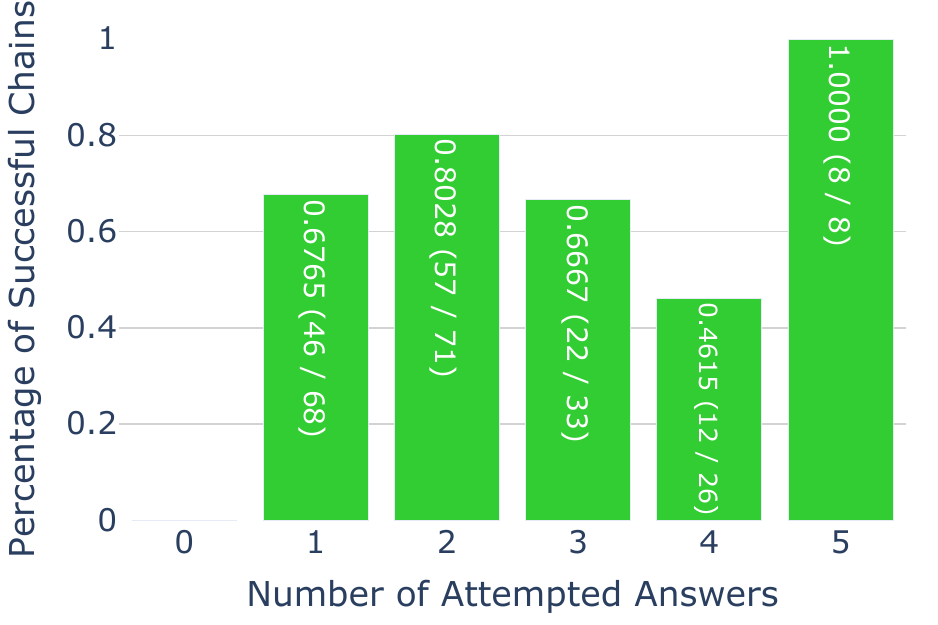}
    \caption{\revision{Percentage of successful \textit{dynamic} hint chains grouped by number of attempted answers.}}
  \end{subfigure}

  \caption{Distribution of hint chains (top), corresponding satisfaction ratings (middle) and \revision{percentage of successful interactions (bottom)} for \textit{static} and \textit{dynamic} settings across attempted answers. \revision{Both static and dynamic hints have similar distribution of attempted answers, however, satisfaction of dynamic hints for instances without interactions is significantly lower than that of static hints. But as the number of attempted answers increase, the satisfaction ratings become similar. After four attempts, the interactions with four attempted answers yielded the least success rate compared to other interactions. This figure only considers interactions where hints were asked.}}
  \label{fig:engagement}
\end{figure*}

%\textbf{Hints helpfulness vary with familiarity.} 
\textbf{The impact of domain familiarity on hint helpfulness.}
Participants described the usefulness of hints to be closely tied to whether the concepts referenced were already familiar to them. If the participants were familiar with the concept, then hints were often helpful in reaching the answer, especially when hints drew on colloquial expressions or widely known facts. For instance, for question "\textit{What is the most abundant metal of the earth's crust?}", the hint "\textit{Consider metals that are commonly used in everyday items like foil and cans.}" was very effective in aiding the participants to answer the question. At the same time, the hints proved less effective when the concepts themselves were unfamiliar. In those cases, hints did not provide sufficient grounding context, leaving participants uncertain about how to proceed. Participants even reported feeling reluctant to request hints in such cases, as P-21 reflected: "\textit{Some questions I don't know about the words, it doesn't make sense to take hints cuz i still don't know the words...}". Using the number of attempted answers as a proxy for engagement (and to some extent familiarity), we observed similar trends, higher engagement instances yielding higher satisfaction ratings (see Figure \ref{fig:engagement}). Therefore, while familiarity could amplify the helpfulness of a hint by triggering recognition and recall, it could also hinder the learning, as without prior knowledge to anchor to, the same hints risked compounding confusion rather than resolving it.

In contrast, several participants emphasized that hints were unhelpful when they already recognized the concept but could not recall the precise term. As P-31 explained, "\textit{If i remember the answer but it was on the tip of my tongue and cant remember the exact one, the hints do not help at all.}".  This highlights a broader constraint of our short-form factoid question-answering setup: the system requires participants to supply the exact target term, meaning that hint effectiveness was often judged not only by whether it supported conceptual understanding, but also by whether it guided them to the expected answer\footnote{While we adopted a fuzzy match for our underlying answer assessment implementation, using in-context learning with an LLM (we used OpenAI's \texttt{gpt-3.5-turbo} for the answer assessment task.), there were instances where it expected the exact lexical form (\textit{e.g.,} for the correct answer \textit{hydroxide ions}, the submission \textit{OH$^-$} was marked incorrect). To ensure the efficacy of our assessment prompt for a functioning study, we test out the quality of the assessment model from a previous pilot study with 697 answers over the same 30 questions. While assessment model (see Figure \ref{fig:assessment_prompt} for the prompt) is able to achieve 88.95\% accuracy over manually labeled responses, we delineate this limitation for real-world deployment in Section \ref{subsubsec:discussion_6.2.1}.}.

% In contrast to the trend of increased helpfulness with familiarity,  \todo{include nicely}

% Hints help guide to the right answer
%\textbf{Hints that are question-agnostic but (sometimes) helpful.} 
\textbf{The tradeoff between goal reaching and knowledge gaining.}
Participants diverged in their views on utility of answer-oriented hints helped that them answer the question without improving conceptual understanding. % \smara{here you need now to say what are answer-oriented hints, this sentence is too abrupt.}
Some appreciated strategies that narrowed the response space % (e.g., "\textit{Think about the rays named after a Greek letter}") 
or partially disclosed the answer (\textit{e.g.,} the hint, "\textit{The name starts with the letter A}", from Figure \ref{fig:case_study} bottom right interaction), even when these cues offered little additional conceptual knowledge. Others, however, criticized these hints as “misleading” or misaligned with their goals. As P-17 reflected, "\textit{[...] A hint which helped me was good but it was not related to the concept itself. I prefer hints which use my external knowledge, but are also somewhat related to the question asked.}" This divergent preference reflects a difference in underlying learning objectives \cite{hoque2016three, sonmez2017association}. For participants primarily aiming to maximize quiz performance, hints that directly enabled correct responses were welcomed. For others seeking to deepen their conceptual understanding, context-grounded hints were preferred—even at the cost of leaving the immediate question unanswered.

%\textbf{Hints at the extremes: too close to the question or the answer.} 
\textbf{Unhelpful hints: close similarity to the question or the answer.}
Participants disliked hints that fell at the opposite ends of the information spectrum, either repeating the questions or revealing the answer directly. Both scenarios limited opportunities for meaningful reasoning. For instance, when asked "\textit{The energy required to remove an electron from a gaseous atom is called?}", one hint rephrased the question ("\textit{Think about the process of removing an electron.}"), while another gave away the answer ("\textit{The term you're looking for involves "ionization"}."). Participants described such hints as either redundant or uninformative, reducing both challenge and conceptual engagement. While such hints could support learners focused solely on correctness, as evident by a weak correlation between the self-reported satisfaction ratings and answer leakage ($corr_{pearson}=0.2226$), others found them frustrating, underscoring the importance of hints that guide the learners without shortcutting the cognitive effort.
% \smara{can you quantify how many users disliked this in percentage of people who were exposed to this type of hints?} \aj{this is well supported by our engagement study with \#attempted answers. Can refer to that (Fig. \ref{fig:engagement}).}

%\textbf{Hints are simple and concise.} 
\textbf{Hint simplicity and clarity.}
From a stylistic viewpoint, participants consistently noted that the hints were straightforward and easy to understand. The hints provided clear, focused guidance, allowing learners to quickly grasp the intended strategy or concept.

% \begin{figure}[t!]
% \centering
% \includegraphics[width=\textwidth]{Artifacts/Figures/fig_case_study.pdf}
% \caption{Interactions between the participants and \textit{static} (left) and \textit{dynamic} (right) hint generation strategies. The top interactions depict a successful journey, whereas the bottom interactions illustrate failure to get participants to reach the answer. \todo{make more clear and aesthetic.} % \smara{hm, the one on the right the user did not give answers or many hints os it is not a good showcase of the dynamic hints.? this we should discuss if they were in a dynamic setting and they did not anwer but rather just asked for hints?}
% }  \label{fig:case_study}
% \end{figure}

\subsection{When One Hint Is Not Enough: The Dynamics of Chain-of-hints} \label{subsec:finding2}

While individual hints often provided clear and concise guidance, participants’ interactions indicated that a single hint was not always sufficient to successfully answer the question. 
In many cases, participants required additional hints to fully grasp the concept, resolve ambiguities, or correct misconceptions. Across the 468 questions where hints were requested, only 32.5\% were resolved with a single hint, whereas 21.5\% required two hints, 13.9\% required three, and 32.1\% required four. These patterns suggest that learners frequently relied on multiple hints to fully grasp the concept, resolve ambiguities, or correct misconceptions. Our analysis shows how the chain-of-hints shaped their overall effectiveness, revealing the conditions under which the participants benefited from successive guidance as well as the pitfalls that emerged along the way.

\textbf{Content diversity (or lack thereof).} Participants often expected successive hints to provide fresh perspectives rather than reiterations of the same clue. While some participants reported frustration when hints felt repetitive, others valued the consistency, noting that reinforcement sometimes helped clarify the intended direction, as explained by P-26, "\textit{Sometimes, I reached a question where I had no idea what the words even meant - but the hints would help me narrow down my guesses and make much more educated guesses [...] I think the hints were good at not providing overlapping information for the most part, and were also quite good and not completely giving away the answer...}". This tension suggests that diversity across a chain-of-hints is not simply about novelty, but about offering complementary information that sustains engagement while gradually deepening understanding. 

\textbf{Coherence and incremental gain.} Participants emphasized the value of hint sequences that gradually introduced new information, allowing them to stay challenged while being guided toward the correct answer. Just as important as diversity, many appreciated when hints were coherent (\textit{i.e.,} building logically on one another). This continuity helped them maintain a sense of flow in their learning journey. Such step-by-step scaffolding not only sustained engagement but also fostered a sense of accomplishment as learners progressed. 
% Participants spent about 45.6\% more time answering the questions in the treatment sections compared to the control no-hint section, averaging about 82 seconds per question with hints. (there were anomalies in the data, actual numbers are very similar)
Overall, participants rated hints as informative in roughly 40\% of cases at the first position, increasing to about 48\% at positions two and three, with a slight decrease to 45\% for the fourth position hints, underscoring the role of successive hints in enriching the learning process.

These findings highlight the challenge in designing effective hint chains. In the context of factual question answering, successful hint chains had to avoid being too close to the question to offer little support, or too close to the answer to risk oversimplification, while striking a balance between providing diverse yet relevant information.
% Designing such chains of hints proved challenging: given the factual nature of the quiz questions, hints needed to strike a careful balance in providing diverse yet relevant information, avoiding being so close to the question that they offered little support, or so close to the answer that they oversimplified the problem.

\subsection{Static vs Dynamic hints} \label{subsec:finding3}

\begin{figure*}[]
    \centering
    \begin{subfigure}[]{0.65\textwidth}
    \centering
    \includegraphics[width=\textwidth]{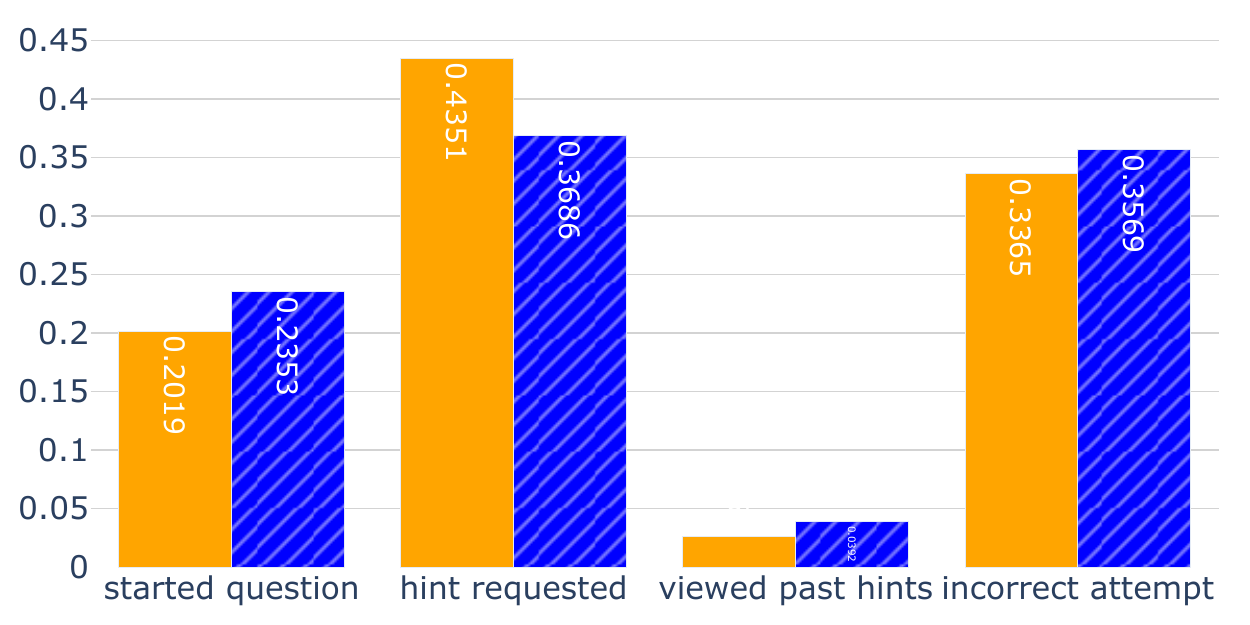}
    \caption{\revision{Distribution of prior actions succeeded by the ``hint request" action.}}
  \end{subfigure}
  % \hspace{3em}
  % \newline
  % \hfill
  \begin{subfigure}[]{0.65\textwidth}
  \centering
    \includegraphics[width=\textwidth]{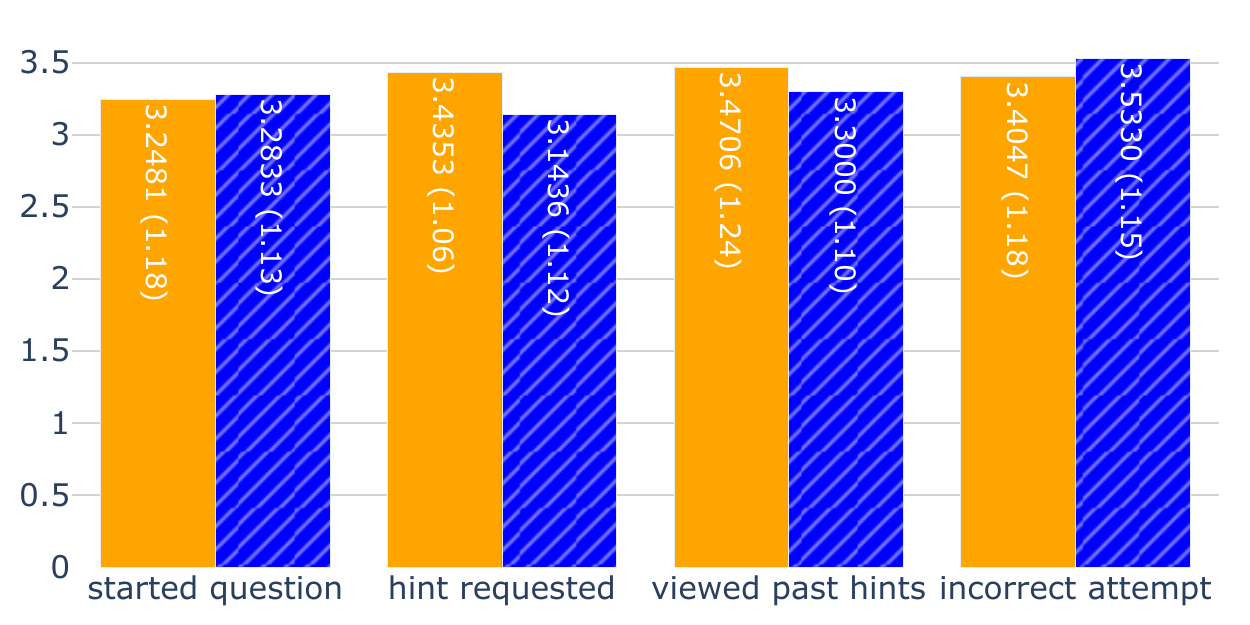}
    \caption{\revision{Average participant reported satisfaction ratings (1-5 Likert scale) grouped by the prior action.}}
  \end{subfigure}
  % \hfill
  % \newline
  \begin{subfigure}[]{0.65\textwidth}
  \centering
    \includegraphics[width=\textwidth]{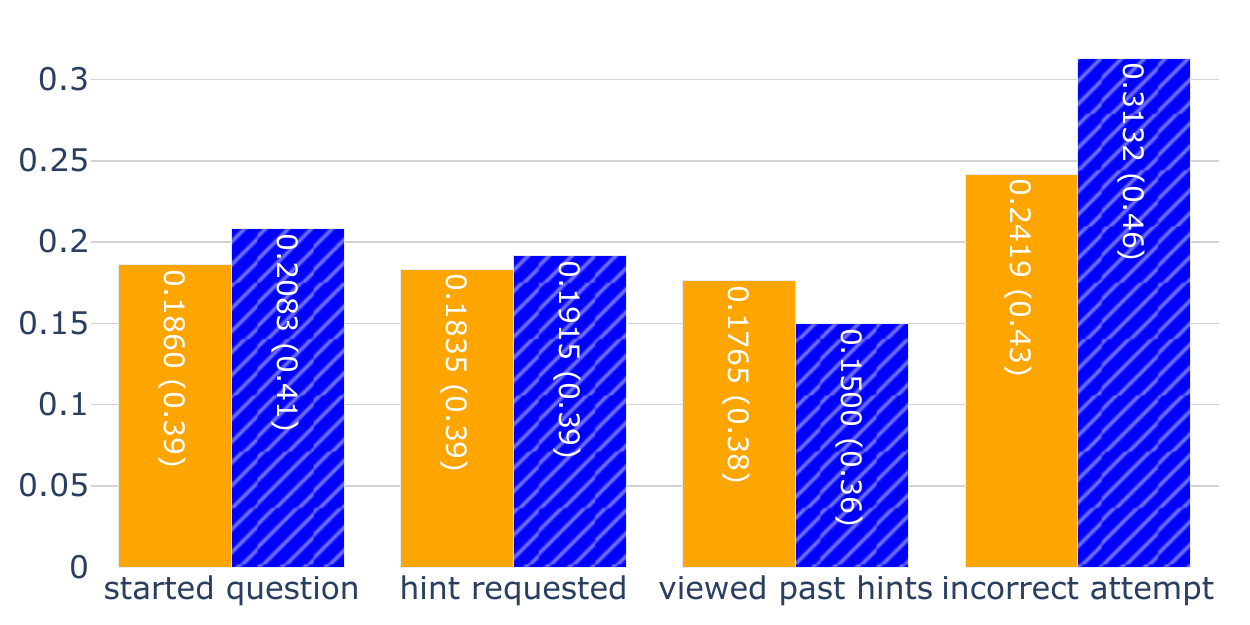}
    \caption{\revision{Percentage of hints yielding immediate success (y-axis) grouped by the prior action (x-axis).}}
  \end{subfigure}

  \caption{\revision{Distribution of the prior actions succeeded by the ``hint request" action (top), average reported satisfaction ratings (on a 1-5 Likert Scale) (middle), and the percentage of the hint request that led to immediate success after that hint grouped by prior actions (bottom). \textcolor{orange}{\textbf{Orange}} (solid) bars denote the \textit{static} hint generation statistics and \textcolor{blue}{\textbf{blue}} (striped) bars denote the \textit{dynamic} hint generation statistics; x-axis denotes the action taken prior to asking a thing, and y-axis denotes the statistic values.
  We observe that dynamic hints succeeding incorrect attempts achieve 31.32\% immediate success, compared to 24.19\% static hints, and obtain slightly higher satisfaction ratings ($\sim$0.13) illustrating that dynamic hints are able to adapt to learners' responses to achieve higher answerability rate.}}
  \label{fig:interaction}
\end{figure*}

We observed minimal differences in participant performance across the static and dynamic hint conditions, with average success of 7.61/10 and 7.88/10 questions respectively, compared to 6.36/10 questions in the no-hint control setting. Participants requested a total of 639 static hints, and 510 dynamic hints. \revision{
Similar interaction patterns emerged across both hint conditions, illustrating substantial engagement with the system, without any design-imposed constraints (see Figure \ref{fig:interaction} top).
% Roughly 22\% of hints were requested at the very start of a question, 34\% after an incorrect attempt, 3\% after reviewing previous hints, and 42\% following a prior hint request (see Figure \ref{fig:interaction} top). 
% In the dynamic-hint condition, participants produced an intervening answer before the next hint in most cases: of 410 interactions, 189 were ultimately answered correctly, 198 included an attempted answer between hints, and only 23 instances (5\%) involved requesting another hint without trying to respond. The static condition showed a nearly identical distribution (172 correct, 211 attempted, 27 unattempted). 
% This indicates that learners typically engaged with the system, even without design-imposed constraints. 
Consequently, the adaptive mechanism in the dynamic condition was regularly engaged, enabling a valid comparison between static and adaptive hinting approaches.
% These patterns suggest that learners generally attempted an answer between hints even without being required to do so, meaning the adaptive mechanism in the dynamic condition was consistently activated—providing a fair basis for comparing dynamic and static hinting strategies.
}
% Similar patterns of learner interactions emerged across both hinting strategies, with $\sim22\%$ hints requested at the beginning of the question, $\sim34\%$ hints requested after an incorrect answer attempt, $\sim3\%$ hints asked after viewing past hints, and $\sim42\%$ hints were requested after having requested a hint already. 
% Across 410 dynamic-hint interactions, participants correctly answered 189 questions, attempted an answer before receiving the next hint in 198 cases, and left only 23 instances (5\%) without an intervening answer attempt. A similar pattern emerged in the static condition (172 correct, 211 attempted, 27 unattempted). These results indicate that learners typically attempted to answer between hints, even without system-imposed constraints. Consequently, the adaptive mechanism in the dynamic condition was regularly engaged, enabling a valid comparison between static and adaptive hinting approaches.

The perceived difficulty of quiz was also rated similarly across all three sections, suggesting that neither hint strategy provided a decisive advantage at the level of overall outcomes. A Mann-Whitney U test \cite{mann1947test} revealed no statistically significant difference between the satisfaction ratings ($U=168351, p=0.365$) with negligible effect size (rank-biserial correlation = -0.033). Even after revealing the corresponding hint strategy to participants in the post-quiz survey, opinions remained divided. 20 out of 41 participants found static hints to be more helpful for answering the questions, while 23 out of 41 participants felt dynamic hints better improved their knowledge about the questions. 
% \smara{Here i I think we want the analysis I asked in the figure captions, what if user keep asking hints without providing incorrect answer then the "dynamic hint" did not really had additional context compared to static... this need to be discussed I think. Maybe in the design implication for future if the user does not attempt to answer always generate static ?Can we look at performance of dynamic when user attemtped to andwrr previous hint vs when they only clicked new hint as is show in Fig 4? }
Point-biserial correlation \cite{kornbrot2014point} between post-section responses and post-quiz responses revealed that the reported quiz difficulty was not correlated with the post-disclosure preference ($corr=0.067, p=0.675$), indicating that the participants independently rated the quiz difficulty agnostic of hint helpfulness. The post-section hint helpfulness was moderately correlated with the post disclosure informativeness preference obtained from the posttest survey ($corr=0.471, p=0.002$), indicating that participants who found one hint strategy more helpful were likely to maintain their preference even after knowing the hint strategy. However, there is no correlation between the performance difference between static and dynamic, and the post-disclosure preference of answering assistance ($corr=0.063, p=0.694$), indicating that participants' preferences were not influenced by their actual performance but rather by their subjective experience with the hints. Although neither strategy prevailed overall, participants described clear differences in how they experienced the two hinting approaches.\looseness=-1

\textbf{Static hints}--pre-generated in chains of four--were rigid in their progression and often provided diverse facts that enriched the context of the question. For example, "\textit{Reflect on the types of landforms created by magma rising to the surface.}", or \textit{"Research the type of radiation that includes visible light."} (see Figure \ref{fig:case_study}). This pattern was reflected in reported informativeness scores: 39\% of static hint chains were labeled informative, compared to only 35\% of dynamic chains. However, Chi-squared test \cite{tallarida1987chi} revealed no statistically significant difference between the two strategies ($\chi^2 = 0.819, p=0.356$) with no association (Cramer's V = 0.027). Providing a breadth of information allowed static hints to support sense-making and knowledge-building, but at times the progression felt misaligned with participants' immediate needs.\looseness=-1

\textbf{Dynamic hints}, in contrast, were answer-fixated and more directly oriented toward guiding participants to the correct response. Because these hints were generated adaptively without awareness of the total number of hints in the chain, they lacked the deliberate pacing and planning of the static setting. However, this immediacy was effective when participants lacked sufficient prior knowledge to answer the question. For instance, hints like, "\textit{Think about the rays named after a Greek letter}", or "\textit{Consider a medication that starts with "a" and is often used for headaches.}" (see Figure \ref{fig:case_study}). At the same time, the absence of chain-level planning made it harder to conceal answers: leakage was reported more often in dynamic hints (18\%) than in static hints (15\%). While Chi-squared test \cite{tallarida1987chi} revealed a statistically significance difference between the leakage responses  ($\chi^2 = 5.35, p=0.021$), there was no association (Cramer's V = 0.068). If the participant's did not engage with the question, (\textit{i.e.,} not attempting any answers) the average satisfaction ratings for dynamic hints was worse than the static hints without engagement. While when they engaged with dynamic hints, their satisfaction ratings were comparable to static hints (see Figure \ref{fig:engagement} middle). 
\revision{While the participant reported satisfaction increased with increased engagement, the success rate of instances for dynamic instances were comparable to static hints (see Figure \ref{fig:engagement} \revision{middle}). However, further investigation at individual hint level revealed that dynamic hints did adapt to incorrect responses. More specifically, the dynamic hints that succeeded an incorrect answer attempt were immediately followed by a successful answer for 31.32\%, compared to 24.19\% times for static hints (see Figure \ref{fig:interaction}) bottom.}
% as well as when participants engaged with the dynamic hints .

Across participants, the overall sentiment remained consistent. Static hints were valued for their contextual richness, while dynamic hints were appreciated for their answer-directed efficiency. Minor exceptions arose due to randomized question distribution, with four participants noting one section to be easier than other, but the broader trend was stable. These findings suggest that static hints may be better suited for fostering understanding and providing transferable knowledge, whereas dynamic hints may be more effective in ensuring task success and engagement when learners are struggling.

This contrasting experience highlights a design trade-off between \textit{insightfulness} and \textit{answer-orientation}. Further exploration is required to find the right balance, potentially by combining the breadth of static hints with the responsiveness of dynamic hints into a hybrid approach that scaffolds learning while still supporting successful task completion. \looseness=-1

\subsection{Assessing effectiveness of automatic evaluation metrics} \label{subsec:autoeval_analysis}

Automatic evaluation metrics are a key step toward scaling hint generation research: they allow us to compare systems without relying exclusively on human annotation. However, their usefulness depends on how well they capture the aspects of hint quality that learners themselves experience. To assess this alignment, we compared a set of automatic evaluation metrics against participant feedback collected in our human study.
In total, 41 participants annotated 1,149 hints across 468 questions. Unlike our standardized automatic evaluation setup with a fixed chain size (see Appendix \ref{sec:autoeval}), the hint chains here varied in length, providing a more realistic but also more heterogeneous basis for evaluation.

We observed only weak correlations between information gain metrics and participants’ reported informativeness of hints for $InfoGain_{mean}$ ($corr_{pearson}=-0.121$) and $InfoGain_{comb}$ ($corr_{pearson}=-0.142$). These negative but shallow correlations highlight the limitations of relying on LLM-based measures of content quality, which often fail to capture the contextual insightfulness valued by participants. Although our study design did not directly test $Redundancy$ or $Consistency$, we also examined their correlations with reported satisfaction. $Redundancy$ showed a weak negative correlation ($corr_{pearson}=-0.245$), while $Consistency$ was essentially uncorrelated ($corr_{pearson}=-0.053$).

For answer leakage, we were able to compare participant reports directly to our automatic metrics. The exact-match metric achieved high precision (0.834) but very low recall (0.075), indicating that it severely underestimated the degree of leakage present. Note that we implemented $leakage_{EM}$ with a simple string matching approach, that is unable to capture the nuances where hints reveal the answer in plural form (\textit{e.g.,} \textit{capillary} vs \textit{capillaries}), or change in part-of-speech (\textit{e.g.,} \textit{volcano} vs \textit{volcanic}). We also identified several indirect oversimplifications from hints that remain undetected by a simple syntactic overlap approach. For example, "\textit{The bohr model works only for which atom?}", participants considered the hint "\textit{It is the first element on the periodic table.}" to have given away the answer, whereas it was undetected by the $leakage_{EM}$ metric. Conversely, the LLM-based leakage metric offered slightly higher recall (0.183) but suffered from low precision (0.266), indicating that while the learners didn't consider certain hints to be oversimplifications, the evaluation LLM considered those hint chains to be so, overshooting the answer leakage predictions. For example, for the question "\textit{What is the most abundant metal of the earth's crust?}", and the provided hints "\textit{Consider which metals are commonly found in rocks and soil.}" and "\textit{Think about the primary component of bauxite, a common ore.}", $leakage_{LLM}$ metric considers the hint chain to have leaked the answer (\textit{Aluminum}), whereas the participant doesn't indicate that in their responses. These results suggest that neither approach yet provides a reliable estimate of answer leakage from a learner’s perspective.\looseness=-1

Taken together, these results indicate that current automatic evaluation metrics only partially reflect learners’ perceptions of hint quality. The weak alignment underscores the need to refine these metrics, either by tuning existing approaches or by developing new measures grounded in interaction data. Our dataset of hint annotations provides a valuable resource for future work toward more learner-centered evaluation.

\section{Discussion} \label{sec:discussion}

Our findings highlight the complexity of automatic hint generation and the limitations of adopting off-the-shelf LLMs as tutoring systems.
% a one-size-fits-all approach to tutoring systems. 
While participants consistently valued certain hint characteristics such as simplicity, novel knowledge perspectives, content diversity, and incremental scaffolding, their preferences diverged on various other aspects. For instance, some learners favored answer-focused hints that guided them toward the answer, while others found such feedback unhelpful in improving their learning. Beyond feedback on the hints generated by our system, participants also emphasized some unmet needs, with the hints unable to help bridge the linguistic barriers or connect well to their prior knowledge. \revision{In Section \ref{subsec:cognitive_offloading}, we discuss the growing challenge of cognitive debt incurred from AI overreliance, and how automatic hint generation can serve as an avenue for addressing this problem. Drawing on our findings, we then discuss the design considerations for advancing hint generation beyond factual question answering in Section \ref{subsec:design_discussion}.}

% We position this research problem as a potential avenue for addressing the growing challenge of cognitive offloading among learners in Section \ref{subsec:cognitive_offloading}). Then in Section \ref{subsec:design_discussion}, we discuss the design considerations for advancing hint generation beyond factual question answering ().

\subsection{Hint Generation as a strategy against Cognitive Offloading} \label{subsec:cognitive_offloading}

Recent years have seen rapid adoption of AI tools across everyday life including education, where students increasingly turn to AI assistance for help with assignments, writing, and problem-solving \cite{pitts2025student, pitts2025students, kanont2024generative, lin2024factors}. Prior work has documented this growing dependence, noting that learners often use these systems not only as aids but also as substitutes for their own cognitive effort \cite{adiyono2025impact}. 
\revision{Recent studies further demonstrate the risks associated with excessive reliance on AI. For example, using electroencephalography signals, \citet{kosmyna2025your} found reduced neural coupling during AI-assisted essay writing. Similarly, utilizing a mixed-method approach over a large set of participants, \citet{gerlich2025ai} observed that heavier AI reliance correlates with weaker critical thinking scores among younger learners.}

\revision{Our findings further contribute empirical evidence to this concern. Learners exhibited behaviors consistent with cognitive shortcuts, favoring hints that simplified answer retrieval rather than those that promoted reflection. This illustrates that over-reliance can arise even when systems are explicitly designed not to give full solutions, suggesting that interface-level decisions and scaffolding structure substantially influence learners’ help-seeking strategies.}

\revision{While these risks are well-documented, recent systems-level approaches propose countermeasures such as systems that emphasize explainability \cite{hoq2025explainable}, promote metacognitive reflection \cite{kumar2024supporting, yuan2024generative}, or deliberately withhold direct answers \cite{gupta2025beyond}. Our findings refine these design arguments; we show that simply withholding solutions is not sufficient as learners may still gravitate toward the most cognitively economical hint type. Instead, our data suggest that how hints are sequenced and when learners choose to request them are critical determinants of whether the system supports or undermines cognitive engagement.
}

% Hint generation represents one such framework. 
\revision{
% Hint generation provides one avenue for maintaining this balance. 
When well-designed, hints can scaffold reasoning while preserving opportunities for productive struggle. Prior work shows that effective hints promote motivation and self-efficacy \cite{rogers2025playing, mohammadi2025effect, vanlehn2011relative, ma2014intelligent}.} 
Our findings reinforce this perspective. Participants generally rated hints as helpful for answering questions, with several explicitly noting their value in promoting deeper thinking. For instance, P-22 described hints as "\textit{hints encouraged to make associations and engage in reflection from a scientific perspective}," P-37 noted they "\textit{encouraged active thinking}," and P-40 highlighted they "\textit{encouraged me to look for other ways to say things}." 
\revision{We found significant correlation between participants' satisfaction with their information gain ($corr_{pearson}=0.4955$), indicating preference towards novel but relevant information conveyed through hints.}
\revision{At the same time, the contrast between static and adaptive hints in our study illustrates that subtle design choices can meaningfully shift the cognitive engagement patterns, and thereby the learning outcomes. % with evidence of diverging cognitive engagement patterns across two comparable strategies.
% This nuance adds to the existing literature by showing that hint systems can influence cognitive engagement.
}

\subsection{Design Implications Beyond Factual Hinting} \label{subsec:design_discussion}

The revised Bloom’s taxonomy of educational goals \cite{bloom1956taxonomy, anderson2001taxonomy} organizes cognitive processes from remember and understand through apply, analyze, evaluate, and create, while distinguishing knowledge goals between factual, conceptual, procedural, and metacognitive dimensions. Our study deliberately focused on factual knowledge at the lower levels of this hierarchy, where hints primarily supported remembering and understanding. This scope allowed us to systematically examine the role of hinting, while revealing several limitations in current technology that impede reliable implementation in authentic educational settings. For instance, challenges in robust answer assessment and misalignment between automatic evaluation metrics and learner perceptions. As one moves upward in Bloom's hierarchy, such limitations become even more consequential: supporting conceptual reasoning, procedural mastery, or metacognitive reflection requires richer forms of adaptation, sensitivity to learner goals, and integration of affective and self-regulatory scaffolding. Here we outline design implications that extend beyond factual hinting, drawing on our findings to point toward future directions for building tutoring systems capable of supporting higher-order learning. 
% As learning science research moves upward in Bloom’s hierarchy, such limitations carry higher stakes. We need richer forms of adaptation, sensitivity to learner goals, and self-regulatory scaffolding to support conceptual reasoning, procedural mastery or metacognitive reflection. 
% Supporting conceptual reasoning, procedural mastery, or metacognitive reflection requires richer forms of adaptation, sensitivity to learner goals, and integration of affective and self-regulatory scaffolding. 
% In the following subsections, we reflect on our results toward outlining design implications that extend beyond factual hinting, and highlighting future directions for building tutoring systems capable of supporting higher-order learning.

\subsubsection{\revision{Building Trustworthy Foundations: Towards Reliable and Interpretable Tutoring Systems}} \label{subsubsec:discussion_6.2.1} \hfill
% line of summary / signposting... what do you mean by trustworthy foundations 

\revision{
Building AI tutoring systems that learners and educators can rely on requires more than producing fluent hints—it demands transparent, predictable, and pedagogically aligned foundations. Trustworthy systems make their reasoning traceable, provide consistent behavior across learners and contexts, and enable researchers to diagnose and mitigate errors. In the context of hint generation, this involves rigorously validating answer assessment for user interaction experience, building pedagogy-driven evaluation measures, and developing steerable and transparent hint generation systems that meet the learner's goals. The following passages outline these challenges and potential pathways towards developing reliable and equitable AI tutors.
}

\textbf{Robust Answer Assessment.} To evaluate the reliability of our answer assessment module for the user study, we manually annotated 697 attempted answers from a pilot study. The in-context learning system achieved an accuracy of 88.95\% in correctly labeling learner responses. While answer assessment was not the central focus of this work, we considered this level of performance sufficient for conducting a controlled user study. Nonetheless, even in the relatively constrained setting of short-form factual questions, our system encountered challenges in reliably assessing some learner responses. For example, while it correctly accepted “CO$_2$” as an equivalent to “carbon dioxide,” it failed on other equivalent forms such as “OH$^-$” for hydroxide ions, and in one case incorrectly marked “ions” as fully correct. Such inconsistencies highlight the fragility of off-the-shelf LLM-based assessment pipelines and underscore how errors, if unaddressed, can affect learner trust and the validity of downstream evaluation.
% Similarly, responses like “isotope” were rejected against the ground truth “isotopes,” despite being logically acceptable. 
% These inconsistencies illustrate the fragility of off-the-shelf LLM-based answer assessment pipelines, and how such failures can undermine both learner trust and the accuracy of downstream evaluation, illustrating a gap in existing technologies for real-world deployment. 
% \smara{As I said I am very concerend about this, unelss we show it was not on a large scale, the question is: how reliable is the study if the users were giving good answers but models said was not correct.... }

The difficulty compounds when moving beyond factual recall, as assessment approaches vary across domains. In subjects such as mathematics, robust evaluation requires formal verification of solutions and scrutiny of intermediate steps to detect logical errors \cite{sangwin2016automation, lan2015mathematical, beevers2003automatic}. In contrast, domains like history or biology demand capabilities such as argumentation mining and semantic overlap analysis \cite{lippi2016argumentation, lawrence2020argument, li2025large}. For instance, in answering the open-ended question “\textit{Why did the Western Roman Empire fall?},” there could be multiple possible responses, but an effective assessment % no single response is sufficient, but a fair assessment 
would align the learner’s reasoning with key factors outlined in reference solutions.
Building reliable assessment modules that are rigorously tested and aligned with the cognitive goals of instruction is therefore both a significant design challenge and a necessary foundation for developing effective, equitable feedback systems that support higher-order learning objectives.\looseness=-1

% Moving beyond factual recall only amplifies these challenges: with approaches varying across subjects. For formal subjects like mathematics, on formal verification of solution while scrutinizing individual steps for logical misakes is important \cite{}, whereas, for informal logic subjects like history and biology, successful assessment systems must be capable of argumentation mining and overlap analysis \cite{}. For instance, for the open-ended question “\textit{Why did the Western Roman Empire fall?}”, multiple plausible answers exist with no single correct response, but a correct assessment would evaluate the attempted answer with the factors in the reference solution that a student must be aware of. Building reliable assessment modules that are rigorously tested and aligned with the cognitive goals of instruction is therefore a major design roadblock, but also a necessary prerequisite for developing effective and fair feedback systems that support higher-order learning objectives. 

\textbf{\revision{Using LLMs for data synthesis and hint evaluation.}}
% \smara{this is good overall but I think we need to say in what percentage there were error, to show it is a small fraction otherwise to me your study is no reliable (if models tell user they are wrong but they are not in many cases, what do we learn? So we need to be clear this is small scale and the assessment was fairly constraint to factual answers where LLMs can judge easier if two phrasing are similar or not?}
% below, we discuss the possibilities / limitations of using large language models to generate data for training / evaluating educational applications. 
\revision{
LLMs are increasingly adopted in two major ways within educational technology research: generating synthetic data and serving as automated evaluators. For data synthesis, recent work shows that role-playing \cite{wang2025training} and simulated tutoring dialogues can be used to fine-tune smaller models \cite{zhang2025sefl} or optimize tutoring policies through LLM-based reward models \cite{dinucu2025problem}. These approaches promise scalability but also inherit the representational biases of the models generating the synthetic learners \cite{gautam2024melting, ghosh2024generative}, which may distort pedagogical signals, especially outside well-structured domains.
}

% \revision{
% Recent work has begun using synthetic student–teacher interactions to train educational models \cite{zhang2024simulating, yuan2025simulating}, offering a potentially scalable but underexplored direction for future systems. \citet{zhang2025sefl} show that role-playing LLMs can generate synthetic tutoring dialogues used to fine-tune smaller models, and complementary efforts such as \citet{dinucu2025problem} use LLM-based reward models and group relative policy optimization \cite{shao2024deepseekmath} over synthetic scenarios to optimize tutoring behavior. While promising, these approaches carry representational risks: synthetic learners reflect the biases and limitations of the models that generate them \cite{gautam2024melting, ghosh2024generative}, which can distort pedagogical signals—particularly outside structured domains like math and science.} 
% As educational LLMs increasingly rely on synthetic data and LLM-as-judge supervision, careful scrutiny is needed to ensure that these systems do not amplify bias or collapse toward self-reinforcing patterns of artificial student behavior.

\revision{
A parallel line of research uses LLMs as evaluators (``LLM-as-a-judge") \cite{gu2024survey} to assess quality of educational content \cite{dinucu2025problem, li2025exploring, benedetto2024using, lu2024generative, mannekote2025can, qian2025dean}. This method offers clear practical advantages over human studies with lower cost, fewer logistical barriers, and rapid iteration. However, educational evaluation is uniquely sensitive, as it involves assessing nuanced learner behaviors, metacognitive processes, and affective states—dimensions that the use of LLMs might oversimplify or misrepresent. Our own findings reflect this tension, the LLM-based metrics for informativeness and answer leakage showed weak alignment with learners’ judgments of hint quality, suggesting that LLM–based judges fail to capture the contextual insightfulness learners valued. Prior work similarly highlights that LLM evaluators overidentify student errors for math tutoring \cite{kakarla2024using}, and while fluent, often misaligned with instructor assessments for project report assessment \cite{dai2023can}.
}

\revision{
Investigations on LLM-based evaluators revealed that they are prone to systemic biases \cite{chen2024humans, ye2024justice, shi2024judging, wang2023not}, and are vulnerable to adversarial manipulation \cite{tong2025badjudge}. Self-assessment studies demonstrate that while LLMs may show high overall agreement with human annotators, they often fail on subsets of tasks they cannot themselves solve \cite{krumdick2025no}. Additionally, LLM-based evaluation performs poorly in expert domains (e.g., dietetics, mental health), with substantially lower agreement than human judges \citet{krumdick2025no}. Beyond performance metrics, qualitative investigations raise broader concerns: \citet{kapania2025simulacrum} showed that when LLMs are used to simulate participants, their outputs often lack contextual depth, obscure marginalized perspectives, and amplify researcher positionality.
}

\revision{
Together, these observations suggest that while LLM-driven synthesis and evaluation can help scale research, they are not yet dependable proxies for real learner experience. Progress will require metrics grounded in authentic interaction data and stronger validation protocols. Our annotated corpus of 1,100+ hints across 450+ questions provides one step toward more trustworthy evaluation frameworks, but future systems must adopt these techniques with caution to avoid reinforcing misaligned pedagogical signals or undermining trust in educational tools.
}

\revision{
\textbf{In-Context \textit{vs.} Post-Training Approaches to Hint Generation.} LLM-based hint generation generally follows two paths: (1) using in-context learning to produce feedback during a learner’s interaction, or (2) post-training models on task-specific corpora to develop specialized tutors. Because high-quality educational datasets are scarce, most prior work—including ours—follows the in-context approach, relying on off-the-shelf reasoning capabilities of LLMs to scaffold learners \cite{gabbay2024combining, dai2023can}. For instance, \citet{pardos2024chatgpt} demonstrated that combining learner-authored prompts with chain-of-thought and self-consistency decoding \cite{wang2022self} can improve algebra and statistics tutoring, though at higher computational cost. Similarly, \citet{wan2024exploring} found that few-shot prompting enabled physics explanations that students preferred and domain experts judged as requiring only minimal edits.
}

\revision{
Our findings align with these observations: in-context hints increased learner answerability by 13.8\% relative to a no-hint control. However, our automatic evaluation over eighteen LLMs also shows substantial variance across model families and sizes--strong feedback is not a universal property of current LLMs (see Figure \ref{fig:scale_autoeval_app}). This motivates attention to the second paradigm: targeted post-training. Prior work demonstrates its potential; for example, \citet{scarlatos2025training} trained a Llama3-8B \cite{grattafiori2024llama} model with direct preference optimization to outperform GPT models on mathematics tutoring benchmarks. Yet post-training is computationally demanding and hinges on carefully curated, pedagogically grounded datasets. As \citet{dinucu2025problem} showed, even within the same domain, different alignment strategies (e.g., supervised fine-tuning vs. RL-based optimization) can yield widely divergent results.
}

\revision{
Taken together, these findings suggest that evaluating in-context learning is necessary but not sufficient for trustworthy educational systems. Reliable scaffolding will require progress on two fronts: building stronger benchmarks for model evaluation and developing principled post-training pipelines grounded in educational theory and interaction data. 
% We encourage future work to advance both paradigms, particularly the relatively underexplored post-training approaches that may ultimately offer more interpretable and pedagogically aligned behavior.
}

\revision{
\textbf{Trade-offs between closed-source and open-source LLMs.} 
A key distinction in our work is an explicit emphasis on open-source models, in contrast to much prior research that has relied on proprietary GPT systems \cite{jukiewicz2024future}. Closed-source models such as ChatGPT offer strong off-the-shelf performance, broad adoption among students, and rapid prototyping benefits for educational applications. However, their opacity limits opportunities for fine-grained analysis, reproducibility, and transparent evaluation of tutoring behavior \citet{balloccu2024leak}. Open-source alternatives, while often less capable without additional tuning, provide inspectability, controllability, and the ability to adapt models to curriculum-specific or domain-specific needs with modest compute budgets. Future work should consider the trade-offs between open-source and closed-source approaches: closed-source models remain valuable for advancing prototype performance, while open-source models are essential for reproducible research, equitable deployment, and sustained advances in educational AI.
}

% A key distinction from prior work is our emphasis on open-source models rather than proprietary GPT systems. Much existing research relies on ChatGPT because it is widely used by students and convenient for prototyping tutoring systems; however, its closed-source nature limits transparency, constrains reproducibility, and raises issues around data privacy. Open-source models, by contrast, offer inspectability and the potential for fine-grained adaptation to learner needs with specialized support using significantly less computational resource, but they require substantial infrastructure and often exhibit weaker out-of-the-box pedagogical performance. By studying hint generation in open-source LLMs, we highlight challenges that may be obscured when relying solely on highly capable proprietary models and surface design considerations necessary for building equitable, verifiable educational tools.
% future work should consider the trade-offs between open-source and closed source models. For advancing model capabilities it is better to use in research... 
% trade-offs and tensions...
% }

\subsubsection{Designing for Learner Diversity and Goals}\hfill

% limits of "one-size-fits-all" tutoring solutions. \smara{Hm, a lot of tutoring systems are designed for personalized feedback and user modeling... at least the older systems... so to me this feels like you ignore all the tutoring systems research to date. mabye here you need to be more nuanced and argue that in the current debate between general-purpose models and specialized models, when student learning is consider we have to opt for specialized user-centric/personalized systems}.  
\textbf{Designing for diverse learners.} Our study highlights the limitations of relying solely on off-the-shelf general purpose models in educational context. Recent years have seen a shift from specialized models towards large, general purpose foundational models \cite{bommasani2021opportunities, nori2023can, li2024multimodal}. While training on massive datasets has produced highly capable systems with emergent abilities \cite{wei2022emergent, berti2025emergent}, this trend risks discarding lessons from decades of research in the ITS community, which emphasized specialized, learner-centric frameworks \cite{vanlehn2011relative, zhang2017evaluating, conati2021toward}. 
Our findings also show that learners have different language proficiencies, educational backgrounds, and goals, which shape how they interpret and act on hints. As one participant (P-22) explained, "\textit{...I know how to answer them in my native language, but in English I get limited by my vocabulary. (I used to be able to spell and handwrite these words, but it’s already been five years since I graduated from high school! I’ve forgotten how to write them, though I can still recognize the words when I see them.)}" P-22's experience exemplifies the importance of designing systems that accommodate linguistic diversity and recognize that learners’ needs vary widely. 
We argue for a balanced approach that combines the reasoning capabilities of LLMs in conjunction with insights from specialized models to better cater to individual learner needs, while adopting non-invasive methods for modeling preferences. One promising direction is to infer learner goals and prior knowledge from interaction data. For example, \citet{liu2022open}, utilizes long short-term memory cells \cite{hochreiter1997long} for knowledge estimation \cite{corbett1994knowledge} of student's current understanding based on their past responses for open-ended program synthesis. We can adopt user behavior modeling techniques from the vast literature of recommendation systems \cite{he2023survey} to measure progress and adapt the hints to the learner's preferences. Beyond personalization, future work should also rigorously evaluate tutoring approaches across more diverse populations to ensure that adaptive systems provide equitable and inclusive support.

\textbf{Aligning with learner goals.} Every learner brings different objectives depending on the task and context \cite{hoque2016three, sonmez2017association}. Our findings reveal a similar learners' preference variability, where some participants valued question-agnostic hints that guided them toward answers, while others who sought deeper understanding found these unhelpful. While providing context agnostic feedback to facilitate successful responses from learners improves short-term satisfaction, it hinders in development of critical thinking. While it is important to identify individual learning goals to provide the most suitable feedback, further caution is required to prevent over-reliance and over simplification. % In Section \ref{subsec:cognitive_offloading}, we discuss this tension of satisfaction with cognitive offloading in depth.
Besides that, even identifying these learning goals isn't a straightforward task, as prior work has shown, learners often lack the metacognitive awareness to articulate their own goals \cite{karpicke2009metacognitive, ku2010metacognitive}, and their learning preferences evolve over time \cite{gurpinar2011learning, andrews2011changing}. Thus, tutoring systems must balance explicit input from learners with adaptive mechanisms that infer goals through interaction patterns. Metrics such as satisfaction ratings or longitudinal engagement may provide richer signals for tailoring support, pointing to the need for nuanced feedback frameworks that remain grounded in pedagogical values.

\textbf{Toward scaffolding self-regulated learning.} 
% While mixed-initiative frameworks can compensate for learners who struggle to recognize when support is needed, c
% \smara{all previous discussion points show some clear connection to your findings but this seems almost future work that can be discussed in general}
We found that even the hint generation systems explicitly designed to promote cognitive engagement can still exhibit a tendency to produce overly answer-oriented hints (Section \ref{subsec:cognitive_offloading}). 
% built to encourage cognitive engagement can suffer from the tendencies to prefer answer-oriented hints (Section \ref{subsec:cognitive_offloading}. 
% Prolonged exposure to this fixed technology can reinforce harmful overreliance on the very hint generation systems built to prevent cognitive offloading. 
Prolonged exposure to such systems may inadvertently reinforce overreliance, amplifying the cognitive offloading they were intended to mitigate.
Therefore, cultivating metacognitive skills is essential for fostering critical thinking \cite{ku2010metacognitive}. Tutoring systems should not only deliver assistance but also nurture learners’ capacity for self-regulation. Two pedagogical paradigms with strong empirical support offer promising directions: self-regulated learning (SRL) \cite{zimmerman1990self, zimmerman2002becoming, prasad2024self} and learning-by-teaching (LBT) \cite{frager1970learning}. Both approaches have long-standing evidence of effectiveness in enhancing metacognition and deep learning \cite{allen1973learning, bargh1980cognitive, chi2001learning, cohen1982educational, cheng2011role, bouchet2012identifying}, and recent advances in AI have opened new possibilities for embedding them in digital environments. For instance, \citet{ge2025srlagent} demonstrate that LLM-assisted SRL can significantly improve regulation skills when scaffolded within a gamified environment. Likewise, \citet{rogers2025playing} show how positioning LLMs as “tutees” enables learners to practice learning by teaching, thereby strengthening conceptual understanding as well as metacognitive reflection. Building on these insights, we envision future hint generation systems that extend beyond factual recall: systems that scaffold reflection, gradually fade support, and empower learners to assume increasing responsibility for their own learning. Such designs not only reduce risks of over-reliance but also promote the development of independent, critical thinkers.

\section{Limitations} \label{sec:limitations}

Our study has several limitations that should be acknowledged when interpreting the findings. \textbf{First}, there are recruitment-related constraints. While our target demographic was high school students, ethical considerations around exposing minors to early-stage AI technology led us to recruit recent high school graduates instead. As a result, our participant pool consisted primarily of university students. \revision{
% We acknowledge that this creates a mismatch between the intended learner population and the actual participant pool, which may limit the generalizability of our findings to secondary-level learners. 
Rather than treating the study as a direct evaluation of high school science learning, we position it as an exploratory investigation of hint–learner interaction dynamics in a controlled scientific question-answering setting. 
% Using factual science questions allowed us to probe model behavior and hint mechanisms in a domain with well-defined correctness criteria, even though our participants were more advanced than the target learners. 
% Future work should extend this analysis to authentic high school populations and to content domains aligned with their background knowledge, such as conceptual science questions, financial literacy, or civic reasoning.
}
% This group may hold more favorable views of AI technologies compared to younger learners, as suggested by prior work [CITE], potentially biasing perceptions of the systems tested in our study. \smara{there is a discrepancy also between first year undergrad and junior or senior. so if you have first year that also might be more relunctant, so to me you }

\textbf{Second}, our study design reflects several methodological choices that constrain the scope of our findings. By focusing on factual questions, we primarily probed for remembering and understanding, rather than deeper levels of cognition such as application or analysis. This limits the generalizability of our insights to higher-order learning tasks. 
% \revision{To maintain the naturalistic flow and minimize cognitive load from the questionnaires, we did not collect qualitative feedback for hint-specific feedback, and relied on categorical responses. This limited our investigation of hint characteristics to the three measures of satisfaction, informativeness and leakage. Future work should design more fine-grained procedures to gather deeper insights about hint utility.}
% Moreover, our learner-driven setup assumes that participants possess sufficient metacognitive awareness to judge when to request hints. Prior research \cite{karpicke2009metacognitive, ku2010metacognitive} suggests that this assumption does not always hold true, as many learners lack these skills, which in turn restricts the broader applicability of our proposed hint generation system. \revision{First-party perspective to gauge learning might not correlate with learning either, as \citet{deslauriers2019measuring} found that learners' feeling of learning might be negatively correlated with actual learning.}
% \revision{[Add ``Measuring actual learning versus feeling of learning in response to being actively engaged in the classroom."]} 
Additionally, our study also doesn't probe the participants' views of AI technologies, a factor that can affect the applicability of the results to broader learner population. 

\textbf{Third}, there are limitations tied to the implementation of our hinting strategies. \revision{Our work does not conduct an exhaustive search in the prompt space for hint generation, and uses same prompts across all models. Due to lack of prior interaction data, we limit the investigation to zero-shot prompts, that lack in-context supervision to improve performance \cite{brown2020language}. For a fair assessment and comparison across several models, the hint generation prompts are quite simple. We leave the investigation of using behavioral signals, learning feedback loops and expert-written instructions in hint generation prompts to future work.} 
Additionally, the dynamic hinting condition was unaware of the hint chain length, making it more difficult to plan hint sequencing and leading to a somewhat uneven comparison with the static condition. Furthermore, our system was agnostic to learners’ backgrounds and prior knowledge: while hints adapted superficially to student responses, they did not provide deeper personalization that could account for individual learning histories or subject expertise. Besides prior knowledge, our design also didn't explicitly take into consideration affective dimensions like frustration, disengagement, or overconfidence, even though they are critical for learning \cite{mao2010agent, cunha2018incorporating, yadegaridehkordi2019affective, hasan2020transition}. 

\revision{
\textbf{Fourth}, our study relies on the \texttt{SciQ} dataset \cite{welbl2017crowdsourcing} as the source of question–answer pairs. Although the dataset is derived from high school science materials, it was originally constructed for benchmarking question-answering systems rather than for evaluating pedagogical scaffolding. As a result, the factual and decontextualized nature of SciQ limits the ecological validity of our setting and may not fully capture the complexity of real instructional interactions. Moreover, our automatic evaluation pipeline for answer assessment and detecting leakage relies on LLM-based judgments rather than human-annotated ground truth. This introduces potential self-consistency bias, reproducibility concerns, and misalignment with human perceptions of hint quality. While we compliment this by conducting a user study, the dependence on LLM-based automatic metrics to select the model for user study remains a limitation. Future work should develop pedagogy-grounded benchmarks and human-validated evaluation protocols tailored specifically for assessing hinting and scaffolding behaviors.
}

Taken together, these limitations highlight important opportunities for future work.  Recruiting learners closer to the intended demographic, \revision{expanding beyond factoid questions} to incorporate higher-order reasoning tasks, and designing interfaces that support diverse help-seeking behaviors will improve ecological validity. \revision{Just as critically, moving beyond QA-oriented datasets like \texttt{SciQ} and reducing reliance on LLM-based self-evaluation will be necessary to build more reliable, pedagogy-grounded benchmarks for assessing hint quality.} Addressing these gaps will help refine automatic hint generation into a more robust and equitable tool for supporting diverse educational contexts.

\section{Conclusion}

This paper examined the capabilities of large language models to generate and evaluate hints for science education. We benchmarked 18 open-source models using five automatic evaluation metrics grounded in learning sciences, each designed to capture a distinct attribute of effective feedback. We investigated two complementary hinting strategies: \textit{static} hints, pre-generated for each problem, and \textit{dynamic} hints, adapted to learners’ progress. Our quantitative study with 41 participants underscores both the promise and complexity of providing automated hints. Participants valued certain attributes that made hints effective, while also identifying aspects that hindered their learning experience. Building on these insights, we propose design principles for future systems that align with Bloom’s taxonomy of educational goals and emphasize learner-centered design. More broadly, we position automatic hint generation as a key research direction to address the growing challenge of cognitive debt among next-generation learners.

% This paper examines the capabilities of large language models to generate and evaluate hints for science education. We benchmark 18 open-source baseline models using five automatic evaluation metrics, designed to quantify distinct characteristics of a successful feedback based on the literature of learning and education sciences. We investigate two complimentary hinting strategies: static hints, pre-generated for each problem, and dynamic hints, adapted to learners progress. Our quantitative study with 41 participants reveals the complexity of providing hints to learners, identifying certain key attributes a successful hint should portray, while highlighting some aspects of hints they disliked. We conclude by highlighting key design principles future work should take into consideration, while building up in Bloom's taxonomy of educational goals. We also position the problem of automatic hint generation as one research direction to tackle the rising cognitive debt in next generation learners. 

% \newpage

% \begin{acks}
% \end{acks}

\bibliographystyle{ACM-Reference-Format}
\bibliography{refs}

\clearpage

\appendix
\section{Automatic Evaluation of Scientific Chain-of-Hints} \label{sec:autoeval}

% Reliable automatic evaluation is critical in advancing research on automatic hint generation, and intelligent tutoring systems broadly \aj{Add citation here}. These automatic evaluation metrics provide a fast feedback cycle in early-stage system development, while safeguarding deployment of underdeveloped systems to learners, which raises ethical concerns. Exposing students to low-quality hint generation systems may not only undermine the learning outcomes, but also puts at risk learner's trust in educational technologies more broadly. Having robust automatic evaluation metrics also helps the community to ensure comparability and reproducibility of research findings across different studies. 

% Therefore, in this section, we establish several automatic evaluation metrics, either by repurposing metrics from prior works, or proposing new ones wherever necessary. To access the quality of a \textit{chain-of-hints} in helping a learner answer a question, we propose automatic evaluation across five different dimensions - \textit{information gain}, \textit{redundancy}, \textit{consistency}, \textit{readability}, and \textit{leakage}. 

We describe and motivate each automatic evaluation metric in Section \ref{subsec:autoeval_metrics}, and discuss how capable various open-source LLMs are in generating scientific chain-of-hints in Section \ref{subsec:autoeval_results}.

% Need to develop appropriate and extensive evaluation metrics for community. our effort to grasp various hint attributes portrayed by human expert tutors in this section. \todo{$\leftarrow$}

% To automatically measure the quality of a sequence of hints to help a user answer a question successfully, we propose to use five evaluation metrics spanning several key aspects of a successful hint - \textit{information gain}, \textit{redundancy}, \textit{consistency}, \textit{readability}, and \textit{leakage}. In this section, we describe these automatic evaluation metrics in detail (Section \ref{subsec:autoeval_metrics}), and discuss how several open-source LLMs perform using this evaluation metrics (Section \ref{subsec:autoeval_results}). 

\subsection{Automatic Evaluation Metrics} \label{subsec:autoeval_metrics}

% \aj{I can also device another informativeness measure that looks at where each hint is situated relative to the QA pair, and see if the degree of informativeness is monotonically increasing. (Using the cosine similarities b/w hints and question answers).}

\begin{figure*}[h]
\centering
\includegraphics[width=\textwidth]{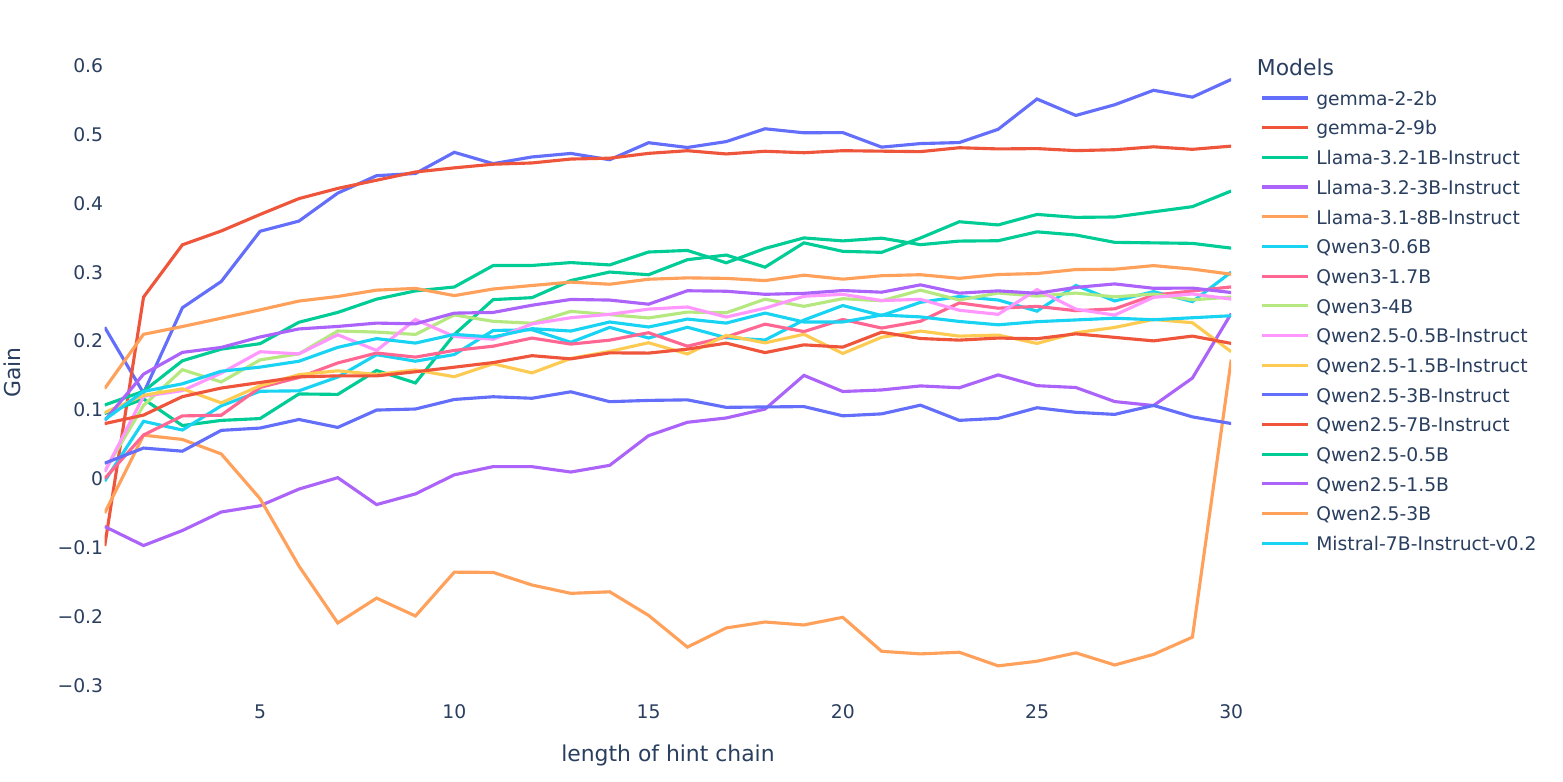}
\caption{Information gain of ROUGE-L on validation set for randomly selected 30 hints using \texttt{Gemma-2}, \texttt{Llama-3}, \texttt{Mistral}, \texttt{Qwen2.5} and \texttt{Qwen3} models.} \label{fig:info_gain}
\end{figure*}

\noindent\textbf{Information Gain.} 
To measure the extent to which a hint can add information to the knowledge base of a user to help answer a question is one of the most important attributes of a hint. To interpret whether a user will be able to gain the necessary information from a hint, we use a large language model to measure the information gain from a hint. More specifically, we measure the difference in performance of the evaluation model with and without the hints. Due to the tendency of LLMs to generate verbose responses, we use \texttt{ROUGE-L} recall \cite{lin2004rouge} to match the generated response to the correct answer as a scalable approximation for assessment. We propose two different variants for this metric: i) average information gain across all hints ($InfoGain_{mean}$), and ii) combined information gain from all hints ($InfoGain_{comb}$), defined as follows - 
% \begin{equation*}\label{eq:info_gain1}
%     InfoGain_{mean}(\textbf{h}) = \frac{1}{k}\sum_{i=1}^{k} ROUGE_{recall}^{L}(LLM(q, h_i), a) - ROUGE_{recall}^{L}(LLM_{QA}(q), a) 
% \end{equation*}
\begin{equation}\label{eq:info_gain1}
\begin{split}
   InfoGain_{mean}(\mathbf{h}) 
   &= \frac{1}{k}\sum_{i=1}^{k} 
      ROUGE_{recall}^{L}\bigl(LLM_{QA}(q, h_i), a\bigr) \\
   &\quad - ROUGE_{recall}^{L}\bigl(LLM_{QA}(q), a\bigr)
\end{split}
\end{equation}

% \begin{equation*}\label{eq:info_gain2}
%     InfoGain_{comb}(\textbf{h}) =  ROUGE_{recall}^{L}(LLM(q, \textbf{h}), a) - ROUGE_{recall}^{L}(LLM_{QA}(q), a) 
% \end{equation*}
\begin{equation}\label{eq:info_gain2}
\begin{split}
   InfoGain_{comb}(\mathbf{h}) 
   &= ROUGE_{recall}^{L}\bigl(LLM_{QA}(q, \mathbf{h}), a\bigr) \\
   &\quad - ROUGE_{recall}^{L}\bigl(LLM_{QA}(q), a\bigr)
\end{split}
\end{equation}

\noindent where $\textbf{h}$ is a chain-of-hints corresponding to question $q$ and answer $a$, and $LLM_{QA}(.)$ is the question answering LLM.

Ideally, the $LLM_{QA}$ model should portray three key characteristics to successfully evaluate the information gained by a chain-of-hints: i) it shouldn't memorize most of the answers (\textit{i.e.,} doesn't know the answer right away), ii) it should still have great reasoning capabilities to answer the question with the help of hints, and iii) should have low computational requirements. To select the appropriate LLM that satisfies these requirements, we conduct a study across the automatic evaluation data split\footnote{We use \texttt{gpt-3.5} \cite{brown2020language} and \texttt{gpt4-turbo} \cite{openai2023gpt} to generate the hints for this experiment.} using 16 models across five model families \texttt{Gemma-2} (2B, 9B) \citep{team2024gemma}, \texttt{Llama-3} (1B, 3B, 8B) \citep{grattafiori2024llama}, \texttt{Mistral} (7B) \cite{jiang2023mistral}, \texttt{Qwen2.5} and \texttt{Qwen2.5-Instruct} (0.5B, 1.5B, 3B, 7B) \cite{qwen2024qwen2}, and \texttt{Qwen3} (0.6B, 1.7B, 4B) \cite{yang2025qwen3}. We iteratively feed $LLM_{QA}$ the following prompt "\textit{Answer the following question succinctly:\textbackslash n Question: [question]\textbackslash n Hint 1: [hint\_1] ... Hint k: [hint\_k]\textbackslash n Answer:}"\footnote{We use multiple hints in the prompt when selecting a LLM for information gain to measure the reasoning capabilities of LLM to identify useful information across multiple hints}. We find that the \texttt{Gemma-2} models have a much higher information gain than other models for the same sequence of k hints across the 1000 instances of the automatic evaluation split. Since our human evaluation study is conducted with a chain of size 4, we select \texttt{Gemma-2-9b} as the LLM used to evaluate the information gain of a chain of hints. % \aj{these 30 hints were randomly extracted from our 80-hint per question dataset we developed for the RAG-based study. There's a slight incompleteness in the current description, which I'm uncertain of how to explain. I can write in passing we generated 30 hints using other model, or explain it in dataset section for completeness. These 80 hints are also the basis of reference-free evaluation of consistency.}

\noindent\textbf{Redundancy.}
Variation theory suggests that contrasting examples and perspectives help learners discern key features of a concept \cite{ling2012variation}. \citet{lomibao2017does}'s findings aligned with the theoretical argument of variation theory, the students exposed to repetition with variation to achieved significantly higher performance and improved knowledge retention in mathematical problem solving. To evaluate the  diversity of perspectives in a chain-of-hints, we measure its self-referenced redundancy  using the Sentence-BERT embeddings \cite{reimers2019sentence} as proposed by \citet{chen2021training} - 
\begin{equation}\label{eq:redundancy}
    Redundancy(\textbf{h}) = \sum_{i=1}^{k} max_{j:i\neq j} Cos_{sim}(h_i, h_j)
\end{equation}
where $j:i\neq j$ denotes that we don't consider similarity of hint j ($\tilde{h_j}$) with itself. % As we desire to maximize the content variation in a chain-of-hints, a 

\noindent\textbf{Consistency.}
To ensure the high quality of information within the hints, we measure the consistency of the hints using \texttt{AlignScore} \citep{zha2023alignscore}, a factual consistency evaluation metric that uses information alignment between context and claims to measure the factual consistency. % If a explanation ($expl$) if available, then we term this as reference-based consistency, otherwise we replace the $expl\rightarrow \mathbf{H} - \tilde{\textbf{h}}$ to obtain a reference-free consistency evaluation metric.

\begin{equation}\label{eq:consistency}
    Consistency(\textbf{h}) =
        AlignScore(context, context) % ; context \in \{expl, \mathbf{H} - \textbf{h}\}
\end{equation}

\noindent\textbf{Leakage.} 
Learners construct cognitive representations when they are guided to derive the solution themselves, molding them into \textit{deep conceptual thinkers} with the ability of question more, seeking to understand rather than to only memorize \cite{rillero2016deep}. However, revealing the correct answer prematurely disrupts this cognitive engagement, hindering their learning journey by oversimplifying the problem \cite{brown2014make}. To capture this, we adopt two metrics to evaluate answer leakage in generated hints - 1) Exact Match (EM), where answer is leaked if the exact answer string is present in the hint, and 2) LLM prompt (LLM), where we use the large language model's in-context learning capabilities to identify presence of answers in the hints, capturing more diverse types of leakages like paraphrased answers (\textit{e.g.,} "volcanic activity" instead of  "volcanoes") or abbreviations (\textit{e.g.,} "CO$_2$" instead of "carbon dioxide"). We use \texttt{Gemma3-27B} model for the \textit{Leakage (LLM)} evaluation metric (see Figure \ref{fig:leakage_prompt} for the exact prompt used by this metric).

% \aj{Where to report readability FK and FRE scores?}
\noindent\textbf{Readability.}
Hints should be expressed in a clear and simple language to ensure that the learners can easily grasp the underlying concept. Avoiding unnecessary complexity in conveying the critical problem solving strategy enhances the usefulness of hints, reinforcing the well-established principle of \textit{direct instructions} \cite{kozloff1999direct, kim2005direct, rosenshine2008five}. To grasp this linguistic clarity in the generated hints, 
we use three different readability measures - i) Flesch-Kincaid (FK) grade level readability score \citep{kincaid1975derivation}, ii) Flesch reading ease (FRE) \citep{flesch1979write}, and iii) Dale Chall (DC) readability score \citep{dale1948formula}. The Flesch-Kincaid grade level score and Flesch reading ease are scored using a linear combination of number of words per sentence and number of syllables per word, whereas the Dale Chall readability score utilizes a list of 769 words that 80\% of fourth-grade students are familiar with to provide a readability score. We implement these readability scores using \texttt{py-readability-metrics}\footnote{\url{https://github.com/cdimascio/py-readability-metrics}}. % \aj{Should I add their formulas here for sake of completeness? It's just polynomials over \#words, \#sentences and \#special words.}

\noindent\textbf{Aggregate Score.} In order to rank the baselines and finalize the best model for our human evaluation study, we formulate a simple aggregate score over the automatic evaluation metrics\footnote{We don't use readability for the aggregate function as all the hints generated by the baselines have similar readability scores, and generate well-formulated sentences.} as follows - 
% \begin{equation*}
%     Aggregate(\textbf{h}) = \frac{1}{4} \{\begin{split}
%         InfoGain_{comb}(\textbf{h}) + Consistency(\textbf{h}) \\
%         + (1-Redundancy(\textbf{h}) + (1 - leakage_{EM}(\textbf{h})
%     \end{split}
%     \}
% \end{equation*}
\begin{equation*}
Aggregate(\mathbf{h}) = \tfrac{1}{4} \left\{
\begin{split}
   & InfoGain_{comb}(\mathbf{h}) + Consistency(\mathbf{h}) \\
   & + \bigl(1 - Redundancy(\mathbf{h})\bigr) 
     + \bigl(1 - leakage_{EM}(\mathbf{h})\bigr)
\end{split}
\right\}
\end{equation*}

% \aj{Should I make use of the informativeness function from previous idea to see how much previous hints build on towards the next hint? How aligned they are. Will think about it once we have more annotation data.}

\begin{table*}[t]
\caption{Automatic evaluation metric results for \textit{static} hint generation setting.} \label{tab:static_autoeval}
\resizebox{\textwidth}{!}{%
\begin{tabular}{lcccccccc}
\toprule
\textbf{Baseline} & $InfoGain_{mean} \uparrow$ & $InfoGain_{comb} \uparrow$ & $Redundancy \downarrow$ & $Consistency \uparrow$ & $Readability_{DC} \downarrow$ & $Leakage_{EM} \downarrow$ & $Leakage_{LLM} \downarrow$ & $Aggregate \uparrow$ \\
\midrule
\texttt{DeepSeek-R1-1.5b} & 0.214 & 0.421 & 0.625 & 0.199 & 10.810 & 0.369 & 0.859 & 0.406 \\
\texttt{DeepSeek-R1-7b} & 0.206 & 0.434 & 0.641 & 0.334 & 10.409 & 0.220 & 0.543 & 0.476 \\
\texttt{DeepSeek-R1-8b} & 0.161 & 0.418 & 0.572 & 0.302 & 10.152 & 0.070 & 0.557 & 0.520 \\
\texttt{DeepSeek-R1-14b} & 0.121 & 0.423 & 0.563 & 0.396 & 9.514 & 0.045 & 0.449 & 0.553 \\
\texttt{DeepSeek-R1-32b} & 0.148 & 0.450 & 0.581 & 0.420 & 10.183 & 0.087 & 0.472 & 0.550 \\ \hline

\texttt{Gemma3-1b} & 0.119 & 0.369 & 0.529 & 0.298 & 10.929 & 0.053 & 0.170 & 0.521 \\
\texttt{Gemma3-4b} & 0.122 & 0.419 & 0.572 & 0.395 & 10.139 & 0.054 & 0.417 & 0.547 \\
\texttt{Gemma3-12b} & 0.047 & 0.393 & 0.524 & 0.382 & 9.959 & 0.019 & 0.337 & 0.558 \\
\texttt{Gemma3-27b} & 0.028 & 0.354 & 0.539 & 0.332 & 10.070 & 0.011 & 0.479 & 0.534 \\ \hline

\texttt{Qwen3-0.6b} & 0.234 & 0.513 & 0.741 & 0.450 & 10.173 & 0.573 & 0.972 & 0.412 \\
\texttt{Qwen3-1.7b} & 0.198 & 0.484 & 0.681 & 0.356 & 10.163 & 0.329 & 0.792 & 0.458 \\
\texttt{Qwen3-4b} & 0.190 & 0.470 & 0.634 & 0.383 & 9.438 & 0.232 & 0.729 & 0.497 \\
\texttt{Qwen3-8b} & 0.168 & 0.473 & 0.613 & 0.393 & 9.508 & 0.178 & 0.628 & 0.519 \\
\texttt{Qwen3-14b} & 0.126 & 0.416 & 0.552 & 0.307 & 9.455 & 0.043 & 0.491 & 0.532 \\
\texttt{Qwen3-30b} & 0.131 & 0.414 & 0.559 & 0.341 & 9.905 & 0.024 & 0.463 & 0.543 \\
\texttt{Qwen3-32b} & 0.138 & 0.436 & 0.577 & 0.354 & 9.454 & 0.034 & 0.664 & 0.544 \\ \hline

\texttt{Mistral-24b} & 0.103 & 0.429 & 0.544 & 0.408 & 9.558 & 0.022 & 0.382 & \textbf{0.568} \\ \hline

\texttt{Phi4-14b} & 0.191 & 0.441 & 0.592 & 0.337 & 10.555 & 0.046 & 0.530 & 0.535 \\
\bottomrule
\end{tabular}
}
\vspace{1em}
\caption{Automatic evaluation metric results for \textit{dynamic} hint generation setting.} \label{tab:dynamic_autoeval}
\resizebox{\textwidth}{!}{%
\begin{tabular}{lcccccccc}
% \vspace{2em}
\toprule
\textbf{Baseline} & $InfoGain_{mean} \uparrow$ & $InfoGain_{comb} \uparrow$ & $Redundancy \downarrow$ & $Consistency \uparrow$ & $Readability_{DC} \downarrow$ & $Leakage_{EM} \downarrow$ & $Leakage_{LLM} \downarrow$ & $Aggregate \uparrow$ \\
\midrule
\texttt{DeepSeek-R1-1.5b} & 0.358 & 0.504 & 0.833 & 0.388 & 10.439 & 0.723 & 0.998 & 0.334 \\
\texttt{DeepSeek-R1-7b} & 0.371 & 0.508 & 0.891 & 0.463 & 10.555 & 0.707 & 0.998 & 0.343 \\
\texttt{DeepSeek-R1-8b} & 0.172 & 0.425 & 0.713 & 0.256 & 10.480 & 0.121 & 0.847 & 0.462 \\
\texttt{DeepSeek-R1-14b} & 0.201 & 0.447 & 0.777 & 0.375 & 10.067 & 0.172 & 0.876 & 0.468 \\
\texttt{DeepSeek-R1-32b} & 0.251 & 0.467 & 0.769 & 0.364 & 10.133 & 0.241 & 0.888 & 0.455 \\ \hline

\texttt{Gemma3-1b} & 0.205 & 0.395 & 0.744 & 0.367 & 10.995 & 0.165 & 0.517 & 0.464 \\
\texttt{Gemma3-4b} & 0.134 & 0.392 & 0.716 & 0.392 & 10.471 & 0.064 & 0.303 & 0.501 \\
\texttt{Gemma3-12b} & 0.093 & 0.413 & 0.648 & 0.336 & 10.097 & 0.063 & 0.245 & 0.509 \\
\texttt{Gemma3-27b} & 0.109 & 0.402 & 0.670 & 0.271 & 10.159 & 0.041 & 0.316 & 0.491 \\ \hline

\texttt{Qwen3-0.6b} & 0.270 & 0.504 & 0.975 & 0.491 & 10.881 & 0.872 & 0.987 & 0.287 \\
\texttt{Qwen3-1.7b} & 0.372 & 0.507 & 0.925 & 0.513 & 10.330 & 0.743 & 0.987 & 0.338 \\
\texttt{Qwen3-4b} & 0.340 & 0.505 & 0.890 & 0.426 & 10.230 & 0.493 & 0.979 & 0.387 \\
\texttt{Qwen3-8b} & 0.223 & 0.487 & 0.828 & 0.439 & 10.507 & 0.272 & 0.861 & 0.457 \\
\texttt{Qwen3-14b} & 0.268 & 0.469 & 0.774 & 0.366 & 10.557 & 0.087 & 0.892 & 0.493 \\
\texttt{Qwen3-30b} & 0.200 & 0.453 & 0.721 & 0.365 & 10.077 & 0.051 & 0.641 & 0.512 \\
\texttt{Qwen3-32b} & 0.263 & 0.472 & 0.791 & 0.362 & 10.614 & 0.095 & 0.955 & 0.487 \\ \hline

\texttt{Mistral-24b} & 0.158 & 0.437 & 0.665 & 0.384 & 9.581 & 0.036 & 0.357 & \textbf{0.530} \\ \hline

\texttt{Phi4-14b} & 0.426 & 0.501 & 0.811 & 0.186 & 10.861 & 0.523 & 0.898 & 0.338 \\
\bottomrule
\end{tabular}
}

\end{table*}

\subsection{Automatic Evaluation Results} \label{subsec:autoeval_results}

We present the performance of all 18 baseline models for \textit{static} and \textit{dynamic} hint generation at Table \ref{tab:static_autoeval} and \ref{tab:dynamic_autoeval} respectively. In this section, we discuss our observations from this automatic evaluation process, and finalize a baseline for the subsequent human evaluation user study. 

\begin{figure*}[h]
  \begin{subfigure}[]{0.48\textwidth}
    \centering
    \includegraphics[width=\textwidth]{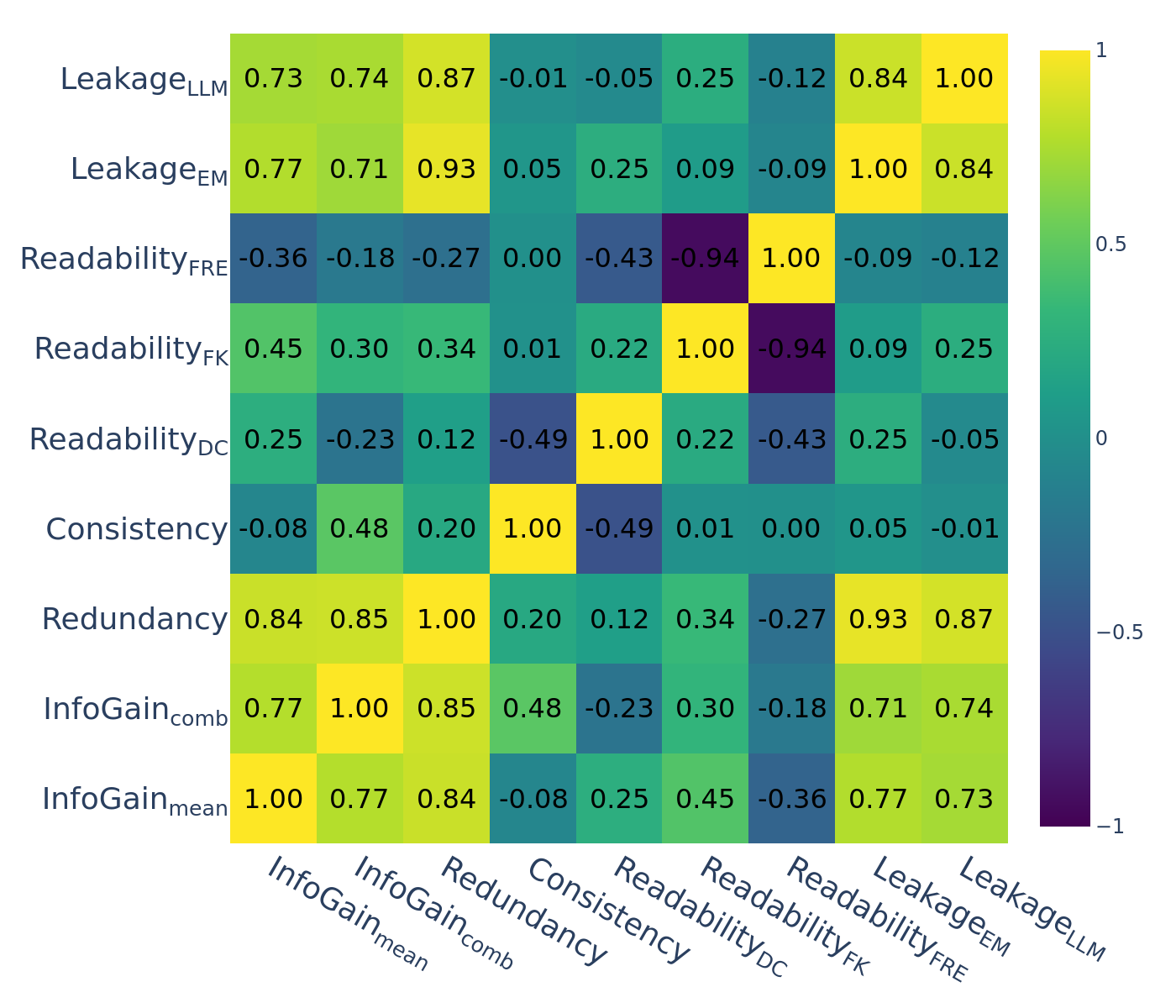}
  \end{subfigure}
  % \hfill
  \begin{subfigure}[]{0.48\textwidth}
    \centering
    \includegraphics[width=\textwidth]{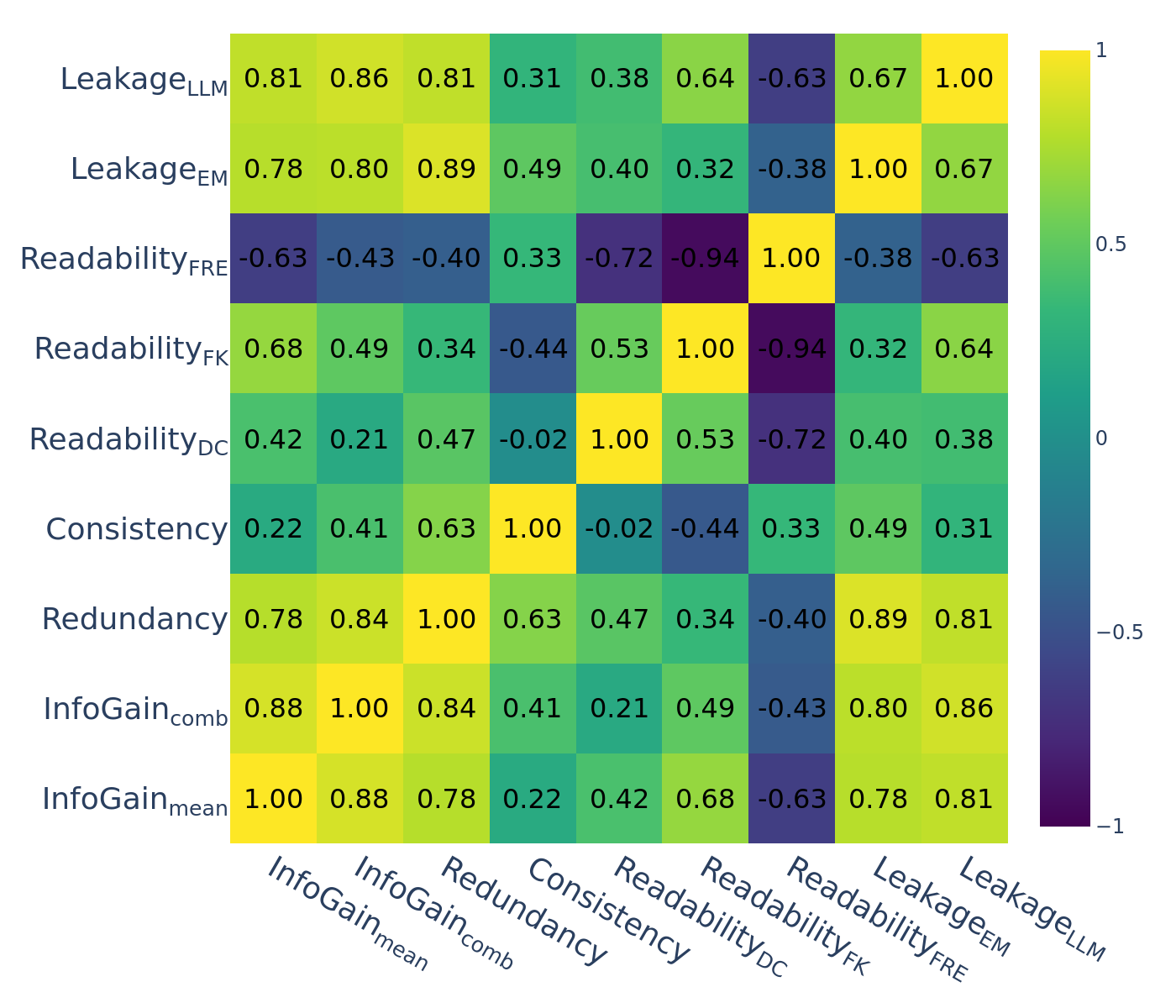}
  \end{subfigure}
  \caption{Pearson's correlation matrix of automatic evaluation metrics for \textit{static} (left) and \textit{dynamic} (right) hint generation.}
  \label{fig:autoeval_corr}
\end{figure*}

\begin{figure*}[t]
  \begin{subfigure}[]{0.4\textwidth}
    \centering
    \includegraphics[width=\textwidth]{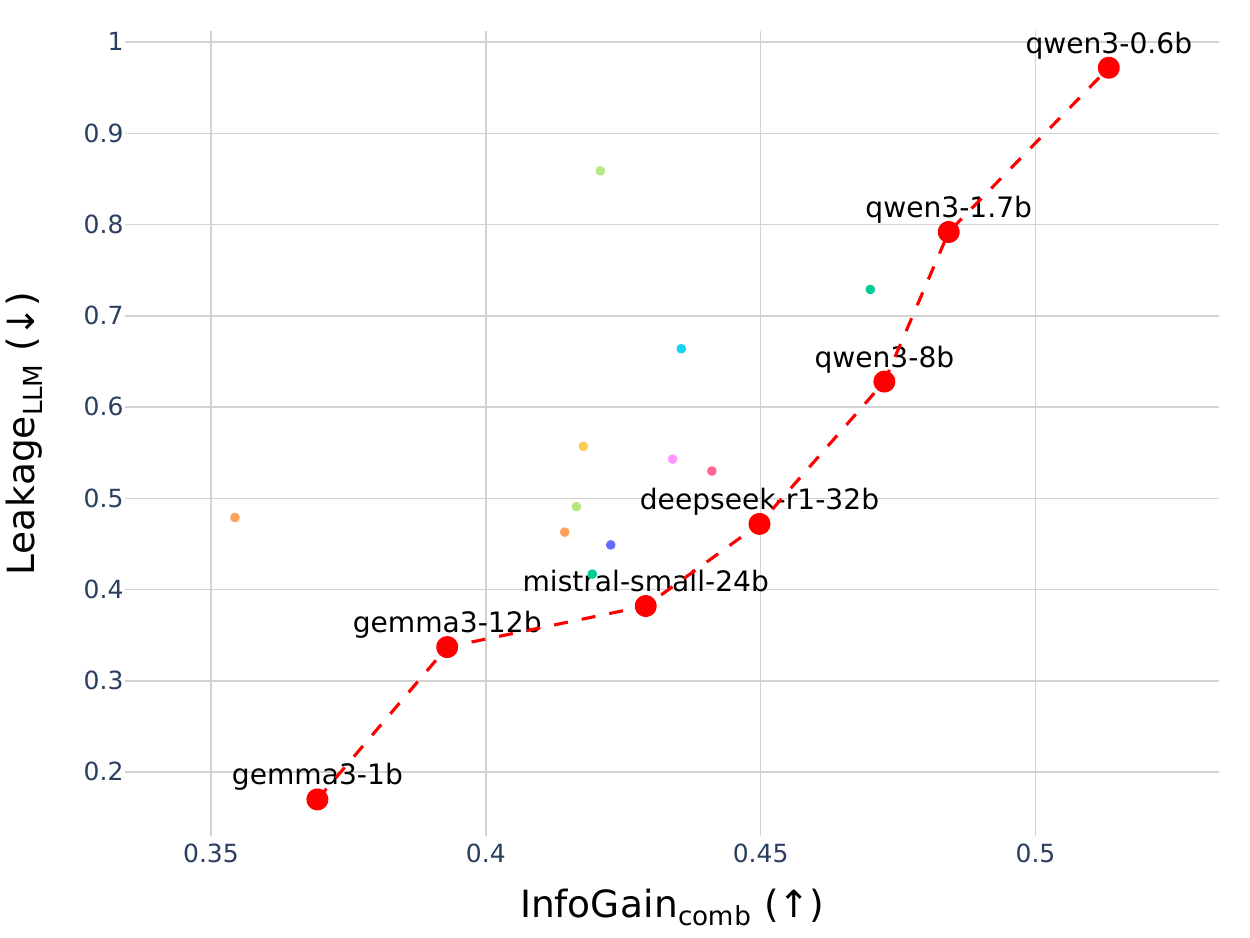}
  \end{subfigure}
  % \hfill
  \begin{subfigure}[]{0.4\textwidth}
    \centering
    \includegraphics[width=\textwidth]{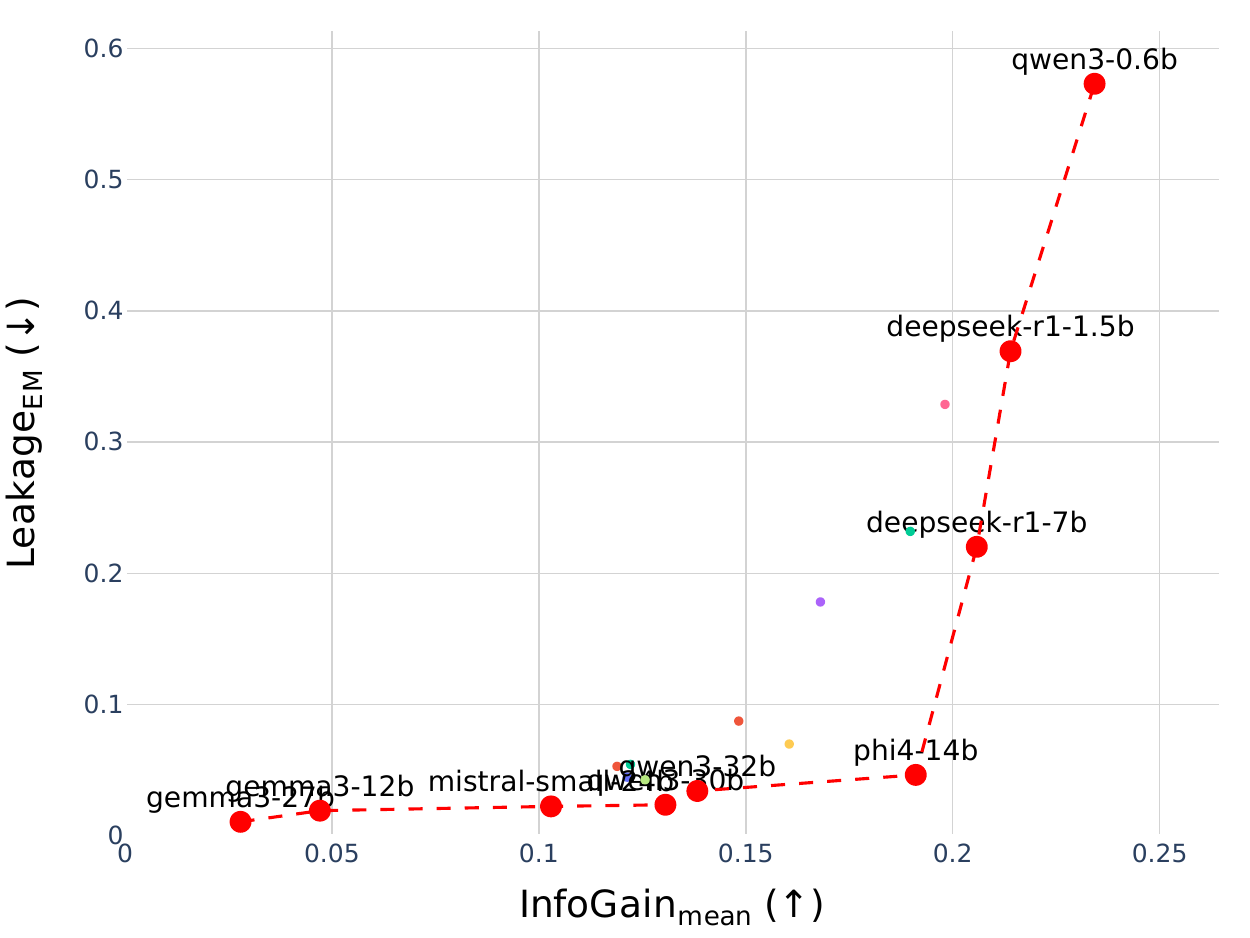}
  \end{subfigure}
  % \hfill
  \begin{subfigure}[]{0.4\textwidth}
    \centering
    \includegraphics[width=\textwidth]{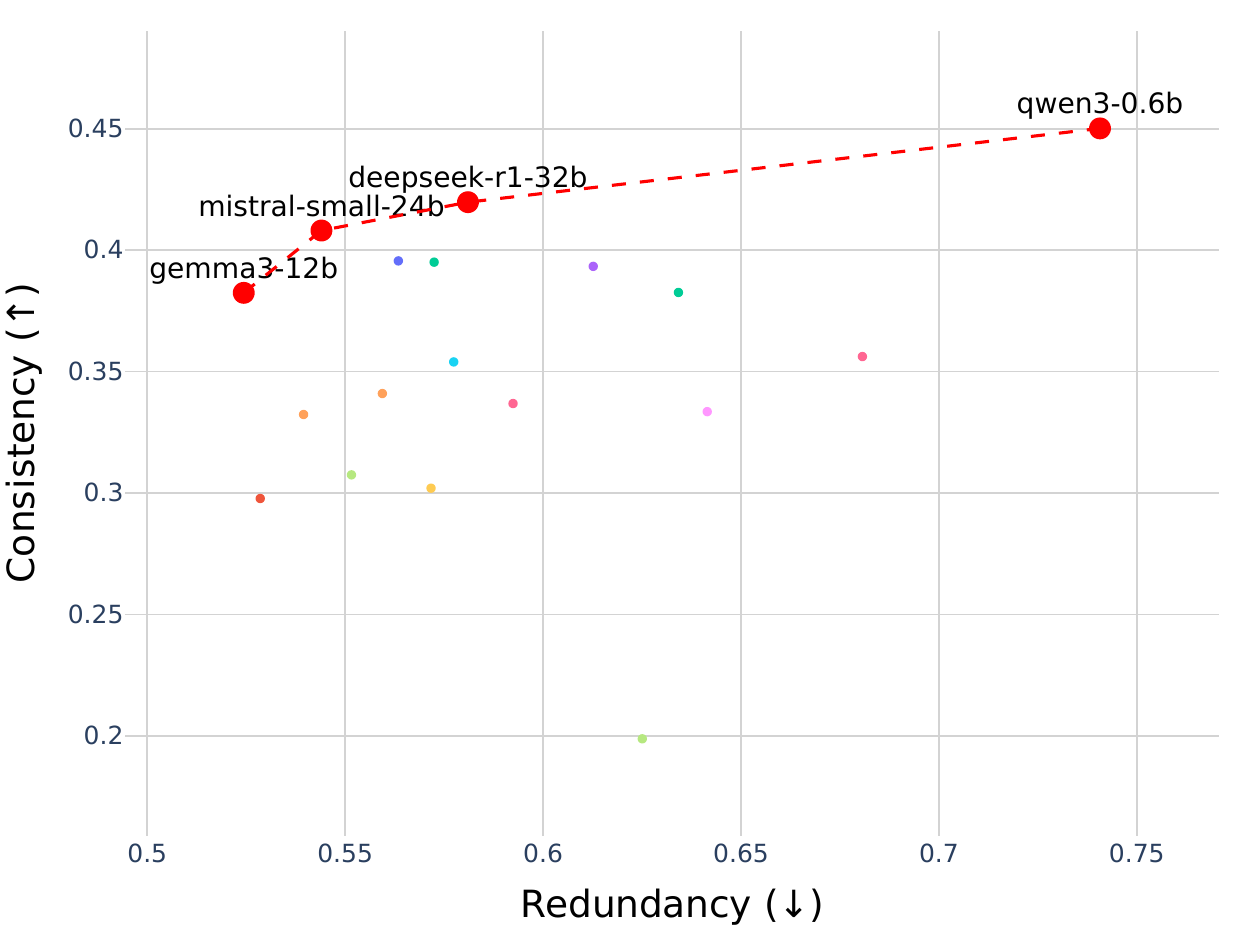}
  \end{subfigure}
  % \hfill
  \begin{subfigure}[]{0.4\textwidth}
    \centering
    \includegraphics[width=\textwidth]{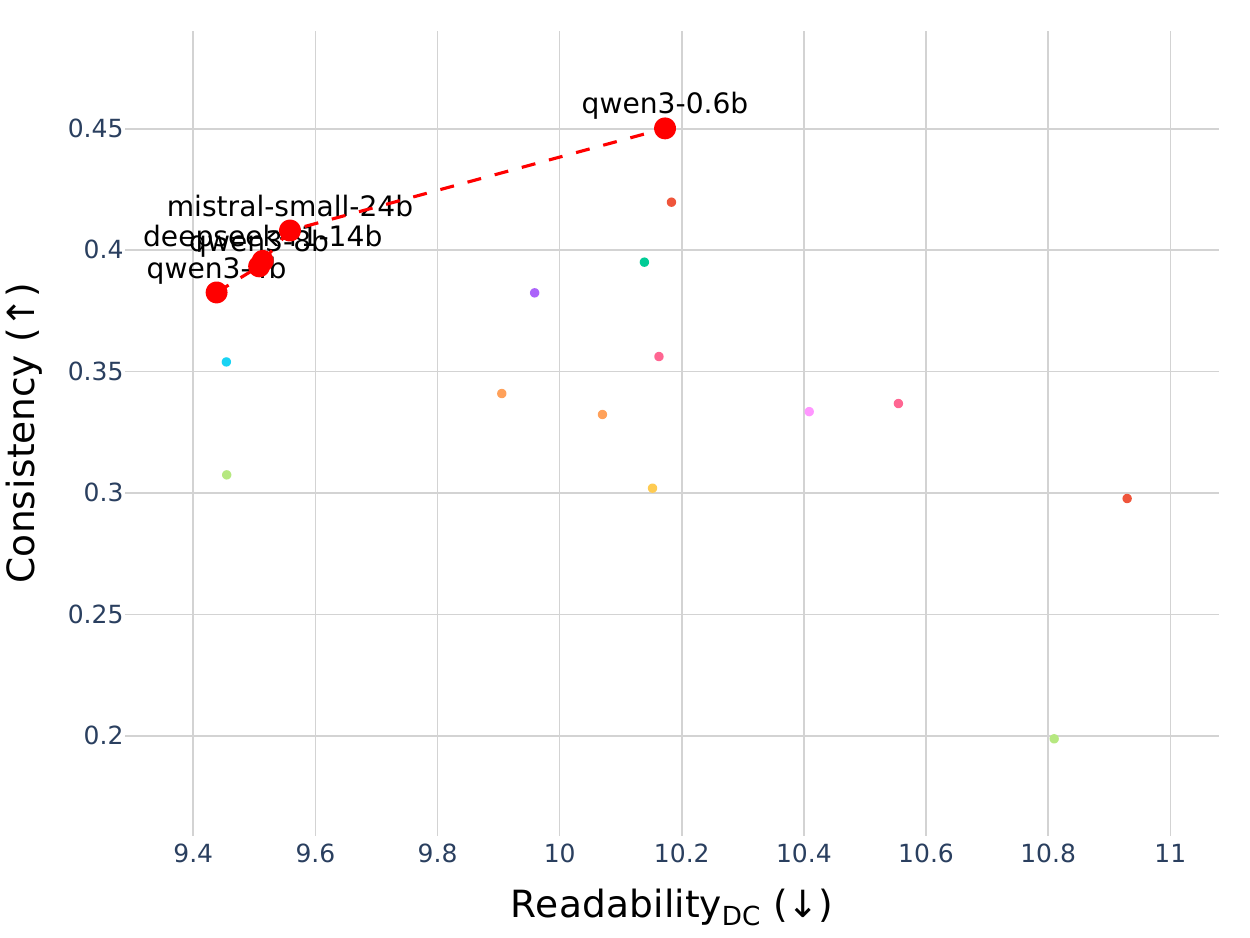}
  \end{subfigure}
  \caption{Pareto-optimal fronts for static hint generation.}
  \label{fig:pareto}
\end{figure*}

\noindent\textbf{Trends across evaluation metrics.} We compute the Pearson's correlation across the automatic evaluation metrics to identify several  trends over the baselines (refer to Figure \ref{fig:autoeval_corr} for the correlation matrices). We observe that the variants of evaluation metrics are highly correlated with each other. \textit{E.g.,} for $InfoGain_{mean}$ and $InfoGain_{comb}$ is 0.77, $Readability_{FRE}$ and $Readability_{FK}$ is -0.94 for static hint generation. $InfoGain$ and $Leakage$ are also heavily correlated ($corr>0.7$), understandably so, since the baselines with tendency to leak the answers oversimplifies the problems for the $InfoGain$ evaluator LM. This is further evident from the pareto-optimal curves between the two metrics (refer to Figure \ref{fig:pareto} top-left and top-right), with a diverse range of models at the pareto-front, with the smaller models with high leakage or low information gain at the extreme ends, and optimal baselines at the elbow of the curve. We also observe redundancy to be highly correlated with both information gain and leakage ($corr>0.8$), likely because the baselines that tend to leak answer lack semantic diversity in their hints, leading to high redundancy along with higher information gain. \looseness=-1

Although $Consistency$ is also a semantic evaluation metric, it doesn't correlate well with other semantic metrics, except $InfoGain_{comb}$ ($corr=0.48$) in \textit{static} setting. We posit this to be likely because $Consistency$ is the only reference-based evaluation metric, aiming to evaluate the factual correctness of the generated hints compared to the context. As $InfoGain_{comb}$ indirectly measures the semantic information captured by a chain-of-hints towards helping answer the question, the moderate correlation between the two metrics is justified. For \textit{dynamic} setting on the other hand, $Consistency$ is correlated with $Redundancy$ ($corr=0.63$) and $Leakage_{EM}$ ($corr=0.49$) as well, most likely because of significantly higher leakage rates and redundant hints compared to the \textit{static} hint generation setting.

We observe limited variability across all three variants of the readability scores, with the results roughly interpretable to be between high school (8.0-8.9) to college (9.0+) level according to the Dale-Chall (high school: 8.0-8.9, college: 9.0-9.9), Flesch reading ease (high-school: 60-50, college: 50-30), and Flesch-Kincaid (high school: 9-12, college: 12-15). Although all readability results follow similar interpretation, we observe that $Readbility_{DC}$ is moderately correlated to the other two metric, likely due to its contrasting strategy of using the ratio of difficult words instead. Being a stylistic measure, readability scores don't correlate well with the other semantic metrics.

\noindent\textbf{Trends across \textit{static} and \textit{dynamic} hints}
We observe some consistent differences in scores between the static and dynamic hints across all baselines. \textit{Static} hints consistently outperformed \textit{dynamic} hints (($\Delta Aggregate=-0.078$)). This was because \textit{dynamic} hints performed worse than \textit{static} hints in all evaluation metrics except readability. Compared to \textit{static} hint generation, the \textit{dynamic} hints had increased redundancy ($\Delta Redundancy = 0.195$), increased answer leakage ($\Delta Leakage_{EM} = 0.169, \Delta Leakage_{LLM} = 0.201$), leading to slightly increased information gain ($\Delta InfoGain_{ind}=0.098$, $\Delta InfoGain_{comb}=0.029$), and marginal change in consistency ($\Delta Consistency=0.019$). We posit the poor quality of distractors used to steer the generation of \textit{dynamic} hints to be the leading cause of this worsened performance. For example, for the question "\textit{What kind of waves are sound waves?}", the dataset provides the distractors "\textit{spinning}", "\textit{external}", and "\textit{internal}" against the correct answer "\textit{mechanical}". Providing pseudo relevant attempted answers forced the baselines to either repeat their point across multiple hints, or reveal the answers directly or indirectly. While using these distractors as attempted might not perfectly represent a learner's attempts, they still act as a useful tool to identify the robustness of model response across different settings, helping us identify models capable of providing useful feedback for real-world adoptability.

% \begin{itemize}
%     \item dynamic hints tended to leak a lot more information than static hints, probably due to poor quality of distractors with irrelevant attempted answers (e.g., ) forcing the models to reveal the answers directly or indirectly. 
%     \item This leak in information leads to higher information gain for dynamic hints.
%     \item Dynamic hints also reiterate the same point multiple times compared to static hints generated in a one-shot manner, leading to a higher degree of redundancy. 
%     \item While consistency of dynamic hints is higher than the static hints in smaller models (below $<$4b parameters), the consistency generally decreases in larger models. This is probably because of poorer long-form generation capabilities of the smaller models \todo{cite a work that shows long-form generation improves with scale}, that are better at generating one hint at a time than multiple hints. 
%     \item While in the same bracket of high-school to college level readability, the $Readability_{DC}$ metric scores are slightly higher for the dynamic hints than static hints across most baselines (about 0.5 point difference) .
% \end{itemize}

\begin{figure*}[]
    \begin{subfigure}[]{0.8\textwidth}
    \centering
    \includegraphics[width=\textwidth]{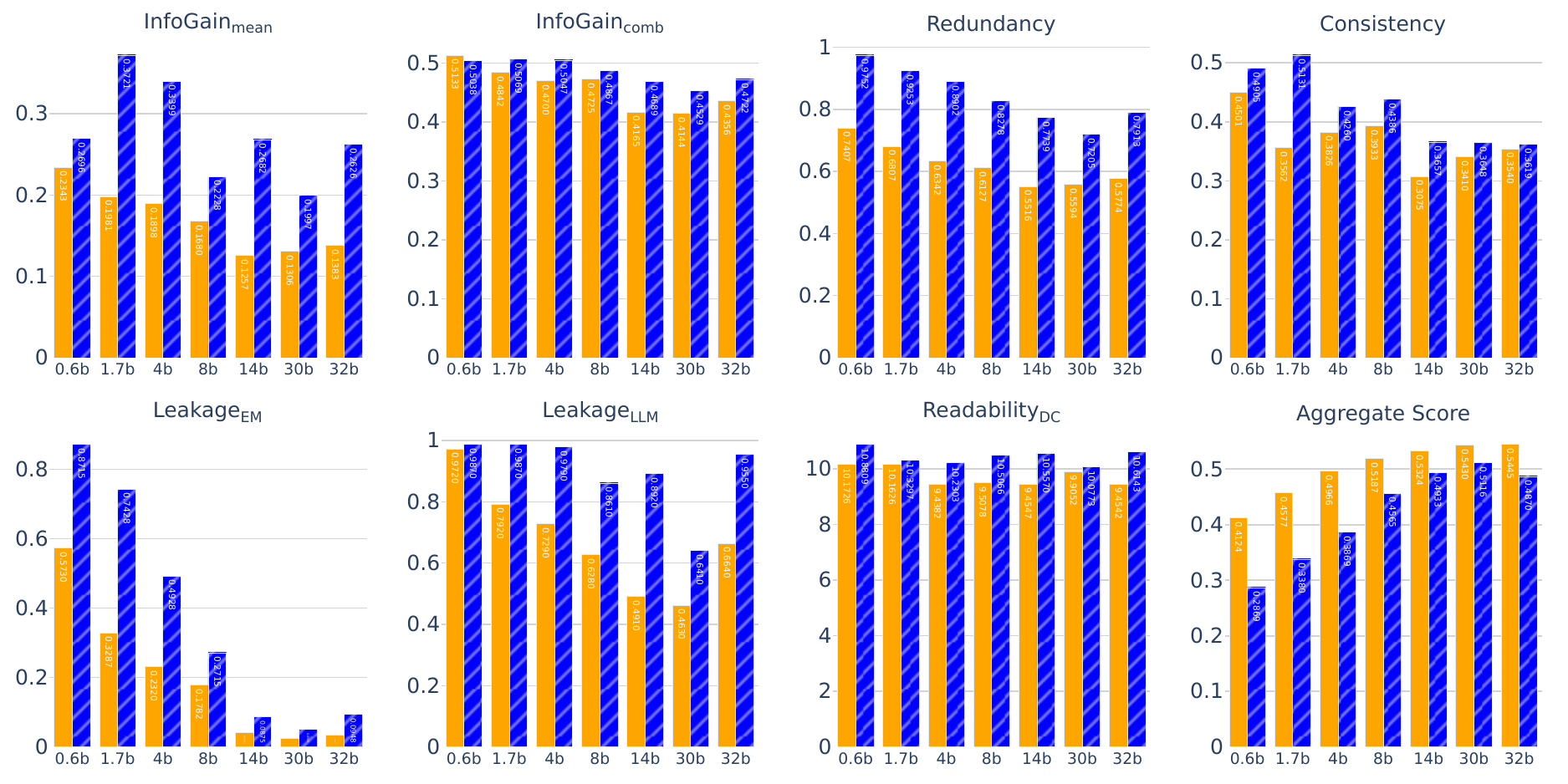}
  \end{subfigure}
  \noindent\makebox[\linewidth]{\rule{0.82\textwidth}{0.4pt}}
  \begin{subfigure}[]{0.8\textwidth}
    \centering
    \includegraphics[width=\textwidth]{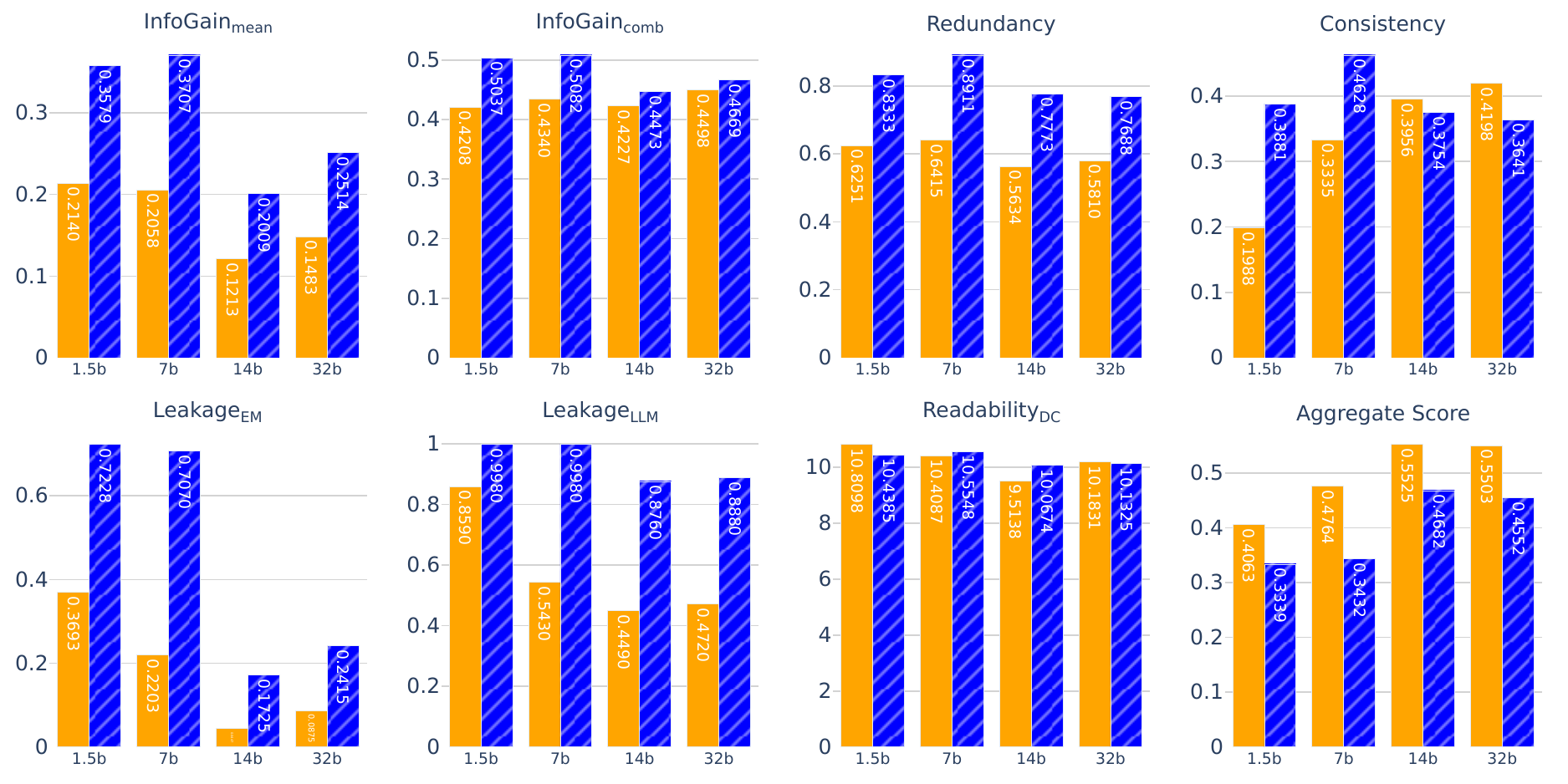}
  \end{subfigure}
  % \hfill
  \noindent\makebox[\linewidth]{\rule{0.82\textwidth}{0.4pt}}
  \begin{subfigure}[]{0.8\textwidth}
    \centering
    \includegraphics[width=\textwidth]{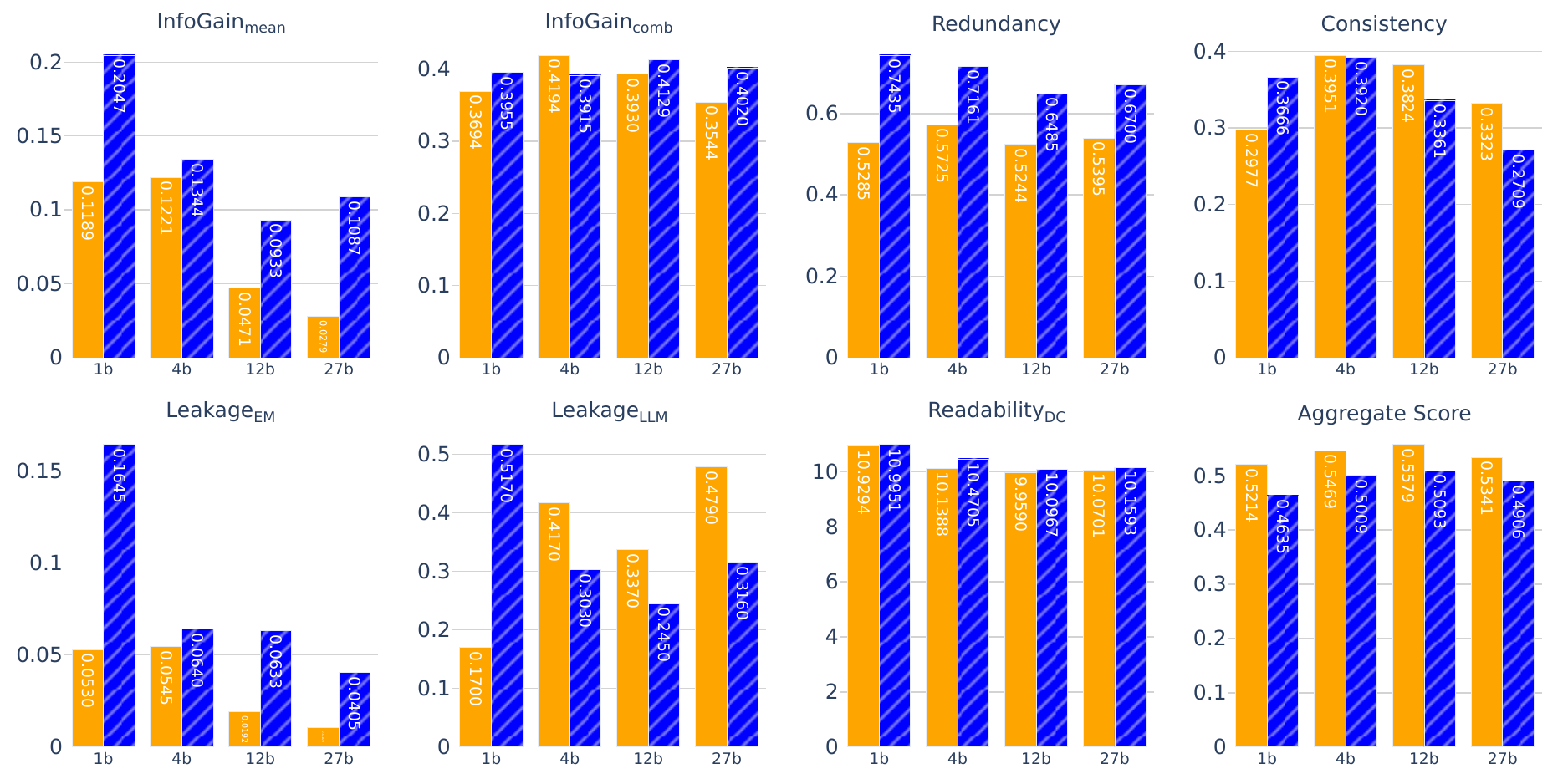}
  \end{subfigure}
  \caption{Performance of \texttt{Qwen3} (top), \texttt{DeepSeek-R1} (middle), and \texttt{Gemma3} (bottom) model families using automatic evaluation metrics. \textcolor{orange}{Orange} (solid) bars denote the static hint generation results and \textcolor{blue}{blue} (striped) bars denote the dynamic hint generation results.}
  \label{fig:scale_autoeval_app}
\end{figure*}

\noindent\textbf{Trends across model size.} To formulate guidelines on what models to use for future research, we explore trends over three model families with multiple variants to observe how their hint generation capabilities varies across scale. We present the results for the three model families \texttt{Qwen3} \cite{yang2025qwen3}, \texttt{DeepSeek-R1} \cite{guo2025deepseek} and \texttt{Gemma3} \cite{team2025gemma} in Figure \ref{fig:scale_autoeval_app}. We observe that the aggregate score increases with the model size, with \textit{static} hints outperforming \textit{dynamic} hints. This gap in performance decreases with the increase in number of parameters, indicating that the larger models are more capable at generating adaptive hints. The biggest contributor to this decrease in performance can be attributed to the answer leakage, where smaller models ($<$4b parameters) have the exact answer string for over 20\% of instances for \texttt{Qwen3} and \texttt{DeepSeek-R1} models. This leakage oversimplifies the problem, leading to an observable difference in the $InfoGain_{mean}$ metric, which measures the average information gain across a chain-of-hints. This effect is not as evident in the $InfoGain_{comb}$ metric, likely because it more holistically gauges the information gain of the entire chain, which even for larger models with less leakage, provide ample information to help the information gain evaluator LM. Smaller models are also not very expressive, with more redundant hint sequences compared to larger models, that provide more diverse perspectives due to their better memorization and reasoning capabilities. While there's a definitive trend of improved hint quality with an increase in size, for a simpler task of factual scientific question answering, the performance plateaus after a certain size, with models between 12-14b parameters performing sufficiently well off-the-shelf. For a resource constrained domain of education, we believe further finetuning these medium size models to be the most pragmatic next direction. % \aj{need to rephrase last 2-3 lines, especially as we choose 24b model in next passage, kind of contradicting. But maybe the post-processing aspect for live experimentation is enough justification? Thoughts?}

% \noindent\underline{Selecting a baseline for human evaluation study}. We select \texttt{Mistral-Small-24b} model to conduct the human evaluation study for two main reasons - i) it achieved the highest aggregate score in both static and dynamic hint generation settings, and ii) due to it's great instruction following capabilities, the hints generated from \texttt{Mistral-Small-24b} followed our desired structure in the output with minimal post-processing, an important property for the human evaluation study with online hint generation for the dynamic hint generation setting.
% \newpage
\section{Prompts} \label{app:prompts}

\begin{figure*}[t!]
\centering
\includegraphics[width=0.7\textwidth]{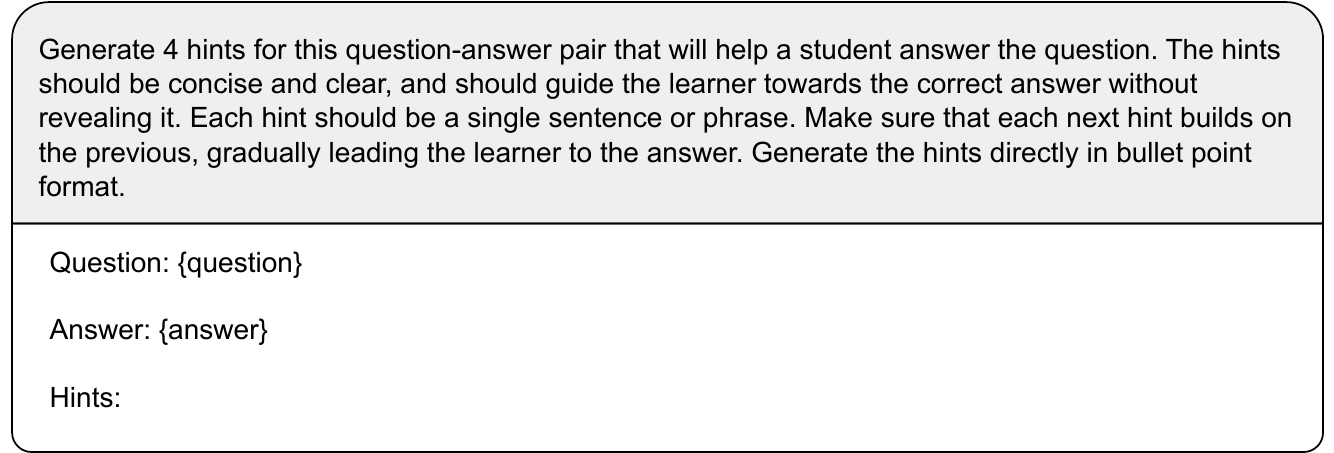}
\caption{Prompt used to generate static hints.} \label{fig:static_prompt}
\end{figure*}

\begin{figure*}[]
\centering
\includegraphics[width=0.7\textwidth]{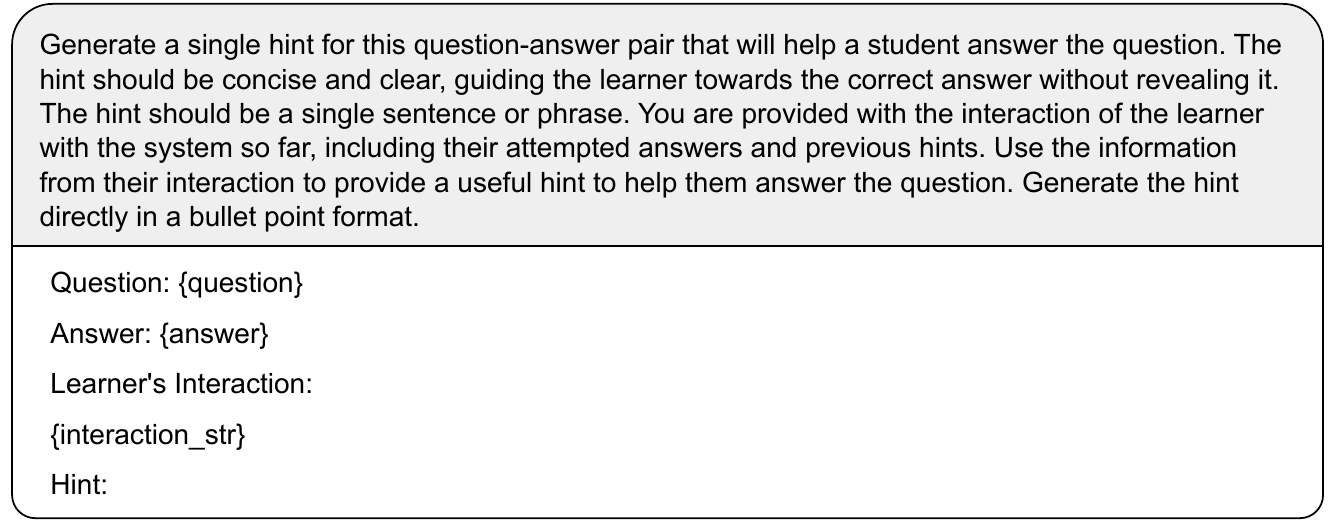}
\caption{Prompt used to generate dynamic hints.} \label{fig:dynamic_prompt}
\end{figure*}

\begin{figure*}[]
\centering
\includegraphics[width=0.7\textwidth]{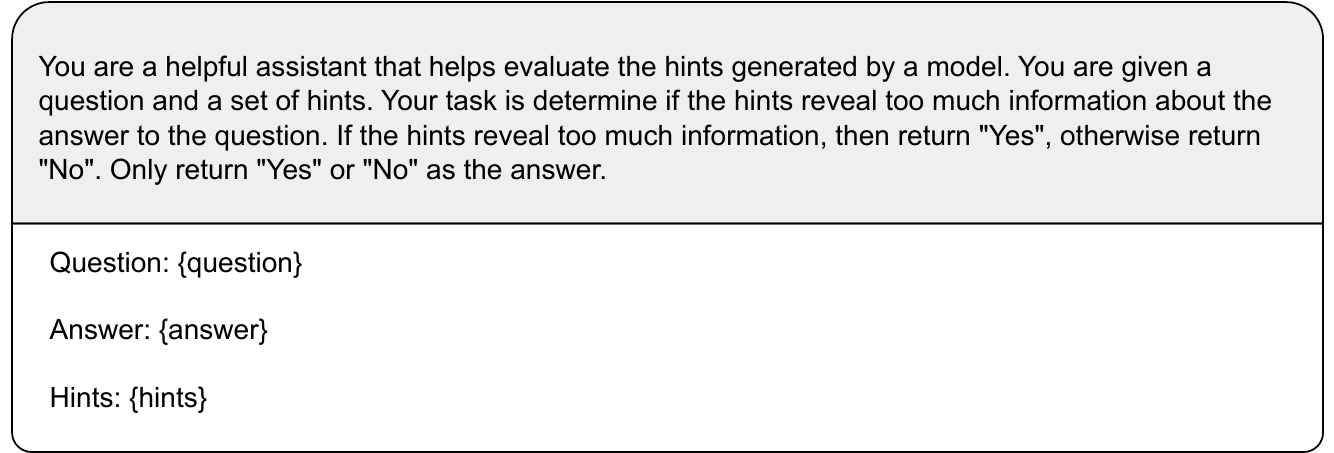}
\caption{Prompt used for $Leakage_{LLM}$ evaluation metric.} \label{fig:leakage_prompt}
\end{figure*}

\begin{figure*}[]
\centering
\includegraphics[width=0.7\textwidth]{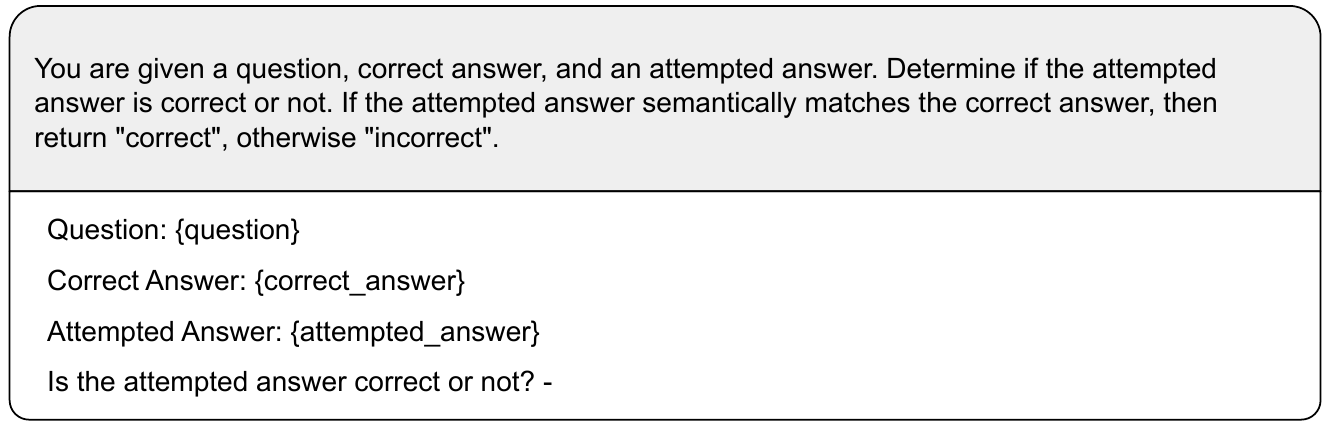}
\caption{Prompt used for answer assessment in our user interface backend.} \label{fig:assessment_prompt}
\end{figure*}

We provide the prompts we used throughout our study in this section. Figures \ref{fig:static_prompt} and \ref{fig:dynamic_prompt} describe the static and dynamic hint generation prompts we used for both automatic evaluation benchmarking and the human evaluation study. Figure \ref{fig:leakage_prompt} contains the prompt we used to measure the LLM-based answer leakage in $leakage_{LLM}$ evaluation metric (using \texttt{Gemma3-27B} model \cite{team2025gemma}). Figure \ref{fig:assessment_prompt} comprises of the answer assessment prompt we used to resolve fuzzy matching of participants' attempted answer with the ground truth answer (using OpenAI's \texttt{o4-mini} model). The answer assessment was preceded by an exact string match to increase the assessment performance due to the non-deterministic nature of LLM responses. While the figures present the prompt in two parts, a system prompt and the user prompt, in practice, we treated them as a single string to ensure a fair comparison across all models. We leave the finetuning of these prompts to future works.
\clearpage

\end{document}